\documentclass[11pt,a4paper,notitlepage]{article}
\usepackage[utf8]{inputenc}
\usepackage[english]{babel}
\usepackage[T1]{fontenc}
\usepackage{hyperref}
\usepackage{amsmath}
\usepackage{amsfonts}
\usepackage{color} 
\usepackage{caption}
\newcommand{\Tr}{\mathrm{Tr}}

\usepackage{enumerate}
\usepackage{subfig}

\definecolor{mygray}{gray}{0.3}

\newcommand\beq{\begin{equation}}
\newcommand\eeq{\end{equation}}
\newcommand{\bes}{\begin{eqnarray}}
\newcommand{\ees}{\end{eqnarray}}
\def\nn{{\nonumber}}
\newcommand{\one}{\mbox{$1 \hspace{-1.0mm}  {\bf l}$}}

\def\psib{\overline{{\psi}}}
\def\vphi{{\varphi}}
\def\vphib{\overline{{\varphi}}}
\newcommand{\su}{\mathfrak{su}}

\newcommand{\cC}{{\mathcal C}}
\newcommand{\cK}{{\mathcal K}}
\newcommand{\cG}{{\mathcal G}}
\newcommand{\cZ}{{\mathcal Z}}
\newcommand{\cW}{{\mathcal W}}
\newcommand{\cX}{{\mathcal X}}
\newcommand{\cY}{{\mathcal Y}}
\newcommand\restr[2]{{
  \left.\kern-\nulldelimiterspace 
  #1 
  \vphantom{\big|} 
  \right|_{#2} 
  }}
\def\extd{\mathrm {d}}

\newcommand{\SU}{\mathrm{SU}}
\usepackage{multicol}
\usepackage{amssymb}
\usepackage{graphicx}
\usepackage[left=2cm,right=2cm,top=2.5cm,bottom=2.5cm]{geometry}
\begin{document}
\begin{center}
\textbf{\Large{
Asymptotic safety in three-dimensional SU(2) Group Field Theory: evidence in the local potential approximation 
}}
\vspace{15pt}

{\large Sylvain Carrozza$^{a,}$\footnote{\url{scarrozza@perimeterinstitute.ca}} and Vincent Lahoche$^{b,}$\footnote{\url{vincent.lahoche@labri.fr}}}

\vspace{10pt}

$^{a}${\sl Perimeter Institute for Theoretical Physics\\
 31 Caroline St N, Waterloo, ON N2L 2Y5, Canada\\
}
\vspace{3pt}

$^{b}${\sl LaBRI, Univ. Bordeaux\\
351 cours de la Lib\'{e}ration, 33405 Talence, France, EU\\
}

\end{center}

\vspace{5pt}

\begin{abstract}
\noindent We study the functional renormalization group of a three-dimensional tensorial Group Field Theory (GFT) with gauge group $\mathrm{SU}(2)$. This model generates (generalized) lattice gauge theory amplitudes, and is known to be perturbatively renormalizable up to order $6$ melonic interactions. 
We consider a series of truncations of the exact Wetterich--Morris equation, which retain increasingly many perturbatively irrelevant melonic interactions
. This tensorial analogue of the ordinary local potential approximation allows to investigate the existence of non-perturbative fixed points of the renormalization group flow. 
Our main finding is a candidate ultraviolet fixed point, whose qualitative features are reproduced in all the truncations we have checked (with up to order $12$ interactions). This may be taken as evidence 
for an ultraviolet completion of this GFT in the sense of asymptotic safety. Moreover, this fixed point has a single relevant direction, which suggests the presence of two distinct infrared phases. Our results generally support the existence of GFT phases of the condensate type, which have recently been conjectured and applied to quantum cosmology and black holes.
\end{abstract}

\setcounter{tocdepth}{2}
\tableofcontents
\pagebreak

\section{Introduction}

Group Field Theories (GFTs) \cite{Freidel:2005qe, daniele_rev2006, daniele_rev2011, Baratin:2011aa, Krajewski:2012aw} are quantum field theories defined on group manifolds, which have mainly been developed in the context of background-independent approaches to quantum gravity. Like tensor models \cite{Ambjorn:1990ge, Sasakura:1990fs, Gross:1991hx, Gurau:2011xp, Gurau:2016cjo, razvan_book}, they are characterized by a peculiar combinatorial structure of the interactions, which has the effect of upgrading the Feynman diagrams of the ordinary quantum field theory perturbative expansion to more general cellular complexes. As a result, both GFTs and tensor models can be thought of as generalizations of matrix models \cite{DiFrancesco:1993cyw, DiFrancesco:2004qj}, and both aim at reproducing the successes of the latter in dimension $d>2$, including quantum gravity applications.


Tensor models and GFTs have witnessed a wave of new results and applications, following the pioneering work of Gurau, who introduced so-called \emph{colored} models in 2009 \cite{razvan_colors}. Colored tensor models, as well as their closely related \emph{uncolored} versions \cite{Bonzom:2012hw}, have a rather rigid combinatorial structure which is at the source of most recent progress in this field. First and foremost, only in this context could an analogue of the $1/N$ expansion of matrix models be realized in dimension $d >2$ \cite{RazvanN, Gurau:2011xq, Gurau:2012vk, Bonzom:2012hw, Bonzom:2014oua, Tanasa:2015uhr, Carrozza:2015adg}, with a non-topological index known as the Gurau degree replacing the two-dimensional genus. This $1/N$ expansion, which is dominated by so-called \emph{melonic} Feynman diagrams \cite{Bonzom:2011zz, Gurau:2013cbh, MelonsAB}, is at work in recently proposed tensorial versions of the Sachdev-Ye-Kitaev model \cite{Witten:2016iux, Gurau:2016lzk, Klebanov:2016xxf} (see also \cite{bonzom2013universality} for an earlier and related idea). It is also at play in the renormalization group of GFTs \cite{Rivasseau:2013uca, Rivasseau:2014ima, Carrozza_review}, which is the focus of the present article. 

The main advantage of GFTs over tensor models is that, thanks to their additional group-theoretic ingredients, they can naturally accommodate a sum over (generalized) lattice gauge theory amplitudes  or discrete quantum gravity path-integrals (see \cite{daniele_rev2011} for a review). They can in particular be seen as quantum field theory completions of spin foam models \cite{perez_review2012}, and have been proposed as a suitable framework for defining and studying the continuum limit of the latter \cite{daniele_hydro} (see also \cite{Oriti:2013aqa, Oriti:2014uga} for an articulation of the same ideas in the context of canonical loop quantum gravity).

A key aspect of this programme  is to understand the behavior of GFTs under renormalization, both perturbatively and non-perturbatively. Indeed, renormalization group techniques are essential in two respects: first, to determine which GFT models can be consistently defined, at least perturbatively; second, to systematically explore the theory spaces of such models, and study the various phases in which they can be realized. Regarding phase transitions, Bose--Einstein condensation have recently been proposed as a possible mechanism for recovering the effective continuum dynamics of general relativity from GFT quantum gravity models \cite{Gielen:2016dss}, at least in the homogeneous \cite{Gielen:2013kla, Oriti:2016ueo} and spherically symmetric \cite{Oriti:2015rwa} sectors. This particular scenario calls for renormalization techniques which go beyond the perturbative regime, and is one of the main motivations of our work.  

So far, a GFT renormalization programme could only be implemented in the context of so-called \emph{tensorial} GFTs, which are based on the same combinatorial structure as uncolored tensor models. A number of perturbatively renormalizable models have been considered, ranging from simple GFTs on Abelian groups \cite{BenGeloun:2011rc, BenGeloun:2012pu, Geloun:2016bhh}, to toy models on Abelian \cite{Carrozza:2012uv, Samary:2012bw, Samary:2014tja, Samary:2014oya, Lahoche:2015yya, Lahoche:2015zya, Lahoche:2016xiq} or non-Abelian \cite{Carrozza:2013mna, Carrozza:2013wda, Carrozza:2014rba, Carrozza:2014rya} groups directly inspired from spin foam models. In all of these theories, the peculiar tensorial structure of the interactions provides a generalized notion of locality, which we may call \emph{tensorial locality} to distinguish it from the ordinary space-time locality. Crucially, this new notion of locality turns out to be compatible with the renormalization group, in the sense that the ultraviolet divergences can be reabsorbed into (tensorially) local terms. More recently, considerable effort has been invested into extending standard Functional Renormalization Group (FRG) methods to tensor models and GFTs. The FRG is a general formulation of the Wilsonian renormaliation group, and has been applied in a wide range of physical situations \cite{Wetterich:1993ne, Morris:1993qb, Bagnuls:2000ae, Berges:2000ew, Delamotte:2007pf, Rosten:2010vm, Blaizot:2012fe, Demmel:2012ub, Reuter:2012id}. It provides a set of non-perturbative approximation and truncation techniques which, even though they do not rely on a fully controllable expansion in a small parameter, have been well-tested empirically. They are particularly useful to uncover non-trivial fixed points, and more generally to study phase transitions. The FRG has been applied to matrix models in \cite{Eichhorn:2013isa, Eichhorn:2014xaa} and tensorial GFTs in \cite{Benedetti:2014qsa, Benedetti:2015yaa, Geloun:2016qyb, Geloun:2016xep, Lahoche:2016xiq}. Interestingly, the phase diagrams of tensorial GFTs have already a rich structure in the simplest truncations, and feature non-trivial fixed points which are reminiscent of the Wilson-Fisher fixed point of local scalar field theory \cite{wilson_fisher}.



Our purpose is to undertake the FRG analysis of the three-dimensional $\SU(2)$ GFT first introduced in \cite{Carrozza:2013wda}. As far as quantum gravity is concerned, it is arguably the most interesting renormalizable GFT on the market; indeed, it lives in the theory space of 3d Euclidean quantum gravity (see \cite{Carrozza_review} and references therein). With this example, we demonstrate that models based on non-Abelian groups like $\SU(2)$ are amenable to the FRG, which is a necessary step towards the application of comparable techniques in a four-dimensional context \cite{eprl, fk, dl, bo_bc, bo}. Another motivation for looking at this particular model is that its renormalization group is already well understood in the perturbative regime. In particular, it was shown in \cite{Carrozza:2014rba} that generic renormalization group trajectories are repelled from the Gaussian fixed point in the ultraviolet, and are therefore not asymptotically free; this failure of perturbative ultraviolet completeness naturally raised the question of the existence of a non-trivial ultraviolet fixed point. The $\varepsilon$-expansion of \cite{Carrozza:2014rya} provided first hints that such a fixed point may actually be realized, and the FRG will allow us to investigate this question in greater detail. 


\

The paper is organized as follows. After recalling the definition of the $\SU(2)$ GFT of \cite{Carrozza:2013wda} in Section \ref{section1}, we will set-up the FRG framework and study its $\phi^6$ truncation in Section \ref{FRG}. This crude approximation only includes perturbatively renormalizable interactions, but it will already suggest interesting features, such as a candidate ultraviolet fixed point. In Section \ref{sec:order8}, we will refine the analysis through the inclusion of order $8$ perturbatively irrelevant interactions in the renormalization group ansatz, which will result in a nine-dimensional truncated theory space. We will again find an ultraviolet fixed point, and hence provide more evidence in favor of an asymptotic safety scenario. Finally, we will remark in Section \ref{sec:non_branching} that this fixed point lives in a restricted sector of the theory space, generated by a small subset of all possible tensorial interactions. This will allow to push the analysis to even higher orders, and confirm the qualitative features of the ultraviolet fixed point in truncations capturing the effect of up to $\phi^{12}$ interactions. Finally, we will comment on the relevance and interpretation of our results in the conclusion, with an emphasis on possible relations to GFT condensates.

\section{Three-dimensional tensorial GFT on SU(2)}
\label{section1}

\subsection{Perturbative definition and ultraviolet regularization}

\begin{figure}
\centering
\includegraphics[scale=1]{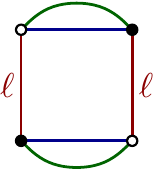} 
\captionof{figure}{Bipartite $3$-colored graph associated to the interaction kernel  $\cW^{(\ell)}$.}\label{fig1}
\end{figure}

In the present article, we are interested in the non-perturbative ultraviolet properties of the TGFT originally introduced in \cite{Carrozza:2013wda, Carrozza:2014rba, Carrozza:2014rya}, whose construction we briefly recall. This three-dimensional model, based on the group manifold $\SU(2)$, is defined by a partition function of the form
\begin{equation}\label{partitionfunction}
\cZ_\Lambda := \int \extd \mu_{C_\Lambda} [\psib,\psi] \, e^{-S_{\Lambda}[\bar{\psi},\psi]}\,,
\end{equation}
where $\psi$ (resp. $\bar{\psi}$) are complex fields over three copies of $\SU(2)$\footnote{We will use the vector notation $\mathbf{g} = (g_1 , g_2 , g_3)$ throughout the paper. Similarly, $\extd \mathbf{g}$ will be short-hand for the product of Haar measures $\extd g_1 \extd g_2 \extd g_3$.}:
\begin{align}
\psi : \SU(2)^3 & \to \mathbb{C}\\
\textbf{g}\equiv (g_1,g_2,g_3 ) & \mapsto \psi(g_1,g_2,g_3) \nn
\end{align}
The UV regularized Gaussian measure $\extd \mu_{C_\Lambda}$, which encodes the kinetic part of the classical action, is characterized by the covariance $C_\Lambda$:
\begin{equation}\label{propagator}
\int \extd \mu_{C_\Lambda}(\psi,\bar{\psi}) \, \psi(\textbf{g}) \bar{\psi}(\textbf{g}')= \int_{\SU(2)} \extd h \int_{1/\Lambda^2}^{\infty} \extd \alpha \, e^{-\alpha m^2(\Lambda)} \prod_{\ell=1}^3 K_{\alpha}(g_\ell h g_\ell^{\prime -1}),
\end{equation}
where $K_{\alpha}$ is the heat kernel on $\SU(2)$ at time $\alpha$, and $\Lambda > 0$ is an ultraviolet regulator (imposing a smooth cut-off on large spin labels in the harmonic expansion of $\psi$ and $\psib$). The integral on $h$ implements the \textit{closure constraint} (see e.g. \cite{daniele_rev2011}) and is responsible for the $\SU(2)$ lattice gauge theory form of the Feynman amplitudes. Finally, the classical (interaction part of the) action $S_{\Lambda}$ is given by:
\begin{align}\label{int1}
S_{\Lambda}[\psi, \psib]=& \frac{\lambda_4 (\Lambda)}{2} \sum_{\ell = 1}^3 \int [\prod_{j=1}^4 \extd\mathbf{g}_j] \, \cW^{(\ell)}(\mathbf{g}_1, \mathbf{g}_2, \mathbf{g}_3, \mathbf{g}_4) \psi(\mathbf{g}_1)\bar{\psi}(\mathbf{g}_2)\psi(\mathbf{g}_3)\bar{\psi}(\mathbf{g}_4) \\
&+ \frac{\lambda_{6,1} (\Lambda)}{3} \sum_{\ell = 1}^3 \int [\prod_{j=1}^6 \extd\mathbf{g}_j] \, \cX^{(\ell)}(\mathbf{g}_1, \mathbf{g}_2, \mathbf{g}_3, \mathbf{g}_4, \mathbf{g}_5, \mathbf{g}_6) \psi(\mathbf{g}_1)\bar{\psi}(\mathbf{g}_2)\psi(\mathbf{g}_3)\bar{\psi}(\mathbf{g}_4) \psi(\mathbf{g}_5)\bar{\psi}(\mathbf{g}_6) \nn \\
&+ \lambda_{6,2} (\Lambda) \sum_{\ell = 1}^3 \int [\prod_{j=1}^6 \extd\mathbf{g}_j] \, \cY^{(\ell)}(\mathbf{g}_1, \mathbf{g}_2, \mathbf{g}_3, \mathbf{g}_4, \mathbf{g}_5, \mathbf{g}_6) \psi(\mathbf{g}_1)\bar{\psi}(\mathbf{g}_2)\psi(\mathbf{g}_3)\bar{\psi}(\mathbf{g}_4) \psi(\mathbf{g}_5)\bar{\psi}(\mathbf{g}_6) \nn
\end{align}
where the symbols $\mathcal{W}^{(\ell)}$, $\mathcal{X}^{(\ell)}$ and $\mathcal{Y}^{(\ell)}$ are products of delta functions associated to tensor invariant interactions, and $\lambda_b (\Lambda)$ ($b \in \{ 4, (6,1), (6,2)\}$) are running coupling constants. For instance:
\begin{equation}
\mathcal{W}^{(\ell)}(\mathbf{g}_1,\mathbf{g}_2,\mathbf{g}_3,\mathbf{g}_4) = \delta(g_{1\ell}g_{4\ell}^{-1})\delta(g_{2\ell}g_{3\ell}^{-1})\prod_{j\neq \ell}\delta(g_{1j}g_{2j}^{-1})\delta(g_{3j}g_{4j}^{-1})\,.
\end{equation}
Each term in the action $S_\Lambda$ is conveniently indexed by a so-called \emph{bubble}, which is a bipartite \emph{$3$-colored graph} (see \cite{Ryan:2016emo} and references therein for more on colored graphs). For instance, the kernel $\mathcal{W}^{(\ell)}$ is associated to the graph shown in Figure \ref{fig1}. Black and white vertices respectively represent fields $\psi$ and $\bar{\psi}$, while a line with color index $\ell \in \{ 1, 2 , 3 \}$ pictures the convolution of two fields with respect to the $\ell^{\mathrm{th}}$ $\SU(2)$ copy of the base space. 
More generally, the interactions involved in the action $S_{\Lambda}$ are associated to the following bubbles\footnote{In the rest of the paper we will freely substitute bubble drawings for the interactions they represent.}:
\begin{align}\label{bubble}
\vcenter{\hbox{\includegraphics[scale=0.8]{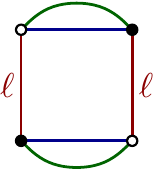}}} &\;\longleftrightarrow\; \int [\prod_{j=1}^4 \extd\mathbf{g}_j] \, \cW^{(\ell)}(\mathbf{g}_1, \mathbf{g}_2, \mathbf{g}_3, \mathbf{g}_4) \psi(\mathbf{g}_1)\bar{\psi}(\mathbf{g}_2)\psi(\mathbf{g}_3)\bar{\psi}(\mathbf{g}_4) \\
\vcenter{\hbox{\includegraphics[scale=0.8]{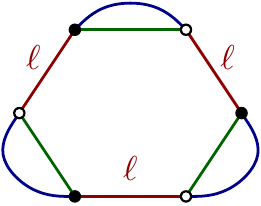}}} &\;\longleftrightarrow\; \int [\prod_{j=1}^6 \extd\mathbf{g}_j] \, \cX^{(\ell)}(\mathbf{g}_1, \mathbf{g}_2, \mathbf{g}_3, \mathbf{g}_4, \mathbf{g}_5, \mathbf{g}_6) \psi(\mathbf{g}_1)\bar{\psi}(\mathbf{g}_2)\psi(\mathbf{g}_3)\bar{\psi}(\mathbf{g}_4) \psi(\mathbf{g}_5)\bar{\psi}(\mathbf{g}_6) \\
\vcenter{\hbox{\includegraphics[scale=0.8]{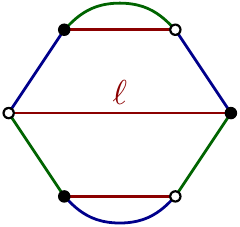}}} &\;\longleftrightarrow\; \int [\prod_{j=1}^6 \extd\mathbf{g}_j] \, \cY^{(\ell)}(\mathbf{g}_1, \mathbf{g}_2, \mathbf{g}_3, \mathbf{g}_4, \mathbf{g}_5, \mathbf{g}_6) \psi(\mathbf{g}_1)\bar{\psi}(\mathbf{g}_2)\psi(\mathbf{g}_3)\bar{\psi}(\mathbf{g}_4) \psi(\mathbf{g}_5)\bar{\psi}(\mathbf{g}_6)   
\end{align}
where the index $\ell$ runs from $1$ to $3$ and characterizes each bubble (up to automorphisms). Note that the action \eqref{int1} is invariant under color permutations, and that the coupling constants $\lambda_4$, $\lambda_{6,1}$ and $\lambda_{6,2}$ have been normalized by the number of automorphisms of the bubble they parametrize\footnote{As usual, this facilitates the computation of combinatorial factors in the Feynman expansion. See \cite{Carrozza:2014rba} for details about the definition of colored graph automorphism we are using.}.

\begin{figure}
\centering
\includegraphics[scale=1]{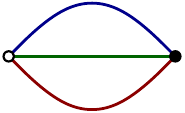} 
\captionof{figure}{The $2$-valent melonic bubble.}\label{figelem}
\end{figure}

These three types of interactions are examples of \emph{melonic} bubbles \cite{Bonzom:2011zz, Gurau:2011xp, Bonzom:2012hw}, which form a recursively generated subclass of colored graphs. They play a central role in tensorial theories in that they generically dominate their large $\Lambda$ regime. In $d=3$, the smallest melonic bubble (with two vertices) is pictured in Figure \ref{figelem}; it represents a mass term. 
All higher order melonic bubbles can be obtained by successive insertions of melonic $2$-point subgraphs, as represented in Figure \ref{figelem2}. In particular, the 4-valent (resp. 6-valent) melonic bubbles are generated by all possible insertions of one (resp. two) melonic $2$-point subgraph(s) in the $2$-valent bubble of Figure \ref{figelem}. 
\begin{center}
\includegraphics[scale=1]{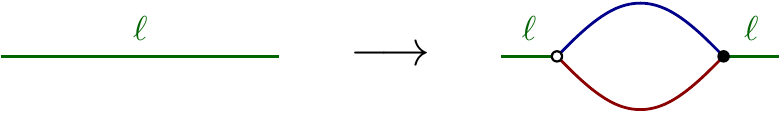} 
\captionof{figure}{Melonic insertion on a line of color $\ell \in \{ 1, 2, 3\}$.}\label{figelem2}
\end{center}

\begin{figure}[ht]
\centering
\includegraphics[scale=1]{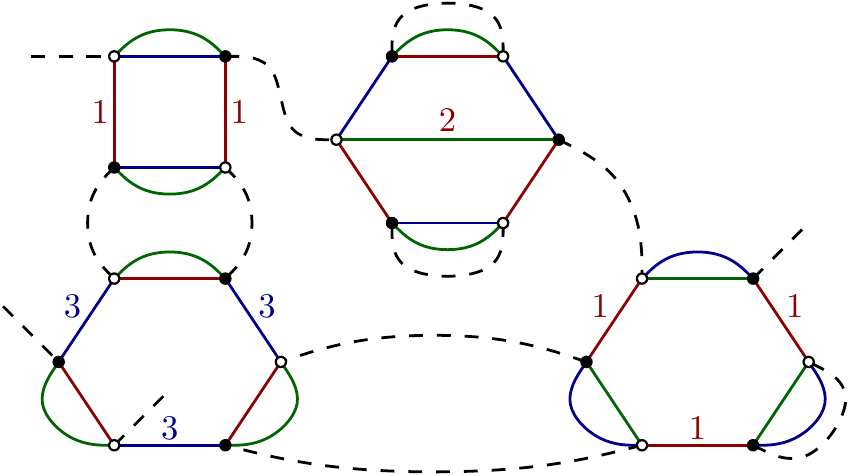}
\captionof{figure}{Feynman graph with $4$ external legs and $4$ vertices.}\label{fig2} 
\end{figure}

The Schwinger $N$-point functions $\mathsf{S}_N$ can be formally defined by their perturbative expansions in powers of $\lambda_b$ ($b \in \{ 4, (6,1), (6,2)\}$), which are indexed by Feynman diagrams $\cG$:
\begin{equation}
\mathsf{S}_N = \sum_{\cG \vert N(\cG) = N}\left( \prod_{b} (-\lambda_b )^{n_b (\cG)} \right) \frac{1}{s(\mathcal{G})}\mathcal{A}_{\mathcal{G}}\,.
\end{equation}
For any Feynman graph $\cG$, $N(\cG)$ denotes its number of external legs, $n_b (\mathcal{G})$ its number of vertices of type $b$, $s(\mathcal{G})$ the order of its automorphism group\footnote{This is the primary reason why normalization factors have been introduced in the definition of $S_{\Lambda}$.}, and $\mathcal{A}_{\mathcal{G}}$ its Feynman amplitude. The Feynman graphs of this model have connected $3$-colored graphs as vertices, which are then connected by additional propagator lines. As is customary in the literature, the latter are represented by dashed lines in order to distinguish them from the colored lines which encode the internal structure of the vertices. Attributing a color (conventionally $0$) to the dashed lines, a Feynman graph is a $4$-colored graph with possibly open half-lines of color $0$. An example is provided in Figure \ref{fig2}. From the point of view of lattice gauge theory and spin foams, an interesting property of these Feynman diagrams is that they are naturally equipped with a notion of \emph{face}. A face is defined as a maximal collection of color-$0$ lines, whose elements lie in a single bicolored connected component of the graph; it is furthermore said to be \emph{closed} when the bicolored connected component is a cycle, and \emph{open} otherwise. In this way, each Feynman diagram generated in perturbative expansion can equivalently be interpreted as a particular gluing of faces, edges and vertices, i.e. a $2$-complex. Furthermore, in the model we are considering, a given Feynman amplitude takes the form of a generalized $\SU(2)$ lattice gauge theory on its associated $2$-complex. For instance, a graph $\cG$ with only closed faces is weighted by the (bare) amplitude:
\beq
\mathcal{A}_\mathcal{G} =  \prod_{l \in L(\cG)} \int_{1/\Lambda^2}^{+ \infty} \extd \alpha_l  \, e^{- m^2 \alpha(\Lambda)} \prod_{f \in F(\cG)} K_{\alpha(f)} \left( \underset{l \in f}{\overrightarrow{\prod}} h_l \right) \,,
\eeq
where $L(\cG)$ (resp. $F(\cG)$) is the set of lines (resp. faces) of $\cG$, $\alpha(f):=\sum_{l \in f} \alpha_l$, $K_\alpha$ is the heat kernel on $\SU(2)$ at time $\alpha$, and $\underset{l \in f}{\overrightarrow{\prod}} h_l$ is the (oriented) $\SU(2)$ holonomy around the face $f$.

Finally, it is convenient for the rest of this paper to introduce the operator $\hat{P}$, acting on the GFT fields as:
\begin{equation}\label{proj}
\hat{P} \psi(g_1,g_2,g_3) = \int_{\SU(2)} \extd h \, \psi(g_1 h,g_2 h,g_3 h)\,.
\end{equation}
$\hat{P}$ is the projector onto the subspace of \emph{gauge invariant fields} $\psi$, which verify:
\begin{equation}
\psi(g_1,g_2,g_3)=\psi(g_1h,g_2h,g_3h)\,, \qquad \forall h\in \SU(2)\,.
\end{equation}

\subsection{Canonical dimensions}

Another important quantity for our purpose is the \emph{canonical dimension}. The reader interested in more detail may consult \cite{Carrozza:2014rba, Benedetti:2015yaa}, we only focus on the key arguments here. As proven in \cite{Carrozza:2013wda}, the divergence degree $\omega(\mathcal{G})$ of a leading order graph $\mathcal{G}$ is given by
\begin{equation}
\omega(\mathcal{G})= 3 - \frac{N(\cG)}{2} - 2 n_2 (\cG) - n_4 (\cG)\,,
\end{equation}
where $n_k (\cG)$ is the number of bubble vertices of valency $k$ in $\mathcal{G}$. For $N=4$, $\omega$ is bounded by $1$, and melonic graphs containing only order-$6$ interactions have $\omega = 1$. This indicates that, perturbatively, $\lambda_4 (\Lambda)$ naturally scales as $\Lambda$ in the ultraviolet. We therefore attribute it a canonical dimension $[\lambda_4]=1$. More generally, a bubble interaction $b$ of order $N_b$ is associated to a coupling constant $\lambda_b$ of dimension:
\begin{equation}
[\lambda_b] := 3 - \frac{N_b}{2}\,,
\end{equation}
in such a way that renormalizable interactions are characterized by $[\lambda_b] \geq 0$. Including the mass term in this definition, we have:
\beq
[m^2] = 2\,, \qquad [\lambda_4] = 1\,, \qquad [\lambda_{6,1}] = [\lambda_{6,2}] = 0 \,.
\eeq
Hence, the \textit{dimensionless coupling constants} $u_2$, $u_4$, $u_{6,1}$ and $u_{6,2}$ are defined as:
\begin{equation}
u_2 (\Lambda) := \frac{m^2 (\Lambda)}{\Lambda^{2}}\,,\qquad u_4 (\Lambda) :=  \frac{\lambda_4 (\Lambda)}{\Lambda}\,, \qquad  u_{6,1} (\Lambda) := \lambda_{6,1} (\Lambda)\,, \qquad u_{6,2} (\Lambda) := \lambda_{6,2} (\Lambda)\,.
\end{equation}
We will use this dimensionless parametrization from section \ref{FRG} onwards, in order to obtain well-defined autonomous systems of renormalization group equations.

\section{Functional renormalization group: order 6 truncation}\label{FRG}

In this section, we introduce the effective average action associated to the field theory defined by \eqref{partitionfunction}. It satisfies a Functional Renormalization Group (FRG) or Wetterich--Morris equation \cite{Wetterich:1993ne, Morris:1993qb}, which can be viewed as a particular realization of Wilson's renormalization group.
After recalling the general construction of the FRG in the context of tensorial field theories \cite{Benedetti:2014qsa, Benedetti:2015yaa, Geloun:2016qyb, Geloun:2016xep, Lahoche:2016xiq}\footnote{See also \cite{Reiko_thomas1, Reiko_thomas2, Krajewski:2016svb} for a FRG based on the Polchinski equation \cite{Polchinski:1983gv}.}, we will use a $\phi^6$ truncation of the Wetterich--Morris equation to extract approximate renormalization group equations. For general reviews ans applications of the FRG, we refer the reader to \cite{Bagnuls:2000ae, Berges:2000ew, Delamotte:2007pf, Rosten:2010vm, Blaizot:2012fe, Demmel:2012ub, Reuter:2012id}.

\subsection{Effective average action and truncation}

The FRG formalism is based on a deformation of the original generating functional \cite{Bagnuls:2000ae, Berges:2000ew, Delamotte:2007pf}. Starting from the partition function of equation \eqref{partitionfunction}, we define the following one-parameter family of generating functionals:
\begin{equation}\label{family}
\mathcal{Z}_{\Lambda, k}[J,\bar{J}]:=\int \extd\mu_{C_\Lambda}(\bar{\psi},\psi)e^{-S_{\Lambda}(\bar{\psi},\psi)-\Delta S_{k}[\bar{\psi},\psi]+(\bar{J},\psi)+(\bar{\psi},J)},
\end{equation}
where $(\bar{J},\psi):=\int \extd\textbf{g} \, \bar{J}(\textbf{g})\psi(\textbf{g})$ and $0 < k \leq \Lambda$. The new contribution $\Delta S_k$ plays the role of infrared cut-off and is chosen to be of the form:
\begin{equation}
\Delta S_{k}[\bar{\psi},\psi]:=\int \extd\textbf{g}_1 \extd\textbf{g}_2 \, \bar{\psi}(\textbf{g}_1)R_{k}(\textbf{g}_1\textbf{g}_2^{-1})\psi(\textbf{g}_2)\,,
\end{equation}
where $R_{k}(\textbf{g})$ is a cut-off function depending only on the geodesic lengths $\vert g_i \vert$ from the identity to $g_i$. It is moreover required to verify:
\begin{enumerate}[a)]
\item $R_{k}(\textbf{g})\geq 0$ for all $\textbf{g}\in \SU(2)^3$ and all $0 < k \leq \Lambda$ ;
\item $\underset{k \to 0}{\lim} \,R_k = 0$ ;
\item $\underset{k \to \Lambda}{\lim} \, R_k (\mathbf{g}) = + \infty$ for any $\mathbf{g} \in \SU(2)^3$ ;
\item at fixed $k$, $R_k ( \vert g_i \vert \lesssim k) \ll 1$ and $R_{k}(|g_i| \gtrsim k) \sim 1$;
\item at fixed $\mathbf{g}$, $\frac{\extd}{\extd k} R_k (\mathbf{g}) \geq 0$.
\end{enumerate}
The second condition ensures that the original generating function is recovered when all the fluctuations are integrated out:
\beq
\cZ_{\Lambda, k= 0} = \cZ_\Lambda\,.
\eeq
The role of the third condition is to impose $S_\Lambda$ as an ultraviolet boundary condition for the renormalization group flow of the effective average action (defined in equation \eqref{effectiveaction} below). The fourth condition guarantees that: on the one hand the ultraviolet modes are almost unaffected by the additional cut-off term; and on the other hand the infrared modes have a large mass, which effectively decouples them from degrees of freedom with small momenta. 

\

The general derivation of the Wetterich--Morris equation for tensorial field theories was first described in \cite{Benedetti:2014qsa, Benedetti:2015yaa} (for Abelian groups), and as announced, we recall only the essential results here. The \emph{effective average action} $\Gamma_{k}$ is defined as the Legendre transform of $W_k[J,\bar{J}]:=\ln(\mathcal{Z}_{k, \Lambda})$ (with the infrared regulator correctly subtracted):
\begin{equation}\label{effectiveaction}
\Gamma_k[\bar{\phi},\phi]+\Delta S_k [\bar{\phi},\phi]=(\bar{J},\phi)+(\bar{\phi},J)-W_k [J,\bar{J}]\,,
\end{equation}
where the mean field $\phi$ is defined as
\begin{equation}
\phi(\textbf{g}):=\frac{\delta W_k}{\delta \bar{J}(\textbf{g})}\,. 
\end{equation}
Deriving the effective average action with respect to $k$, and remarking that the mean field $\phi$ is gauge invariant \cite{Benedetti:2015yaa}, one arrives at the \emph{Wetterich--Morris equation}: 
\begin{equation}\label{Wettericheq}
\partial_k \Gamma_k [\bar{\phi},\phi]=\int \extd\textbf{g}_1 \extd\textbf{g}_2 \extd\textbf{g}_3 \, \partial_k R_k(\textbf{g}_1\textbf{g}_2^{-1})(\Gamma_k^{(2)}+R_k)^{-1}(\textbf{g}_2,\textbf{g}_3)\hat{P}(\textbf{g}_3,\textbf{g}_1)\,,
\end{equation}
where
\begin{equation}
\Gamma^{(2)}_k [\bar{\phi},\phi]:=\frac{\delta^2 \Gamma_k}{\delta\phi\delta\bar{\phi}} [\bar{\phi},\phi]\,.
\end{equation}
The Wetterich--Morris equation describes the evolution of the effective average action $\Gamma_k$ as modes are integrated out from higher to lower scales. Due to property c), the Legendre transform $\Gamma_{k}[\vphi, \vphib]$ reduces to the bare action $S_{\Lambda}[\vphi, \vphib] + \vphib \cdot \left(m^2(\Lambda) - \sum_\ell \Delta_l \right) \cdot \vphi$ in the limit $k \to \Lambda$. Therefore, the functional $\Gamma_k$ interpolates between the classical action and the full \emph{effective action} $\Gamma_0$, which is also the generating functional of one-particle irreducible Feynman graphs. \\

Extracting non-perturbative information from the exact flow equation (\ref{Wettericheq}) requires an appropriate approximation scheme, generally consisting in a projection to a finite-dimensional functional space. A simple strategy -- called \textit{truncation method} -- amounts to: 1) a choice of ansatz for $\Gamma_k$; and 2) a projection of the right-hand side of the Wetterich equation down to the subspace of functionals generated by this ansatz. In this paper we adopt a \emph{local potential approximation}, meaning that we do not include any derivative couplings in the potential. Note that the qualifier 'local' should again be understood in the tensorial sense, which means that the potential is a weighted sum of bubble interactions (and not a sum of local interactions in the ordinary space-time sense). We will moreover restrict our attention to melonic bubbles \cite{Bonzom:2012hw} because they dominate the ultraviolet regime. The first truncation we investigate is limited to the three types of perturbatively renormalizable interactions introduced in the previous section:
\begin{align}\label{ansatz}
\Gamma_k [\bar{\phi},\phi]&=\int \extd\textbf{g} \,\bar{\phi}(\textbf{g}) \left(-Z(k) \sum_\ell \Delta_\ell + Z(k) m^2(k)\right)\phi(\textbf{g})\\
&+ Z(k)^2 \frac{\lambda_4 (k)}{2} \sum_{\ell = 1}^3 \int [\prod_{j=1}^4 \extd\mathbf{g}_j] \, \cW^{(\ell)}(\mathbf{g}_1, \mathbf{g}_2, \mathbf{g}_3, \mathbf{g}_4) \phi(\mathbf{g}_1)\bar{\phi}(\mathbf{g}_2)\phi(\mathbf{g}_3)\bar{\phi}(\mathbf{g}_4) \nn \\
&+ Z(k)^3 \frac{\lambda_{6,1} (k)}{3} \sum_{\ell = 1}^3 \int [\prod_{j=1}^6 \extd\mathbf{g}_j] \, \cX^{(\ell)}(\mathbf{g}_1, \mathbf{g}_2, \mathbf{g}_3, \mathbf{g}_4, \mathbf{g}_5, \mathbf{g}_6) \phi(\mathbf{g}_1)\bar{\phi}(\mathbf{g}_2)\phi(\mathbf{g}_3)\bar{\phi}(\mathbf{g}_4) \phi(\mathbf{g}_5)\bar{\phi}(\mathbf{g}_6) \nn \\
&+ Z(k)^3 {\lambda_{6,2} (k)} \sum_{\ell = 1}^3 \int [\prod_{j=1}^6 \extd\mathbf{g}_j] \, \cY^{(\ell)}(\mathbf{g}_1, \mathbf{g}_2, \mathbf{g}_3, \mathbf{g}_4, \mathbf{g}_5, \mathbf{g}_6) \phi(\mathbf{g}_1)\bar{\phi}(\mathbf{g}_2)\phi(\mathbf{g}_3)\bar{\phi}(\mathbf{g}_4) \phi(\mathbf{g}_5)\bar{\phi}(\mathbf{g}_6) \nn
\end{align}
where $\phi$ is a gauge invariant field, and only the five parameters $m^2$, $Z$ and $\lambda_b$ depend on $k$. Since the effective average action coincides with the bare action at $k = \Lambda$, the boundary conditions for the coupling constants are consistently specified by the perturbative definition of the theory (\ref{partitionfunction}). In particular, the wave--function renormalization parameter has boundary condition $Z(\Lambda)= 1$.

\

We now specify our choice of regulating function $R_k$. Note that, even if the FRG equation \eqref{Wettericheq} is (formally) exact, the truncation generally introduces a spurious dependence on the choice of regulator. Scheme dependence, reliability and optimization are very important challenges of the FRG method; the interested reader may consult e.g. \cite{Litim:2000ci, Litim:2001up}. By focusing on a single regulating function $R_k$, we will leave these questions aside for the time being, but they will certainly deserve more attention in the future. The base space of the GFT studied in this paper being a non-Abelian compact Lie group, and in particular not a linear space, we propose to rely on a heat kernel regularization also in the infrared sector. This will dispense us with intricate $\SU(2)$ recoupling theory computations; we will instead invoke the Laplace approximation to evaluate the amplitudes in integral form in the regime $1 \ll k \ll \Lambda$ (see \cite{vm1, vm2, vm3} for an in-depth discussion of such methods). Let us define the operators
\beq
\cC_k := - Z(k) \sum_{\ell=1}^3 \Delta_\ell + R_k
\eeq
and
\beq
\cK_k :=  \cC_k + Z(k) \, m^2 (k) = - Z(k) \sum_{\ell=1}^3 \Delta_\ell + Z(k)\, m^2(k) + R_k \,.
\eeq
$\cK_k$ is the effective kernel associated to the kinetic term of the effective average action, and in the computations below, $\cK_k^{-1}$ will play the role of effective propagator. It is convenient to adjust $R_k$ in such a way that both $\cC_k^{-1}$ and $\cK_k^{-1}$ are expressible in term of the heat kernel on $\SU(2)$. More precisely, we require:
\beq\label{k-1}
\cC_k^{-1} (\mathbf{g}, \tilde{\mathbf{g}}) = \frac{1}{Z(k)} \int_{\Lambda^{-2}}^{k^{-2}} \extd \alpha \prod_{\ell = 1}^3 K_\alpha (g_\ell \tilde{g}_\ell^{-1})\,.
\eeq
This is achieved by the regulator
\beq
\hat{R}_k (\{j_\ell , m_\ell , n_\ell\}) := Z(k) \, k^2 \, r_{\Lambda, k}\left( \frac{\sum_{\ell=1}^3 j_\ell (j_\ell + 1) }{k^2} \right) \,,
\eeq
where $\hat{R}_k$ is the Fourier transform of $R_k$, $j_\ell$ are spin labels, $m_\ell$ and $n_\ell$ the associated magnetic indices, and the function $r_{\Lambda, k}$ is defined as
\beq
r_{\Lambda, k} (z) = \frac{z}{e^{- z k^2 / \Lambda^2} - e^{-z}} - z \underset{k \ll \Lambda}{\sim} \frac{z}{e^z - 1}\,.
\eeq
The specific form of $r_{\Lambda, k}$ is not important, we only need to know that it exists and that it leads to an admissible cut-off function $R_k$, which can be easily checked. An important property of the operator $R_k$ we will use in the sequel is that it is diagonal in the spin basis $\{j_\ell , m_\ell , n_\ell\}$.

\

We now move on to the extraction of the truncated flow equations for $m^2$, $Z$ and $\lambda_b$ from the full Wetterich--Morris equation \eqref{Wettericheq}. We write the second derivative of $\Gamma_k$ as:
\begin{equation}
\nonumber\Gamma_k^{(2)}[\bar{\phi},\phi](\textbf{g},\textbf{g}') =Z(k)\left( -\Delta+m^2(k)\right)\hat{P}(\textbf{g},\textbf{g}') + F_{k,(1)} [\bar{\phi},\phi](\textbf{g},\textbf{g}') + F_{k,(2)} [\bar{\phi},\phi](\textbf{g},\textbf{g}')\,,
\end{equation}
where all the field-dependent terms of order $2n$ have been gathered in $F_{k, (n)}$. 
In particular, $F_{k,(1)}$ depends on $\lambda_4 (k)$, while $F_{k,(2)}$ depends on $\lambda_{6,1} (k)$ and $\lambda_{6,2} (k)$. We can pictorially represent these quantities as:
\beq\label{pict_F1}
F_{k,(1)} [\bar{\phi},\phi](\textbf{g},\textbf{g}') = Z(k)^2 \, \lambda_4 (k) \sum_{\ell=1}^3 \vcenter{\hbox{\includegraphics[scale=0.8]{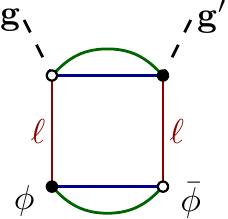}}} + \cdots
\eeq
and
\begin{align}
F_{k,(2)} [\bar{\phi},\phi](\textbf{g},\textbf{g}') &= Z(k)^3 \, \lambda_{6,1} (k) \sum_{\ell=1}^3 \, \vcenter{\hbox{\includegraphics[scale=0.8]{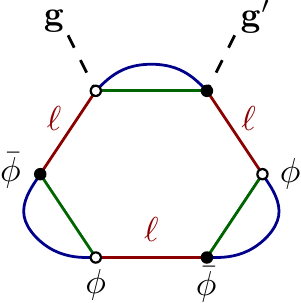}}} \\
&+ Z(k)^3 \, \lambda_{6,2} (k) \sum_{\ell=1}^3 \left( \, \vcenter{\hbox{\includegraphics[scale=0.8]{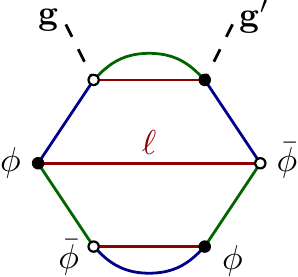}}} \, +  \vcenter{\hbox{\includegraphics[scale=0.8]{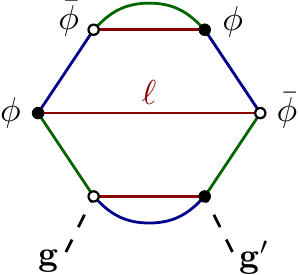}}} \right) + \cdots \nonumber
\end{align}
where the external variables $\mathbf{g}$ and $\mathbf{g}'$ are represented as dashed half-lines, and will be traced over in the Wetterich--Morris equation. Furthermore, we have only represented those diagrams which will eventually contribute to the large $k$ limit. 

Expanding the right-hand side of the Wetterich equation \eqref{Wettericheq}, and denoting by $\Gamma_{k,(n)}$ the field-dependent terms of order $2n$ in the ansatz \eqref{ansatz}, one may identify terms with the same number of fields $\phi$ on both sides of the equality to obtain the following equations (in matricial notation):
\begin{align}
\partial_k\Gamma_{k,(1)}&=-\Tr\big[\partial_k R_k \mathcal{K}_k^{-1} F_{k,(1)}\mathcal{K}_k^{-1}\hat{P}\big]\,, \label{twopoints} \\
\partial_k\Gamma_{k,(2)}&=-\Tr\big[\partial_k R_k \mathcal{K}_k^{-1} F_{k,(2)}\mathcal{K}_k^{-1}\hat{P}\big] + \Tr\big[\partial_k R_k \mathcal{K}_k^{-1} (F_{k,(1)} \mathcal{K}_k^{-1})^2\hat{P}\big]\,, \label{fourpoints} \\
\partial_k\Gamma_{k,(3)} &= \Tr\big[\partial_k R_k \mathcal{K}_k^{-1} F_{k,(1)} \mathcal{K}_k^{-1}F_{k,(2)}\mathcal{K}_k^{-1}\hat{P}\big] + \Tr\big[\partial_k R_k \mathcal{K}_k^{-1} F_{k,(2)} \mathcal{K}_k^{-1}F_{k,(1)}\mathcal{K}_k^{-1}\hat{P}\big] \nn \\
& \qquad - \Tr\big[\partial_k R_k \mathcal{K}_k^{-1} (F_{k,(1)} \mathcal{K}_k^{-1})^3\hat{P}\big]\,. \label{sixpoints}
\end{align}
Next, we need to project back the right-hand side on the finite dimensional subspace of theories associated to the ansatz \eqref{ansatz}, and further identify terms according to their combinatorial structure. For instance, we have to extract the $\beta$-functions of both $\lambda_{6,1}$ and $\lambda_{6,2}$ from equation (\ref{sixpoints}).

\subsection{Approximation of the flow equations in the ultraviolet regime}

In order to simplify our analysis, and because we are primarily interested in the ultraviolet regime of the theory, we assume:
\beq
1 \ll k \ll \Lambda\,.
\eeq
Imposing $k \ll \Lambda$ is equivalent to taking the limit $\Lambda \to + \infty$, which we assume has been done once and for all\footnote{Note that the Wetterich--Morris equation itself is well defined in this limit, even though the initial definition of the partition function is not.}. The ultraviolet condition $k\gg 1$ will be instrumental for finding tractable approximations of the intricate heat kernel integrals entering the flow equations. It is also only in this limit that the renormalization group flow can be approximated by a homogeneous differential system; in general, the non-linear structure of the group makes the flow equations explicitly scale-dependent.

A lot is already known about the divergent structure of this theory in the large $k$ limit and in the vicinity of the Gaussian fixed point \cite{Carrozza:2013mna, Carrozza:2013wda, Carrozza:2014rba}. In particular, the Feynman graphs which dominate in this regime -- the so-called \emph{melonic graphs} -- have been fully characterized. In particular, one finds out that a $1$-loop (non-vacuum) Feynman diagram is melonic if and only if it has exactly two closed faces. Since the Wetterich--Morris equation depends on $1$-loop diagrams only, this is all we need to know to determine its leading-order contributions. 

As long as the anomalous dimension is small enough, that is as long as the scaling dimensions of the coupling constants are not too different from their canonical dimensions, we are guaranteed that $1$-loop melonic diagrams are the leading contributions to the flow. However, it is not a sufficient criterion to check whether a truncation captures the relevant physics. For instance, at an asymptotically safe fixed point, a truncation including up to the canonically marginal couplings is only sufficient if none of the canonically irrelevant couplings are shifted into relevance. Thus, a second check consists in investigating the value of the critical exponents and study how much they depart from the canonical dimensionality. If the departure is large, then a truncation including all couplings up to the canonically marginal ones is presumably insufficient. Henceforth, the reliability of the truncation can be checked by extending it until all further operators that are added remain irrelevant. In addition, the critical exponents for the relevant couplings should show convergent behaviour under the extension of the truncation (as an example, such a procedure has been applied to asymptotically safe quantum gravity in \cite{Falls:2013bv}).

\

Before moving to the actual derivation of the flow equations, let us focus on the expression of the operator $\cK_k^{-1}$. We can express it as an infinite power-series in $\cC_k^{-1}$:
\begin{align}
\cK_k^{-1} &= \left( \cC_k + Z(k) \, m^2(k) \right)^{-1} = \cC_k^{-1} \sum_{n=0}^{+ \infty} \left( -Z(k) \, m^2 (k) \right)^n \cC_k^{-n} =: \cC_k^{-1} G_k [\cC_k] \nn \\ 
&= \frac{1}{k^2 Z(k)} \sum_{n=0}^{+ \infty} \left( - u_2(k) \right)^n  \left[ \prod_{i=0}^n \int_{0}^{1} \extd u_i \right] \exp\left( - \sum_{i = 0}^n \frac{u_i}{k^2} \, \Delta \right)
\end{align}
in which by definition
\beq
\exp\left( - \sum_{i = 0}^n \frac{u_i}{k^2} \, \Delta \right)  (\mathbf{g}, \tilde{\mathbf{g}}) = \prod_{\ell=1}^3 K_{( u_0 + \ldots + u_n) / k^2 } \left( g_\ell \tilde{g}_\ell^{-1} \right)\,.
\eeq

Note that $R_k$ and $\cK_k$ commute as they are both diagonal in the spin basis. Moreover, $\cC_k^{-1}$ commutes with $\hat{P}$, as can be directly checked from expressions (\ref{proj}) and (\ref{k-1}), and hence $\cK_k$ also commutes with $\hat{P}$. This means that we will be able to reorder terms as we like in the traces we will need to compute, except for $F_k$ contributions (but we will eventually see that the relevant contributions from $F_k$ will also be easy to tackle once projected on the local sector). In particular, one will be able to gather a factor
\beq\label{relderiv}
\cC_k^{-1}\partial_k R_k \cC_k^{-1} = - \partial_k \cC_k^{-1} + \partial_k Z(k) \cC_k^{-1}\Delta \cC_k^{-1} 
\eeq
in each of the traces we will have to compute. 

In addition, since we are only interested in the leading order contribution in a $1/k$ expansion, we will be able to approximate the $\SU(2)$ heat kernels by their Abelian counterpart on $\mathbb{R}^3$. This will directly lead to integral expressions of the coefficients entering the flow equations. In the small $\alpha$ limit, the heat kernel on $\SU(2)$ admits the asymptotics
\beq\label{asymptotic}
\extd g \, K_\alpha ( g ) \sim \frac{\extd^3\textbf{X}}{16\pi^2}\frac{2 \sqrt{\pi}}{\alpha^{3/2}} e^{- \| \mathbf{X} \|^2 /4\alpha} \sim \extd^3\textbf{X} \, K_\alpha^{\mathbb{R}^3} ( \mathbf{X} )\,,
\eeq 
where, given a fixed orthonormal basis $\{ \tau^1, \tau^2, \tau^3 \}$ of $\su(2)$ (with respect to the Killing form), $\mathbf{X} = (X_1 , X_2 , X_3 ) \in \mathbb{R}^3$ is the smallest vector such that $e^{X_i \tau^i}=g$, $\extd^3\textbf{X}$ is the Lebesgue measure on $\mathbb{R}^3$, and $K_\alpha^{\mathbb{R}^3}$ is the heat kernel on $\mathbb{R}^3$ at time $\alpha$. The factor $1/16\pi^2$ comes from the normalization of the Haar measure on $\SU(2)$, while the factor $2 \sqrt{\pi}$ appears in the asymptotic evaluation of $K_\alpha (g)$. We are now in position to extract the flow equations in the deep ultraviolet limit from the set of truncated equations \eqref{twopoints}, \eqref{fourpoints} and \eqref{sixpoints}.

\subsubsection{Flow equations for $u_2$  and $Z$}

The flow equation \eqref{twopoints} may be written as: 
\begin{equation}\label{flow2}
\partial_k\Gamma_{k,(1)}=-\sum_{\ell=1}^3\Tr[D(k)F_{k,(1)}^{(\ell)}]\,,
\end{equation}
where
\begin{equation}
D(k):=\hat{P}\mathcal{K}_k^{-1} \partial_k R_k\mathcal{K}_k^{-1} \hat{P}\,,
\end{equation}
and $F_{k,(1)}^{(\ell)}$ denotes the term associated to the kernel $\cW^{(\ell)}$ in the expression of $F_{k,(1)}$, in such a way that $F_{k,(1)}=\sum_{\ell=1}^3 F_{k,(1)}^{(\ell)}$. For each $\ell$, the trace is a sum of two terms, which can be graphically represented as in Figure \ref{fig5} below, where now dashed lines are associated to the effective propagator $D(k)$. The left diagram is melonic (it has two closed faces) and hence has the largest contribution in $k$, while the graph on the right, being non-melonic (it has only one closed face), can be neglected in the large $k$ limit. 
\begin{center}
\includegraphics[scale=1]{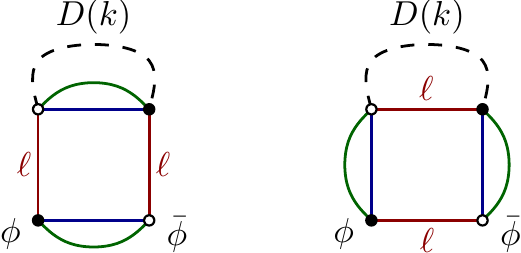} 
\captionof{figure}{Graphical representation of the traces involved in \eqref{flow2}.}\label{fig5}
\end{center}
Hence, we obtain the approximation:
\begin{equation}\label{flow2melo}
\partial_k\Gamma_{k,(1)}\approx -\sum_{\ell=1}^3\Tr[D(k)F_{k,(1),\,\mathrm{melo}}^{(\ell)}]\,,
\end{equation}
where
\begin{align}
F_{k,(1),\,\mathrm{melo}}^{(\ell)}(\{g_l,\bar{g}_l\})&:=Z(k)^2\lambda_4(k) \prod_{l\neq \ell} \delta(g_l\bar{g}_l^{-1}) \nonumber\\
\times &\int \prod_{j=1}^3 [\extd g'_j \extd g''_j] \phi(g'_1,g'_2,g'_3)\bar{\phi}(g''_1,g''_2,g''_3)\delta(g''_\ell g_\ell^{-1})\delta(g'_\ell \bar{g}^{-1}_\ell) \prod_{l \neq \ell} \delta(g'_l g^{\prime\prime\,-1}_l) \\
&:=Z(k)^2 \lambda_4(k) \prod_{l\neq \ell} \delta(g_l\bar{g}_l^{-1})f_{k,(1),\,\mathrm{melo}}^{(\ell)}(g_\ell,\bar{g}_\ell) \nn
\end{align}
Diagrammatically, equation \eqref{flow2melo} may be rewritten as
\beq\label{flow2_graph}
\partial_k ( Z(k) m^2(k) ) \, \vcenter{\hbox{\includegraphics[scale=0.8]{Figures/int2.pdf}}} - \partial_k Z(k) \sum_{\ell=1}^3 \, \vcenter{\hbox{\includegraphics[scale=0.8]{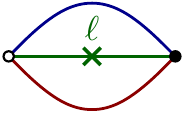}}} \approx - Z(k)^2 \, \lambda_4 (k) \, \sum_{\ell=1}^3 \, \vcenter{\hbox{\includegraphics[scale=0.8]{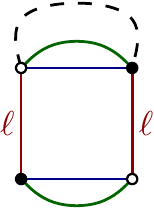}}}\,,
\eeq
in which a crossed colored line represents an insertion of a Laplace operator $\Delta$.

\

Now that we have isolated the dominant traces, we need to explicitly evaluate their leading-order contribution in $k$. By \eqref{relderiv}, the operator $D(k)$ can be expressed as
\begin{align}
D(k) &= - \hat{P} \partial_k \cC_k^{-1} (G_k [\cC_k])^2 + \partial_k Z(k)\, \hat{P} \cK_k^{-1} \Delta \cK_k^{-1} \\
&=  \hat{P} \left( \frac{\partial_k Z(k)}{Z(k)} \cC_k^{-1} + \frac{2}{k^3 Z(k)} \exp\left( - \frac{\Delta}{k^2}\right)\right) (G_k [\cC_k])^2 + \partial_k Z(k)\, \hat{P} \cK_k^{-1} \Delta \cK_k^{-1} \nn
\end{align}
Henceforth we have\footnote{For definiteness we focus on the term associated to $\ell =1$, the computation is identical for other colors.}
\beq
\Tr[D(k)F_{k,(1),\,\mathrm{melo}}^{(1)}]=-T_1+\partial_kZ(k)T_2
\eeq\label{rhs}
with
\begin{align}
T_1 &:=\Tr \left( \hat{P} \partial_k\mathcal{C}_k^{-1} (G_k[\mathcal{C}_k])^2 F_{k,(1),\,\mathrm{melo}}^{(1)} \right)\,, \label{T1} \\
T_2 &:=\Tr \left( \hat{P} \cK_k^{-1} \Delta \cK_k^{-1} F_{k,(1),\,\mathrm{melo}}^{(1)} \right)\,. \label{T2}
\end{align}
These two traces have the following structure:
\begin{align}
\nonumber T_1&:=\int \prod_{l=1}^3dg_ld\bar{g}_ldh \, D_1(\bar{g}_1hg^{-1}_1,\bar{g}_2h{g}^{-1}_2,\bar{g}_3h{g}^{-1}_3)F_{k,(1),\,\mathrm{melo}}^{(1)}(\{g_l,\bar{g}_l\})\\
&=\int dg_1d\bar{g}_1dh \, D_1(g_1h\bar{g}^{-1}_1,h,h)f_{k,(1),\,\mathrm{melo}}^{(1)}(g_1,\bar{g}_1) \label{T12} \\
\nonumber T_2&:=\int \prod_{l=1}^3dg_ld\bar{g}_ldh \, D_2(\bar{g}_1hg^{-1}_1,\bar{g}_2h{g}^{-1}_2,\bar{g}_3h{g}^{-1}_3)F_{k,(1),\,melo}^{(1)}(\{g_l,\bar{g}_l\})\\
&=\int dg_1d\bar{g}_1dh \, D_2(g_1h\bar{g}^{-1}_1,h,h)f_{k,(1),\,\mathrm{melo}}^{(1)}(g_1,\bar{g}_1) \label{T22}
\end{align}
where $D_1$ and $D_2$ are class functions on $\SU(2)^3$ 
defined as: 
\begin{align}\label{defD1D2}
\nonumber D_1(g_1h\bar{g}^{-1}_1,g_2h\bar{g}^{-1}_2,g_3h\bar{g}^{-1}_3)&=-\frac{\partial_k Z(k)}{k^2 Z(k)^2}\sum_{n,m \in \mathbb{N}}(-u_2(k))^{n+m}\bigg[\prod_{i=0}^{n+m} \int_{0}^1du_i\bigg]\prod_{l=1}^3K_{(\sum_{i=0}^{n+m}u_i)/k^2}(g_lh\bar{g}^{-1}_l)\\
& - \frac{2}{k^3 Z(k)}\sum_{n,m \in \mathbb{N}}(-u_2(k))^{n+m}\bigg[\prod_{i=1}^{n+m}\int_{0}^1du_i\bigg]\prod_{l=1}^3K_{(1+\sum_{i=1}^{n+m}u_i)/k^2}(g_lh\bar{g}^{-1}_l),\\
\nonumber D_2(g_1h\bar{g}^{-1}_1,g_2h\bar{g}^{-1}_2,g_3h\bar{g}^{-1}_3)&= \frac{1}{Z(k)^2 k^2}\sum_{n,m}(-u_2)^{n+m}\bigg[\prod_{i=1}^{n+m+2}\int_{0}^1du_i\bigg]\\
&\qquad \times\frac{\extd}{\extd u_1}\prod_{l=1}^3K_{(\sum_{i=1}^{n+m+2}u_i)/k^2}(g_lh\bar{g}^{-1}_l).
\end{align}
Note that, in the second line, we have used the fact that the heat kernel satisfies the heat equation. 

\

We now need to re-express the right-hand side of equation \eqref{flow2_graph} in terms of generalized tensorial interactions (that is, with derivative operator insertions allowed), and conserve only the operators which already appear on the left-hand side. This can be achieved by means of a Taylor expansion of the traces $T_1$ and $T_2$. More precisely, we introduce the parametrized group element   
\beq
g_1 (t) = g_1 \exp(t X_{g_1^{-1} \bar{g}_1})\,,
\eeq
where $X_g\in \mathfrak{su}(2)$ is the smallest Lie algebra element such that $\exp(X_g)=g$, and then write:
\begin{equation}
f_{k,(1),\,\mathrm{melo}}^{(\ell)}(g_1,\bar{g}_1)=\sum_{n=0}^{+ \infty}\frac{1}{n!}\frac{\extd^{(n)}}{\extd t^n}f_{k,(1),\,\mathrm{melo}}^{(\ell)}(g_1,{g}_1(t))\bigg|_{t=0}\,.
\end{equation}
After some algebra (see e.g. \cite{Carrozza:2013wda} for a similar calculation), we deduce that
\begin{align}
T_a&=Z(k)^2\lambda_4(k) \int d{g}dh D_a({g},h,h)\langle\bar{\phi},\phi\rangle+Z(k)^2 \lambda_4(k) \frac{1}{6}\int dgdh D_a(gh,h,h)|X_g|^2\langle\bar{\phi},\Delta_{g_1}\phi\rangle+\cdots\\
&:=T_a^{(0)}+T_a^{(2)}+\cdots \nn
\end{align}
where $\langle\bar{\phi},\phi\rangle:=\int d\textbf{g}\bar{\phi}(\textbf{g})\phi(\textbf{g})$ and $|.|$ denotes the normalized quadratic norm inherited from the Killing form. The zeroth order term $T_a^{(0)}$ contributes to the flow of the mass, while the second order quantity $T_a^{(2)}$ yields a Laplace operator and hence enters in the flow of $Z(k)$. Given our truncation, the next orders are simply discarded. We are in this way led to the identifications:
\begin{equation}\label{eqm}
\partial_k(Z(k)m^2(k))=3Z(k)^2\lambda_4(k)\int d{g}dh D_1({g},h,h)-3Z(k)^2\lambda_4(k) \partial_kZ(k)\int d{g}dh D_2({g},h,h)\,,
\end{equation}
and 
\begin{equation}\label{eqZ}
\partial_kZ(k)= -Z(k)^2\lambda_4(k) \frac{1}{6}\int dgdh D_1(gh,h,h)|X_g|^2 + Z(k)^2\lambda_4(k) \partial_kZ(k) \frac{1}{6}\int dgdh D_2(gh,h,h)|X_g|^2\,.
\end{equation}
The combinatorial factor $3$ in front of the right-hand side of the first equation comes from the sum over the color index $\ell$ in \eqref{flow2melo}. In the large $k$ limit, the asymptotics of the heat kernel allows to explicitly perform the group integrals, leading to:
\begin{align}\label{D1D2_final}
\nonumber \int d{g}_1dh D_1({g}_1,h,h)&=-2 \sqrt{2} k \frac{\partial_k Z(k)}{Z(k)^2}\int_{0}^{\infty} dx \left(1-e^{-x^2}\right) \left( \frac{x^2}{x^2+u_2(k) (1-e^{-x^2})}\right)^2\\
&\qquad-\frac{4 \sqrt{2}}{Z(k)} \int_{0}^{\infty} dx \, e^{-x^2}  \left( \frac{x^3}{x^2+u_2(k) (1-e^{-x^2})}\right)^2\, ,\\
\nonumber \int d{g}_1dh D_2({g}_1,h,h)&= - \frac{2 \sqrt{2} k}{Z(k)^2} \int_{0}^{\infty} dx \left(1-e^{-x^2}\right)^2\left(\frac{x^2}{x^2+ u_2(k) (1-e^{-x^2})}\right)^2\,,
\end{align}
and
\begin{align}\label{D1D2W_final}
\nonumber \int d{g}_1dh D_1({g}_1,h,h)|X_g|^2&=-\frac{9 \sqrt{2} \partial_k Z(k)}{k Z(k)^2}\int_{0}^{\infty} dx \left(1-e^{-x^2}\right) \left(\frac{x}{x^2+ u_2(k) (1-e^{-x^2})}\right)^2\\
&\qquad-\frac{18 \sqrt{2}}{k^2 Z(k)} \int_{0}^{\infty} dx \, e^{-x^2} \left( \frac{x^2}{x^2+ u_2(k) (1-e^{-x^2})}\right)^2\,, \\
\int d{g}_1dh D_2({g}_1,h,h)|X_g|^2&= - \frac{9 \sqrt{2}}{k Z(k)^2} \int_{0}^{\infty} dx \left(1-e^{-x^2}\right)^2 \left( \frac{x}{x^2+ u_2 (k) (1-e^{-x^2})}\right)^2\,. \nn
\end{align}
Details of the calculation, which relies on mere Gaussian integrations, can be found in Appendix \ref{app_traces}.

\

Finally, we conclude that, within our truncation and in the $k\gg 1$ regime:
\begin{align}
\left(k\partial_k+2+\eta(k)\right) u_2(k)&=-3 \, \bigg[2f_1(u_2(k))+\eta(k) g_1(u_2(k))\bigg] \, u_4(k) \,, \label{eqm2} \\
\eta(k)&=\frac{1}{6}\, u_4 (k) \bigg[2f_w (u_2(k))+\eta(k) g_w(u_2(k))\bigg]\,, \label{eqZ2}
\end{align}
where we have introduced the anomalous dimension
\begin{equation}
\eta(k):=k\partial_k\ln(Z(k)) \,,
\end{equation}
and the functions
\begin{align}
f_1(u_2)&:= 2 \sqrt{2}\int_0^{\infty} \extd x \, e^{-x^2} \left(\frac{x^3}{x^2+ u_2 (1-e^{-x^2})}\right)^2 \,, \label{f1} \\
g_1(u_2)&:= 2 \sqrt{2} \int_0^{\infty} \extd x \, e^{-x^2}\left(1-e^{-x^2}\right) \left( \frac{x^2}{x^2+ u_2 (1-e^{-x^2})}\right)^2\,, \label{g1}\\
f_w(u_2)&:=9 \sqrt{2} \int_0^{\infty} \extd x \, e^{-x^2} \left( \frac{x^2}{x^2+ u_2  (1-e^{-x^2})}\right)^2\,, \label{fw}\\
g_w( u_2) &:= 9 \sqrt{2} \int_0^{\infty} \extd x \, e^{-x^2}\left(1-e^{-x^2}\right) \left( \frac{x}{x^2+ u_2 (1-e^{-x^2})}\right)^2\,. \label{gw}
\end{align}

In the following, we will rely on equation \eqref{eqZ2} to consider $\eta$ as a function of $u_2$ and $u_4$ rather than $k$:
\beq\label{eta_u2u4}
\eta(u_2 , u_4 ) = \frac{1}{3} \frac{f_w (u_2) \, u_4}{1 - \frac{1}{6} g_w (u_2) \, u_4}\,.
\eeq
We will also use the short-hand notation
\beq
L_1 ( u_2 , u_4 ) := 2 f_1 (u_2 )+\eta(u_2 , u_4) g_1 (u_2)\,,
\eeq
which will more generally encapsulate the contributions of melonic loops of length $1$ (i.e. tadpoles).

\

\begin{figure}
\centering
\includegraphics[scale=.5]{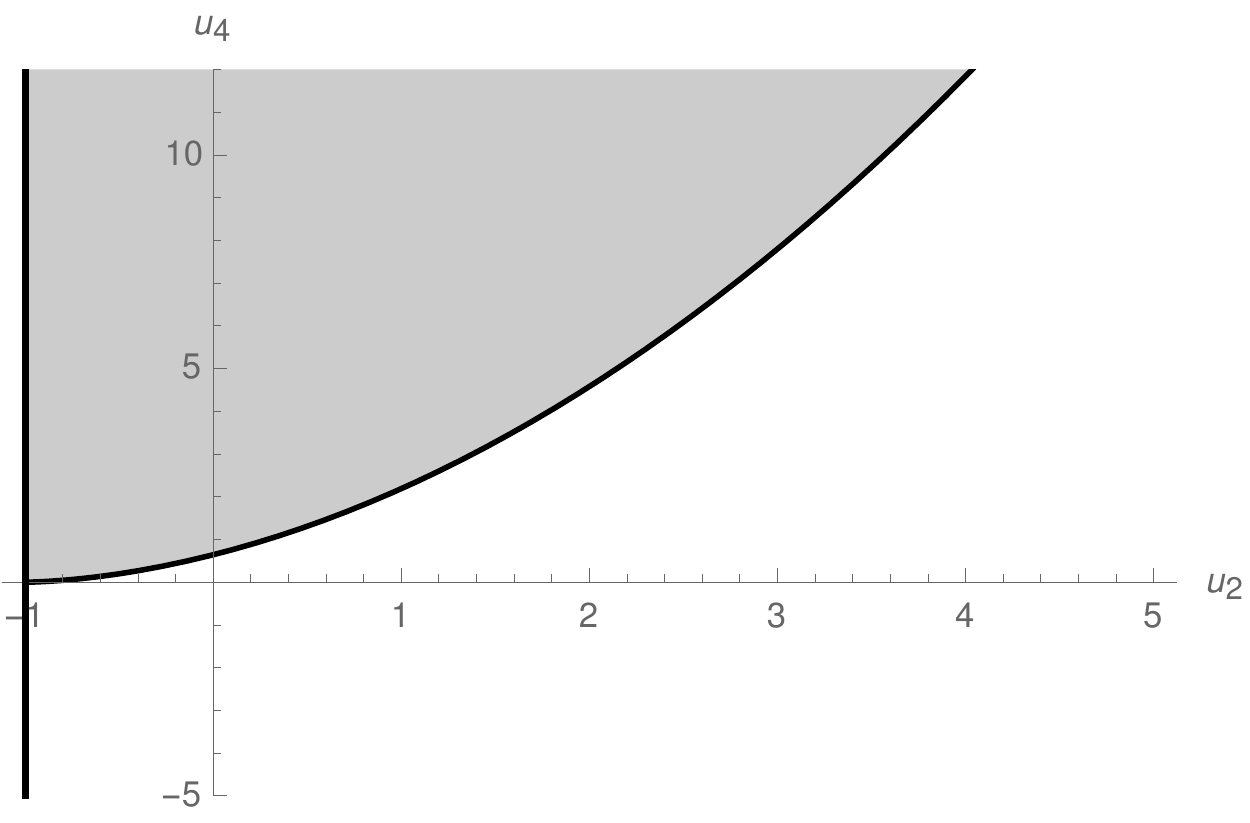} 
\captionof{figure}{The flow equations become singular in the subspaces $\{  g_w (u_2) \, u_4 = 6 \}$ and $\{ u_2 = - 1 \}$. The shaded region, which is separated from the Gaussian fixed point by a singular subspace, is therefore out of reach in our framework.}\label{excluded_region}
\end{figure}

Note that the integrals (\ref{f1}-\ref{gw}) are convergent for $u_2 > - 1$ only. Moreover, from \eqref{eta_u2u4}, we notice that $\eta$ becomes singular when its denominator vanishes. This denominator being strictly positive in the vicinity of the Gaussian fixed point, our flow equations will only be trustworthy in the region $\{  g_w (u_2) \, u_4 < 6 \, , \, u_2 > - 1 \}$. See Figure \ref{excluded_region}. 

\subsubsection{Equation for $u_4$}

The only melonic graphs contributing to the flow equation \eqref{fourpoints} are pictured in Figure \ref{phi4}. As before, the dashed lines are associated to the effective propagator $D(k)=\hat{P}\mathcal{K}_k^{-1}\partial_kR_k\mathcal{K}_k^{-1}\hat{P}$. The dotted line in the rightmost diagram is on the other hand associated to the operator $\mathcal{K}_k^{-1}$. \\

\begin{center}
\includegraphics[scale=1]{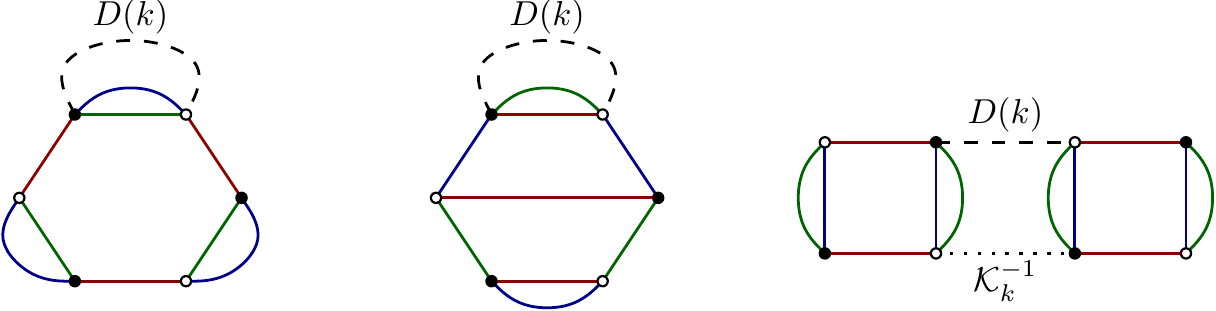} 
\captionof{figure}{Melonic graphs contributing to the flow of $\lambda_4$.}\label{phi4}
\end{center}

We therefore obtain the following equation:
\begin{align}\label{fourpointmelo}
\partial_k \left( Z(k)^2 \frac{\lambda_4 (k)}{2} \right) \, \sum_{\ell = 1}^3 \vcenter{\hbox{\includegraphics[scale=0.8]{Figures/int4.pdf}}} &\approx - Z(k)^3 \, \lambda_{6,1} (k) \, \sum_{\ell=1}^3 \, \vcenter{\hbox{\includegraphics[scale=0.8]{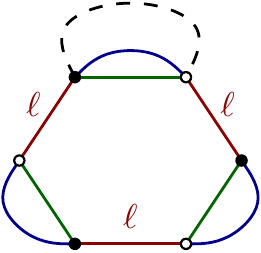}}} \nn \\
&- Z(k)^3 \, \lambda_{6,2} (k) \, \sum_{\ell=1}^3 \, \left( \vcenter{\hbox{\includegraphics[scale=0.8]{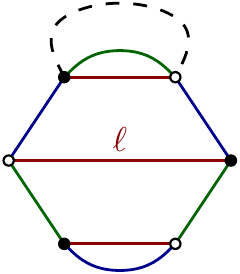}}} + \vcenter{\hbox{\includegraphics[scale=0.8]{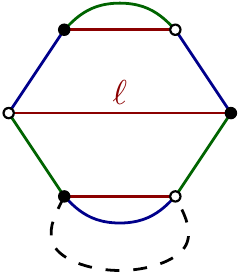}}} \right) \\
&+ \left( Z(k)^2 \lambda_4 (k) \right)^2 \, \sum_{\ell=1}^3 \, \vcenter{\hbox{\includegraphics[scale=0.8]{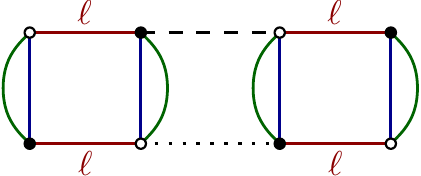}}} \nn
\end{align}
Since we did not include any derivative coupling in our truncation, we need to evaluate each diagram appearing on the right-hand side at zeroth order in a Taylor expansion with respect to its external variables. 

The diagrams appearing in the first two lines of \eqref{fourpointmelo} have the same loop structure as the mass counter-terms computed in the previous subsection, we may therefore proceed identically. We find that:
\begin{align}
\vcenter{\hbox{\includegraphics[scale=0.8]{Figures/phi61_4.pdf}}} &\approx 
\frac{1}{Z(k)} L_1 ( u_2 (k) , u_4 (k) ) \times  \vcenter{\hbox{\includegraphics[scale=0.8]{Figures/int4.pdf}}}\,, \label{approx_phi4_6-1} \\
\vcenter{\hbox{\includegraphics[scale=0.8]{Figures/phi62_4.pdf}}} &\approx 
\frac{1}{Z(k)} L_1 ( u_2 (k) , u_4 (k) ) \times  \vcenter{\hbox{\includegraphics[scale=0.8]{Figures/int4.pdf}}}\,. \label{approx_phi4_6-2}
\end{align}

The contribution of the last line of \eqref{fourpointmelo} can be computed following the same strategy. We have for instance:
\begin{align}
\vcenter{\hbox{\includegraphics[scale=0.8]{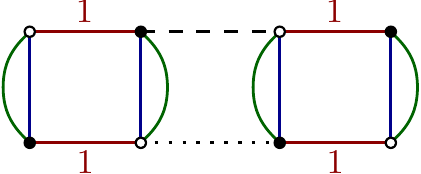}}} &= \int\prod_{l=1}^3 [\extd g_l \extd g'_l \extd\bar{g}_l \extd\bar{g}'_l] \extd h \, D(g_1 h \bar{g}^{\prime\,-1}_1 , g_2 h \bar{g}^{\prime\,-1}_2 , g_3 h \bar{g}^{\prime\,-1}_3 ) \nonumber\\
&\times \delta(g_2 \bar{g}_2^{-1}) \delta(g_3 \bar{g}_3^{-1}) \delta(\bar{g}_2' g_2^{\prime\,-1}) \delta(\bar{g}_3' g_3^{\prime\,-1}) \\
&\times f_{k,(1),\,\mathrm{melo}}^{(1)}(\bar{g}^{\prime}_1,g_1') \mathcal{K}_k^{-1}(g_1' \bar{g}^{-1}_1,g_2' \bar{g}^{-1}_2,g_3' \bar{g}^{-1}_3)f_{k,(1),\,\mathrm{melo}}^{(1)}(\bar{g}_1,{g}_1)\,. \nonumber
\end{align}
In order to project it down onto the basis of (generalized) tensor interactions, one may use:
\beq
f_{k,(1),\,\mathrm{melo}}^{(1)}(\bar{g}^{\prime}_1,g_1') f_{k,(1),\,\mathrm{melo}}^{(1)}(\bar{g}_1,{g}_1) =\sum_{n=0}^{+ \infty}\frac{1}{n!}\frac{\extd^{(n)}}{\extd t^n}f_{k,(1),\,\mathrm{melo}}^{(1)}(\bar{g}^{\prime}_1,g_1'(t)) f_{k,(1),\,\mathrm{melo}}^{(1)}(\bar{g}_1,{g}_1(t))\bigg|_{t=0}\,,
\eeq
with
\begin{equation}
g_1(t):=g_1' \exp(tX_{g_1^{\prime \, -1} g_1})\,,\qquad g_1'(t):= \bar{g}_1 \exp(tX_{\bar{g}_1^{-1} g_1'})\,.
\end{equation}
In accordance with our truncation, we retain only the zeroth order term in this Taylor expansion. Recognizing that 
\begin{equation}
\int \extd \bar{g}_1' \extd \bar{g}_1 f_{k,(1),\,\mathrm{melo}}^{(1)}(\bar{g}^{\prime}_1,\bar{g}_1) f_{k,(1),\,\mathrm{melo}}^{(1)}(\bar{g}_1,\bar{g}_1')= \vcenter{\hbox{\includegraphics[scale=0.8]{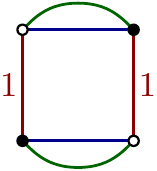}}} \,,
\end{equation}
we obtain:
\begin{align}\label{approx_phi4_4}
\vcenter{\hbox{\includegraphics[scale=0.8]{Figures/phi4_4_1.pdf}}} &\approx \int \extd h \prod_{l=1}^3 [ \extd g_l \extd g'_l ] D(g_1,g_2hg^{\prime\,-1}_2,g_3hg^{\prime\,-1}_3)\mathcal{K}_k^{-1}(g'_1,g^{\prime}_2{g}^{-1}_2,g^{\prime}_3{g}^{-1}_3) \times \vcenter{\hbox{\includegraphics[scale=0.8]{Figures/int4_1.pdf}}}\, \nonumber\\
&\approx \frac{1}{k^2 Z(k)^2} L_2 ( u_2 (k) , u_4 (k) ) \times \vcenter{\hbox{\includegraphics[scale=0.8]{Figures/int4_1.pdf}}}\,,
\end{align}
where
\beq
L_2 ( u _2 , u_4 ) := 2 f_2(u_2 )+\eta( u_2 , u_4) g_2 (u_2) 
\eeq
and
\begin{align}
f_2(u_2) & := 2 \sqrt{2} \int_{0}^{\infty} dx\frac{x^6\, e^{-x^2} \, (1-e^{-x^2})}{\left( x^2+ u_2(1-e^{-x^2})\right)^3}\,,\\
g_2( u_2) & := 2 \sqrt{2} \int_{0}^{\infty} dx\frac{x^4\, e^{-x^2} \, \left(1-e^{-x^2}\right)^2}{\left( x^2+ u_2(1-e^{-x^2})\right)^3}\,.
\end{align}
We refer the reader to Appendix \ref{app_traces} for more detail.

\

All in all, equations \eqref{fourpointmelo}, \eqref{approx_phi4_6-1}, \eqref{approx_phi4_6-2} and \eqref{approx_phi4_4} imply the flow equation:
\begin{align}
(k\partial_k+1+2\eta(k)) u_4(k)&=-2 L_1 ( u_2 (k) , u_4 (k) ) \, \left( u_{6,1}(k)+2 u_{6,2}(k)\right) \\
&\qquad +2 L_2 ( u_2 (k) , u_4 (k) )  \, u_4(k)^2\,. \nn
\end{align}

\subsubsection{Equations for $u_{6,1}$ and $u_{6,2}$}

We now identify the leading melonic contributions to the flow of the marginal couplings $u_{6,1}$ and $u_{6,2}$. 

Let us start with $u_{6,2}$, which follows a simpler equation than $u_{6,1}$. The melonic approximation yields:
\begin{align}\label{u62}
\partial_k \left( Z(k)^3 \lambda_{6,2} (k) \right) \, \sum_{\ell = 1}^3 \vcenter{\hbox{\includegraphics[scale=0.8]{Figures/int62.pdf}}} &\approx 2 Z(k)^5 \, \lambda_{4} (k) \lambda_{6,2} (k) \, \sum_{\ell=1}^3 \, \left( \vcenter{\hbox{\includegraphics[scale=0.8]{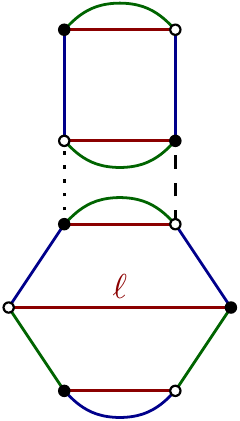}}} + \vcenter{\hbox{\includegraphics[scale=0.8]{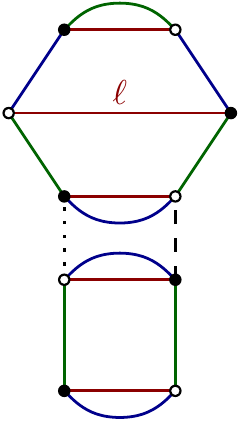}}} \right)\,.
\end{align}
Similarly to \eqref{approx_phi4_4}, we can then derive the tensorial approximation:
\begin{align}\label{approx_phi62_62}
\vcenter{\hbox{\includegraphics[scale=0.8]{Figures/phi62_62.pdf}}} &\approx \frac{1}{k^2 Z(k)^2} 
L_2 (u_2(k) , u_4 (k) ) \times \vcenter{\hbox{\includegraphics[scale=0.8]{Figures/int62.pdf}}}\,,
\end{align}
and conclude that:
\begin{align}
(k\partial_k+3\eta(k)) u_{6,2}(k) =4 L_2 (u_2(k) , u_4 (k) ) \, u_4(k)  u_{6,2}(k) \,.
\end{align}\\

The melonic flow equation for $u_{6,1}$ is slightly more involved, as graphs with three vertices also come in: 
\begin{align}\label{u61}
\partial_k \left( Z(k)^3 \frac{\lambda_{6,1} (k)}{3} \right) \, \sum_{\ell = 1}^3 \vcenter{\hbox{\includegraphics[scale=0.8]{Figures/int61.pdf}}} &\approx 2 Z(k)^5 \, \lambda_{4} (k) \lambda_{6,1} (k) \, \sum_{\ell=1}^3 \, \vcenter{\hbox{\includegraphics[scale=0.8]{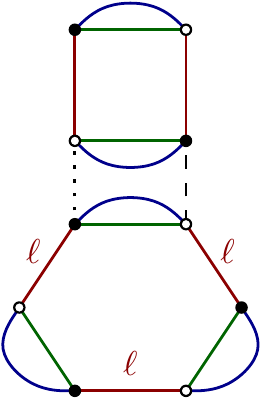}}}\\
&\qquad -  Z(k)^6 \, \lambda_{4} (k)^3 \, \sum_{\ell = 1}^3 \vcenter{\hbox{\includegraphics[scale=0.8]{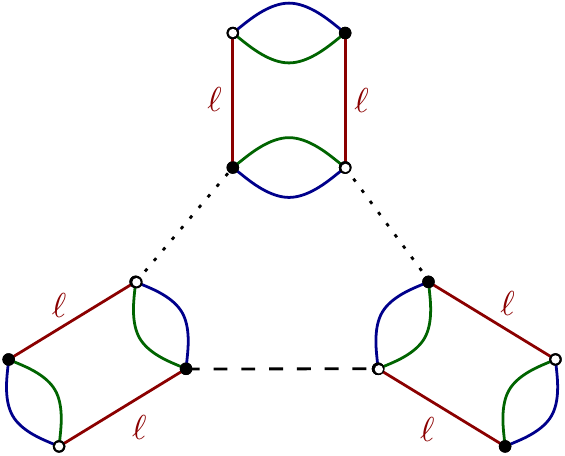}}}. \nn
\end{align}
The local approximation of the diagrams appearing in the first line can be computed again as in \eqref{approx_phi4_4} and \eqref{approx_phi62_62}, yielding:
\begin{align}\label{approx_phi61_61}
\vcenter{\hbox{\includegraphics[scale=0.8]{Figures/phi61_61.pdf}}} &\approx \frac{1}{k^2 Z(k)^2} 
L_2 (u_2(k) , u_4 (k) ) \times \vcenter{\hbox{\includegraphics[scale=0.8]{Figures/int61.pdf}}}\,.
\end{align}
The diagrams appearing in the second line of \eqref{u61} have loops of length three, and must therefore be computed independently. Following the same method as before, we find:
\begin{align}\label{approx_phi61_3}
\vcenter{\hbox{\includegraphics[scale=0.8]{Figures/phi61_3.pdf}}} &\approx \frac{1}{k^4 Z(k)^3} 
L_3 (u_2(k) , u_4 (k) ) \times \vcenter{\hbox{\includegraphics[scale=0.8]{Figures/int61.pdf}}}\,,
\end{align}
with $L_3 ( u_2 , u_4 ) := 2 f_3 (u_2 )+ \eta( u_2 , u_4 ) g_3 ( u_2 )$ and:
\begin{align}
f_3(u_2)& :=2 \sqrt{2}\int_{0}^{+\infty} dx\frac{e^{-x^2} \, x^6 \left(1-e^{-x^2}\right)^2}{\left(x^2+u_2 (1-e^{-x^2})\right)^4}\,,\\
g_3(u_2) &:=2 \sqrt{2}\int_{0}^{+\infty} dx \frac{e^{-x^2} \, x^4 \left(1-e^{-x^2}\right)^3}{\left(x^2+u_2 (1-e^{-x^2})\right)^4}\,.
\end{align}
All in all, we obtain the flow equation:
\begin{align}
(k\partial_k+3\eta(k))u_{6,1}(k)&=6 L_2 (u_2(k) , u_4 (k) ) \, u_4(k) u_{6,1}(k) \\
&\qquad -3 L_3 (u_2(k) , u_4 (k) ) \, u_4(k)^3\,. \nn
\end{align}

\subsubsection{Summary: flow equations for renormalized dimensionless couplings}\label{summarize}

In summary, the $\beta$-functions in the $\phi^6$ melonic truncation and in the large $k$ regime are the following:
\begin{equation}\label{flow_phi6_full}
\left\lbrace\begin{split}
\beta_{2} &= - \left( 2 + \eta \right) u_2 -3 L_1 (u_2 , u_4 ) \, u_4 
\\
\beta_4&= - \left( 1 + 2 \eta \right) u_4 - 2 L_1 (u_2 , u_4 ) \, \left(u_{6,1}+2 u_{6,2} \right)+2 L_2 (u_2 , u_4 ) \, {u_4}^2 
\\
\beta_{6,1} &= - 3 \eta \, u_{6,1} + 6 L_2 (u_2 , u_4 ) \, u_4 \, u_{6,1} - 3 L_3 (u_2 , u_4 ) \, {u_4}^3 
\\
\beta_{6,2} &= - 3 \eta \, u_{6,2} + 4 L_2 (u_2 , u_4 ) \, u_4 \, u_{6,2} 
\end{split}\right.
\end{equation}
where the anomalous dimension is given by
\begin{equation}\label{eqZ23}
\eta =\frac{1}{3} \frac{f_w (u_2) \, u_4}{1 - \frac{1}{6} g_w (u_2) \, u_4}\,.
\end{equation}
The rest of this section is devoted to the analysis of this four-dimensional closed autonomous system.

\subsection{Fixed points and phase diagrams}

\subsubsection{Vicinity of the Gaussian fixed point}

The system \eqref{flow_phi6_full} admits a trivial fixed point at vanishing coupling constants: $u_2 = u_4 = u_{6,1} = u_{6,2} = 0$. This is the standard Gaussian Fixed Point (GFP), which has already been studied at length in the series of papers \cite{Carrozza:2013wda, Carrozza:2014rba, Carrozza:2014rya}. Let us briefly comment on how its properties may be recovered in the present framework. 

Taylor expanding the system \eqref{flow_phi6_full} around the GFP, we obtain the universal perturbative $1$-loop expression of the flow:
\begin{equation}\label{flow_6}
\left\lbrace\begin{split}
\beta_{2} &\approx - 2 u_2 -6 f_1(0)\, u_4
\\
\beta_4&\approx - u_4 - 4 f_1(0)\, \left( u_{6,1}+2 u_{6,2} \right) - \left( \frac{2}{3} f_w (0) - 4 f_2 (0) \right) \, {u_4}^2 
\\
\beta_{6,1} &\approx - \left( f_w (0) - 12 f_2 (0) \right)\, u_4 \, u_{6,1} - 6 f_3(0) \, {u_4}^3 
\\
\beta_{6,2} &\approx - \left( f_w (0) - 8 f_2 (0) \right)\, u_4 \, u_{6,2}
\end{split}\right.
\end{equation}
where the anomalous dimension itself has been expanded at one loop:
\begin{equation}
\eta \approx \frac{1}{3} f_w (0) \, u_4\,.
\end{equation}
Remarking that
\begin{equation}
f_w(0) = \frac{9}{2} \sqrt{2 \pi}\,, \qquad f_1(0) = \sqrt{\frac{\pi}{2}} \,, \qquad f_2(0) = (\sqrt{2} - 1) \sqrt{\pi}\,, \qquad f_3 (0) = (\sqrt{11 - 4 \sqrt{6}} - 1) \sqrt{\pi}\,, 
\end{equation}
this results in the following numerical evaluation of the $\beta$ functions:
\begin{equation}\label{taylor_flow}
\left\lbrace\begin{split}
\beta_{2} &\approx - 2 \,u_2 - 7.5 \, u_4
\\
\beta_4&\approx - u_4 - 5.0 \, \left( u_{6,1}+2 u_{6,2} \right) - 4.6 \, {u_4}^2 
\\
\beta_{6,1} &\approx - 2.5 \, u_4 \, u_{6,1} - 1.0 \, {u_4}^3 
\\
\beta_{6,2} &\approx - 5.4 \, u_4 \, u_{6,2}
\end{split}\right.
\end{equation} 

\

Denoting by $\{g_i\}=\{u_2,u_4,u_{6,1},u_{6,2}\}$ the set of coupling constants, the stability matrix at a fixed point $\{g_i = g_i^*\}$ is defined as 
\beq
\beta_{ij}:=\restr{\frac{\partial\beta_i}{\partial g_j}}{g_l=g_l^*}
\eeq
Its eigenvalues are the opposite of the usual critical exponents. For the GFP, this stability matrix reads: 
\begin{equation}[\beta^{\mathrm{GFP}}_{ij}]=\begin{pmatrix}
-2&-3\sqrt{2\pi}&0&0\\
0&-1&-2\sqrt{2\pi}&-4\sqrt{2\pi}\\
0&0&0&0\\
0&0&0&0
\end{pmatrix},
\end{equation}
with eigenvectors $e_1^{\mathrm{GFP}}= (1,0,0,0)^{\mathrm{T}}$, $e_2^{\mathrm{GFP}}=(-3\sqrt{2\pi},1,0,0)^{\mathrm{T}}$, 
$e_3^{\mathrm{GFP}}=(0,0,-2,1)^{\mathrm{T}}$ and $e_4^{\mathrm{GFP}}=(48 \pi,-16 \sqrt{2 \pi},-9\sqrt{2}+6,1+ 9\frac{\sqrt{2}}{2})^{\mathrm{T}}$; they respectively have eigenvalues $-2$, $-1$, $0$ and $0$. Hence, the first two directions correspond to relevant operators, with canonical dimensions\footnote{Recall that the critical exponent for the GFP coincides with the canonical dimension defined at the end of Section \ref{section1}.} $2$ and $1$, and the two remaining directions correspond to marginal couplings with vanishing canonical dimensions.

\

\begin{figure}
\centering
\includegraphics[scale=.5]{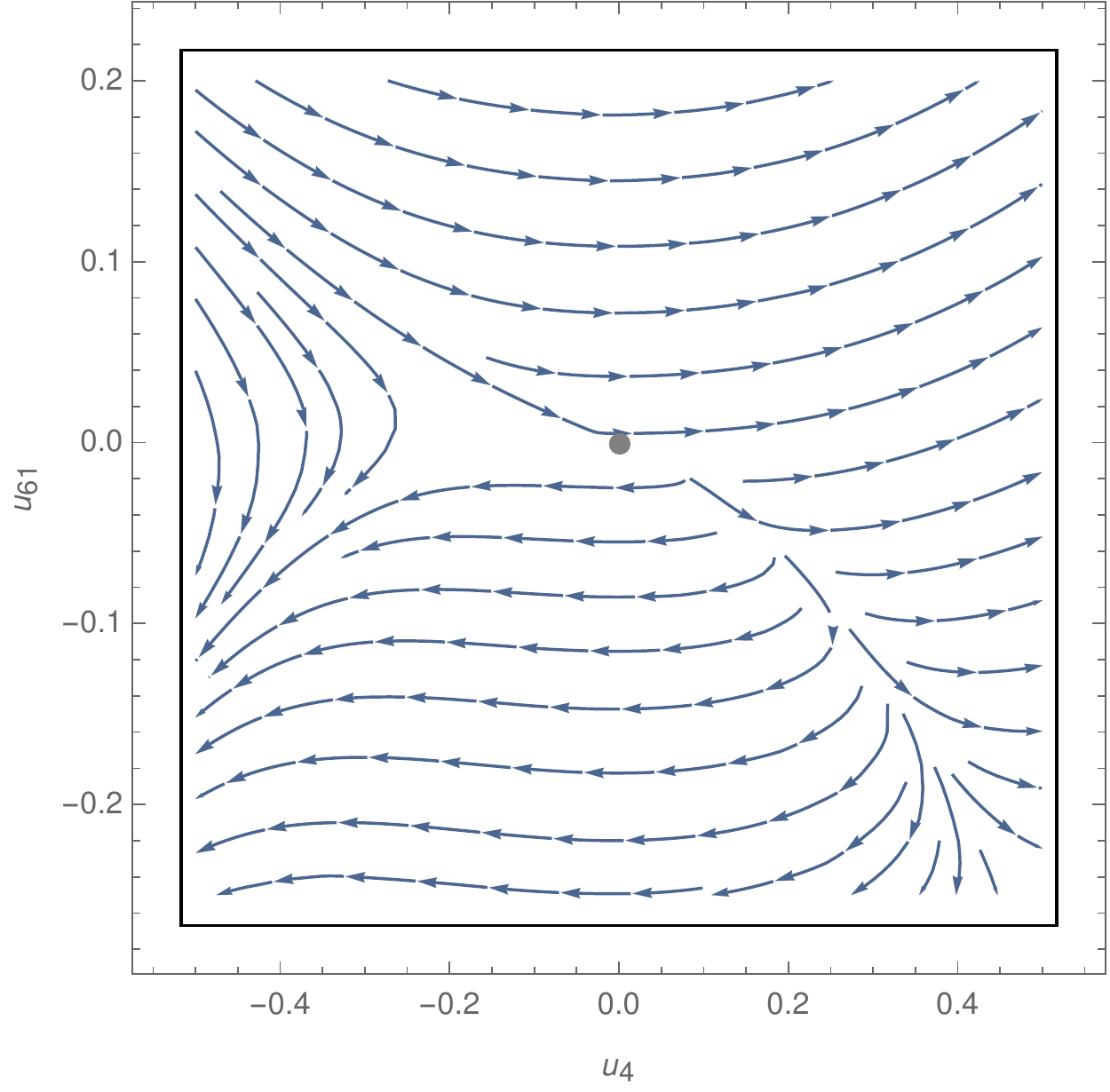} 
\captionof{figure}{Phase diagram around the Gaussian fixed point (represented as a gray dot), in the plane $\{ u_2 = 0 , u_{6,2} = 0 \}$. The arrows point towards the infrared (i.e. low spin modes).}\label{diag_gfp}
\end{figure}

Note that, due to the presence of marginally relevant directions (i.e. those with vanishing critical exponents), the qualitative behaviour of renormalization group trajectories in the vicinity of the origin is not fully determined by the linearized version of \eqref{flow_phi6_full}; one needs to retain the quadratic terms appearing in \eqref{taylor_flow}. This question was first investigated in \cite{Carrozza:2014rba} in the context of a discrete version of the renormalization group flow. It was in particular shown that trajectories with positive coupling constants $u_{6,1}$ and $u_{6,2}$ in the infrared are necessarily repelled from the GFP in the ultraviolet, and are therefore not asymptotically free. This behaviour is recovered in the present realization of the flow equations, as illustrated in Figure \ref{diag_gfp}.  We finally point out that \cite{Carrozza:2014rba} was incorporating $2$-loop radiative corrections which we do not see in the present truncation. As usual, this is due to the fact that the truncation of the Wetterich--Morris equation relies on a Taylor expansion in the mean field $\phi$, rather than on a loop expansion. Hence the perturbative higher loop contributions can only be recovered from a higher-dimensional truncation of the theory space. In particular, we could in principle recover the $2$-loop terms of \cite{Carrozza:2014rba} from the $\phi^8$ truncation we will introduce in the next section.

\subsubsection{One-parameter family of non Gaussian fixed points}

As was already observed in similar $\phi^6$ truncations of tensorial field theories \cite{Geloun:2016xep, Lahoche:2016xiq}, we remark the presence of a one-parameter family of Non Gaussian Fixed Points (NGFPs). In contrast to ordinary local scalar field theories, we have two independent marginal couplings at our disposal ($u_{6,1}$ and $u_{6,2}$), which can compensate each other in the flow equation for $u_4$. Henceforth, for any $s \in \mathbb{R}$ we notice that:
\beq\label{line_NGFP}
u^*(s) := \left( u_2^* , \, u_4^*, \, u_{6,1}^* , \, u_{6,2}^* \right) (s):= \left( 0, \, 0 , \, s , \, - s/2\right)
\eeq 
is a fixed point of the truncated flow \eqref{flow_phi6_full}. This constitutes a one-parameter family of fixed points, which is moreover connected to the GFP $u^{*}(0)$.

\

The stability matrix at $u^*(s)$ is 
\begin{align}
[\beta_{ij}](s)
&=\begin{pmatrix}
-2&-3\sqrt{2\pi}&0&0\\
0&-1&-2\sqrt{2\pi}&-4\sqrt{2\pi}\\
0&-(12 - \frac{15\sqrt{2}}{2})\sqrt{\pi} \, s&0&0\\
0&(8-\frac{7\sqrt{2}}{2}) \sqrt{\pi} \, \frac{s}{2}&0&0
\end{pmatrix} 
\end{align}
For small enough $s$\footnote{That is for $s < \frac{1}{32(2 - \sqrt{2})\pi} \approx 0.017$.}, the critical exponents (which are minus the eigenvalues) are real and equal to:
\begin{align}
\theta_1(s) &= 2 \,,\\
\theta_2(s) &=  \frac{1}{2} \left( 1 + \sqrt{1 - (64 - 32 \sqrt{2}) \pi s} \right) \,,\\
\theta_3(s) &= 0 \,,\\
\theta_4(s) &=  \frac{1}{2} \left( 1 - \sqrt{1 - (64 - 32 \sqrt{2}) \pi s} \right)\,,
\end{align}
with corresponding eigenvectors $e_1(s)= e_1^{\mathrm{GFP}}$, $e_2(s)$, 
$e_3(s)=e_3^{\mathrm{GFP}}$ and $e_4(s)$, such that $e_2(0) = e_2^{\mathrm{GFP}}$ and $e_4(0)=e_4^{\mathrm{GFP}}$. Note that in the small $s$ regime, one furthermore has:
\begin{align}
\theta_2(s) & \underset{s\to 0}{\approx}  1 - \left( 16 - 8 \sqrt{2} \right) \pi \, s \,, \\
\theta_4(s) & \underset{s\to 0}{\approx}  \left( 16 - 8 \sqrt{2} \right) \pi \, s \,.
\end{align}
This means in particular that $e_4 (s)$ is a relevant direction when $s\geq 0$ and irrelevant when $s<0$. 

Interestingly, note that we are also able to determine the sign of the action in the vicinity of $u^{*}(s)$: it is positive (and hence bounded from below) for $s>0$, and negative for $s<0$. Therefore, as far as the convergence of the path-integral is concerned, the region $s>0$ should be favoured. See Appendix \ref{appendix_sign}.

A key question to address is whether such a continuous family of fixed points is realizable in a tensorial field theory, as has been suggested in \cite{Geloun:2016xep}, or if it is on the contrary a mere pathological feature of low order truncations. Similarly to \cite{Lahoche:2016xiq}, we will argue in favor of the latter in section \ref{sec:order8}.

\subsubsection{Isolated non Gaussian fixed points}

We relied on Mathematica to compute numerical solutions of the fixed point equations, and found three candidate fixed points. The technical difficulties encountered in this analysis have to do with the complicated $u_2$-dependence of the loop integrals $L_n (u_2, u_4)$, which in particular prevents a direct use of the built-in Mathematica numerical equation solvers. We instead had to rely on piecewise polynomial interpolations of the loop integrals and on partial solutions to the flow equations in order to select candidate fixed point values of $u_2$. We have in this way systematically scanned the parameter space $\{u_2 > -1\}$. Our findings are summarized in Table \ref{fixedpoints_6_full}, in which $X:= 1 - 1/6 g_w (u_2^*) u_4^*$ is the value of the denominator of $\eta$ at a given fixed point $u^* = ( u_2^* , u_4^* , u_{6,1}^* , u_{6,2}^* )$.

\begin{table}[ht]
\centering
\begin{tabular}{|c||c|c|c|c|c|c|c|}
\hline
Fixed points & $u_2^*$ & $u_4^*$ & $u_{6,1}^*$ & $u_{6,2}^*$ & Critical exponents & $\eta$ & X \\ \hline\hline
$\mathrm{FP}_1$ & $2.7$ & $-2.9$ & $-0.9$ & $0.$ & $( 2.7 , -1.7 , -1.6 , -0.31 )$ & $-0.82$ & $1.4$ \\ \hline
$\mathrm{FP}_2$  & $-0.35$ & $0.063$ & $-0.011$ & $0.$ & $( 3.0 , 1.3 , 0.77 , - 0.51 )$ & $0.60$ & $0.80$ \\ \hline
{\color{mygray}$\mathrm{FP}_3$}  & {\color{mygray}$-0.78$} & {\color{mygray}$0.15$} & {\color{mygray}$-0.11$} & {\color{mygray}$0.$} & {\color{mygray}$( -1.7, -4.8, -5.0, -17 )$} & {\color{mygray}$-3.3$} & {\color{mygray}$-1.9$} \\ \hline
\end{tabular}
\caption{Isolated non-Gaussian fixed points in the $\vphi^6$ truncation.}\label{fixedpoints_6_full}
\end{table}

We first notice that, while $\mathrm{FP}_1$ and $\mathrm{FP}_2$ both have $X > 0$ and are therefore on the same side of the singularity surface as the Gaussian fixed point, $\mathrm{FP}_3$ is not. We have no way of making sense of $\mathrm{FP}_3$ in our parametrization and must therefore discard it (this is why we have used a different nuance of gray for $\mathrm{FP}_3$ in Table \ref{fixedpoints_6_full}). 

We are left with $\mathrm{FP}_1$ and $\mathrm{FP}_2$. Interestingly, both of them lie within the subspace $\{ u_{6,2} = 0 \}$, which is stable under the flow \eqref{flow_phi6_full}. From the signs of the coupling constants (positive $u_2$ and negative $u_4$), we may assume that $\mathrm{FP}_1$ is the $\varepsilon \to 1$ incarnation of the fixed point found in the $\varepsilon$-expansion of \cite{Carrozza:2014rya}. $\mathrm{FP}_1$ has one relevant direction and three irrelevant directions\footnote{Recall that in this paper, the term relevant (resp. irrelevant) is understood as \emph{infrared} relevant (resp. irrelevant).}, while $\mathrm{FP}_2$ has three irrelevant directions and one relevant direction. At this stage, it might be tempting to conjecture that one of them may play the role of ultraviolet fixed point, the other  being an infrared fixed point characterizing one of the possible phases of the theory. However, the only way to support this hypothesis would be to refine the truncation and check that fixed points with similar qualitative features are reproduced. In particular, evidences about the ultraviolet (resp. infrared) nature of a given fixed point may be gathered if its number of relevant (resp. irrelevant) directions is stable under refinement of the truncation. In the next sections, we will produce such evidences for $\mathrm{FP}_1$ but not for $\mathrm{FP}_2$, whose nature will henceforth remain largely open. 

\

\begin{figure}
\centering
\subfloat[]{\label{plane_u2u4}\includegraphics[scale=0.5]{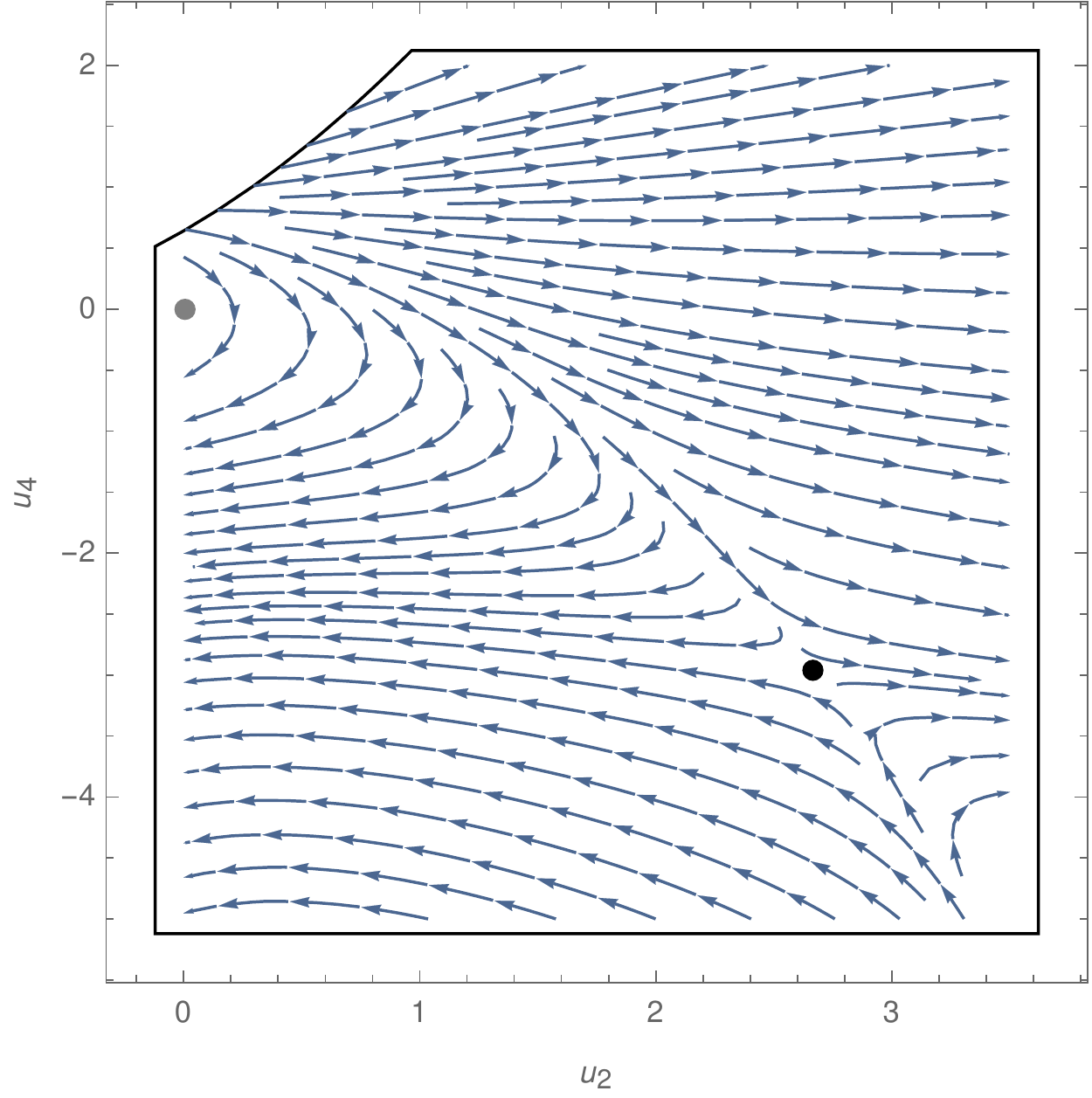}}
\hspace{1cm}\subfloat[]{\label{plane_wz}\includegraphics[scale=0.5]{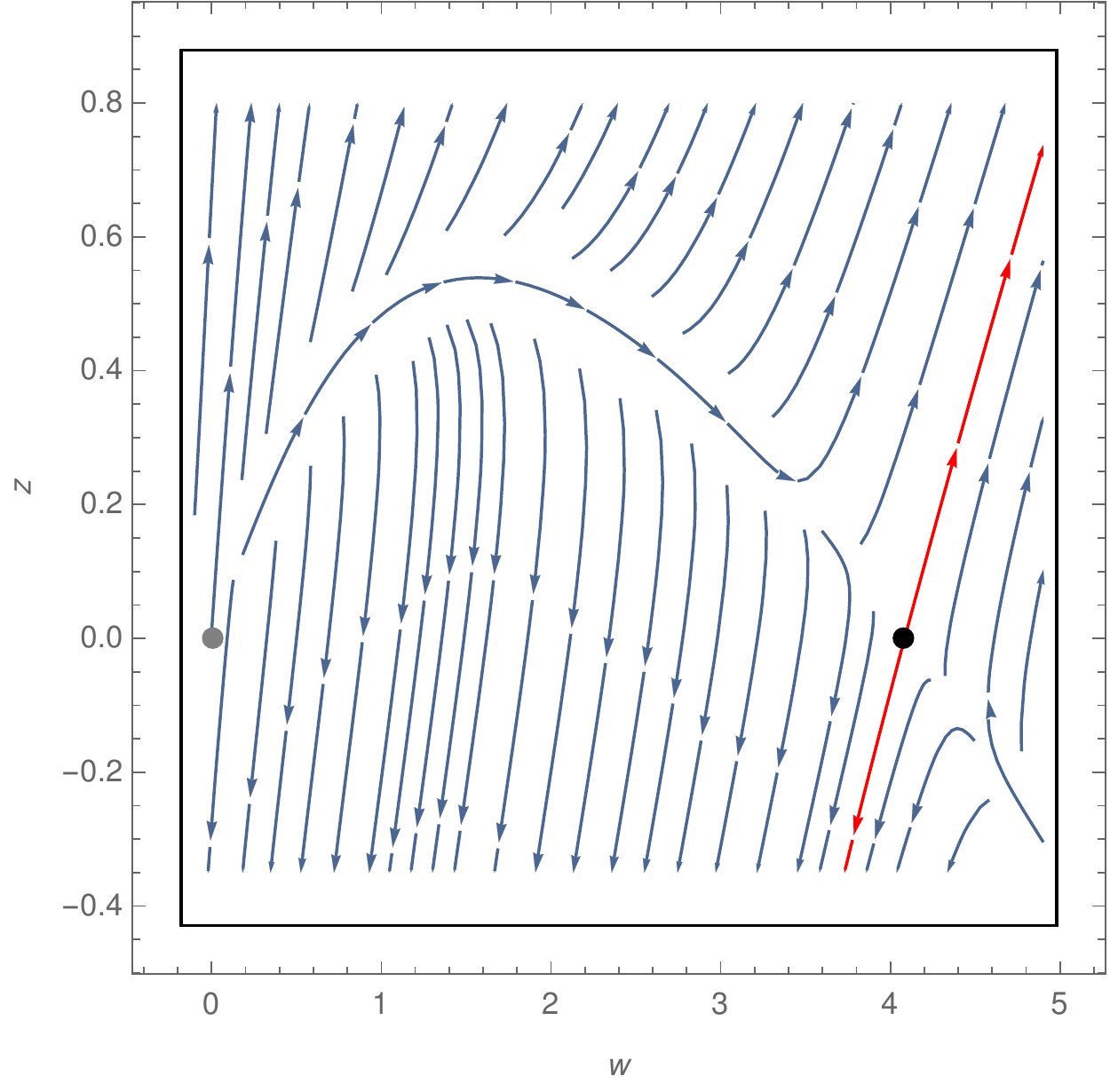}}
\caption{Projections of the renormalization group flow onto particular planes of the four-dimensional $\vphi^6$ theory space. The black (resp. grey) dot represents the non-Gaussian fixed point $\mathrm{FP}_1$ (resp. the Gaussian fixed point), and the arrows point towards the infrared. Figure \ref{plane_u2u4}: projection onto the plane $\{ u_{62} = 0 , u_{61} = u_{61}^* ( \mathrm{FP}_1 )\}$; the boundary in the upper left corner is a singularity of our truncation ($g_w ( u_2 ) u_4 = 6$). Figure \ref{plane_wz}: projection onto the plane containing the origin, the non-Gaussian fixed point $\mathrm{FP}_1$, and its relevant direction $\mathbf{V}$; $w$ is a (normalized) coordinate along the line connecting the origin to $\mathrm{FP}_1$, and $z$ is a parameter along its orthogonal direction; the red trajectories arise from $\mathrm{FP}_1$ in the directions $\pm \mathbf{V}$, and $\mathbf{V}$ is pointing upwards in this diagram.}\label{diag_ngfp}
\end{figure}

Let us focus on the interesting features of $\mathrm{FP}_1$. The vector:
\beq
\mathbf{V} \approx (0.98, -0.18 , 0.14 , 0. )^{\mathrm{T}}
\eeq 
is a normalized relevant eigenvector, with critical exponent approximately equal to $2.7$. In Figure \ref{diag_ngfp} are represented two slices of the vector field $(-\beta_2 , - \beta_4 , - \beta_{6,1} , - \beta_{6,2})$. In particular, Figure \ref{plane_wz} gives a good qualitative picture of the influence of $\mathrm{FP}_1$ in the theory space. A point in the vicinity of $\mathrm{FP}_1$ will be dragged along one of two possible trajectories, which we highlighted in red (the upper part of the diagram corresponds to the direction $\mathbf{V}$, and the lower part to $-\mathbf{V}$). This suggests the existence of two distinct low energy phases, with the one parameter family of models generated by $\pm \mathbf{V}$ interpolating between them. 

\section{Consistency under refinement: order 8 contributions}\label{sec:order8}

We now extend the previous truncation up to order $8$ melonic interactions. This allows to test the robustness of the fixed points found at order $6$. We will confirm in this regard that the one-dimensional set of fixed points found in the previous section is an artefact of the truncation, whereas the isolated fixed points $\mathrm{FP}_1$ and $\mathrm{FP}_2$ are reproduced in the finer truncation. This is particularly true for $\mathrm{FP}_1$, which will be shown to have identical qualitative characteristics in the two truncations.

\subsection{Determination of the flow equations}

\subsubsection{Order $8$ melonic truncation}

One can show that there exists exactly $5$ order-$8$ melonic bubbles up to automorphisms, leading to the order-$8$ melonic truncation of the effective average action:
\begin{align}\label{ansatz8}
\Gamma_k &= -Z(k) \sum_{\ell=1}^3 \vcenter{\hbox{\includegraphics[scale=0.8]{Figures/int2_laplace.pdf}}}  + Z(k) k^2 u_2(k) \vcenter{\hbox{\includegraphics[scale=0.8]{Figures/int2.pdf}}} \nn \\
& \quad + Z(k)^2 k \frac{u_4 (k)}{2}  \sum_{\ell=1}^3 \vcenter{\hbox{\includegraphics[scale=0.8]{Figures/int4.pdf}}} + Z(k)^3 \frac{u_{6,1} (k)}{3} \sum_{\ell=1}^3 \vcenter{\hbox{\includegraphics[scale=0.8]{Figures/int61.pdf}}} + Z(k)^3 {u_{6,2} (k)} \sum_{\ell = 1}^3 \vcenter{\hbox{\includegraphics[scale=0.8]{Figures/int62.pdf}}} \nn \\
& \quad + \frac{Z(k)^4}{k} \left( \frac{u_{8,1} (k)}{4}  \sum_{\ell=1}^3 \vcenter{\hbox{\includegraphics[scale=0.8]{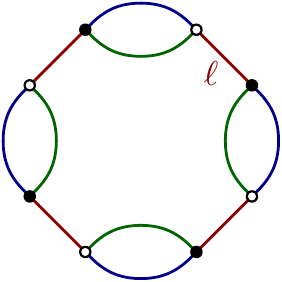}}} + u_{8,2} (k) \sum_{\substack{\ell, \ell'=1\\ \ell\neq \ell'}}^3 \vcenter{\hbox{\includegraphics[scale=0.8]{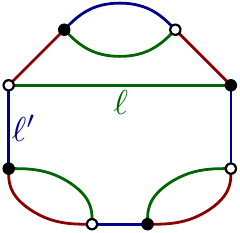}}} + \frac{u_{8,3} (k)}{2} \sum_{\substack{\ell, \ell'=1\\ \ell\neq \ell'}}^3 \vcenter{\hbox{\includegraphics[scale=0.8]{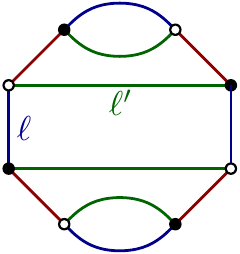}}} \right. \nn \\
& \qquad\qquad\qquad \left. +  \, u_{8,4} (k) \sum_{\ell = 1}^3 \vcenter{\hbox{\includegraphics[scale=0.8]{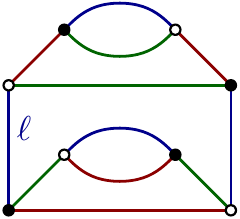}}} + u_{8,5} (k) \vcenter{\hbox{\includegraphics[scale=0.8]{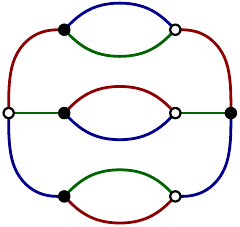}}}
\vphantom{\sum_{\ell=1}^3 \vcenter{\hbox{\includegraphics[scale=0.8]{Figures/int81.pdf}}}} 
\right)
\end{align}   
To make sure that the terms appearing in the last two lines exhaust the set of order-$8$ melonic bubbles, one may resort to the known enumeration of melonic bubbles \cite{razvan_virasoro, Bonzom:2012zf, Geloun:2013kta}. In dimension $3$, melonic bubbles with $2p$ nodes are in one-to-one correspondence with certain equivalent classes of rooted $3$-ary graphs with $p$ vertices, each representative in a given class being associated to a choice of root melon. The total number of rooted $D$-ary graphs with $p$ vertices is known to be:
\beq
C_p^{(D)} = \frac{(Dp)!}{p! [(D-1)p+1]!}\,,
\eeq 
hence $C_4^{(3)} = 55$. One can evaluate the number of bubbles and number of tree representatives associated to each type of order-$8$ interaction appearing in \eqref{ansatz8}, and check that they indeed sum up to $C_4^{(3)} = 3 \times 1 + 6 \times 4 + 6 \times 2 + 3 \times 4 + 1 \times 4$ (see Table \ref{comb_8}). Similarly to order-$4$ and -$6$ melonic bubbles, we have normalized each order-$8$ coupling constant by the number of automorphisms, which is non-trivial for bubbles of types $(8,1)$ and $(8,3)$ only. 
\begin{table}[ht]
\centering
\begin{tabular}{|c|c|c|c|}
\hline
Type & Structure & Number of graphs & Number of $3$-ary representatives\\ \hline\hline
$(8,1)$  & $\vcenter{\hbox{\includegraphics[scale=0.8]{Figures/int81.pdf}}}$ & $3$ & $1$ \\ \hline
$(8,2)$  & $\vcenter{\hbox{\includegraphics[scale=0.8]{Figures/int82.pdf}}}$ & $6$ & $4$ \\ \hline
$(8,3)$  & $\vcenter{\hbox{\includegraphics[scale=0.8]{Figures/int83.pdf}}}$ & $6$ & $2$ \\ \hline
$(8,4)$  & $\vcenter{\hbox{\includegraphics[scale=0.8]{Figures/int84.pdf}}}$ & $3$ & $4$ \\ \hline
$(8,5)$  & $\vcenter{\hbox{\includegraphics[scale=0.8]{Figures/int85.pdf}}}$ & $1$ & $4$ \\ \hline 
\end{tabular}
\caption{Enumeration of order-$8$ melonic interactions and of their $3$-ary tree representatives.}\label{comb_8}
\end{table}

The analogues of equations \eqref{twopoints}, \eqref{fourpoints} and \eqref{sixpoints} in the $\phi^8$ truncation are:
\begin{align}
\partial_k\Gamma_{k,(1)}&=-\Tr\big[\partial_k R_k \mathcal{K}_k^{-1} F_{k,(1)}\mathcal{K}_k^{-1}\hat{P}\big]\,, \label{twopoints8} \\
\partial_k\Gamma_{k,(2)}&=-\Tr\big[\partial_k R_k \mathcal{K}_k^{-1} F_{k,(2)}\mathcal{K}_k^{-1}\hat{P}\big] + \Tr\big[\partial_k R_k \mathcal{K}_k^{-1} (F_{k,(1)} \mathcal{K}_k^{-1})^2\hat{P}\big]\,, \label{fourpoints8} \\
\partial_k\Gamma_{k,(3)} &= -\Tr\big[\partial_k R_k \mathcal{K}_k^{-1} F_{k,(3)}\mathcal{K}_k^{-1}\hat{P}\big] + \Tr\big[\partial_k R_k \mathcal{K}_k^{-1} F_{k,(1)} \mathcal{K}_k^{-1}F_{k,(2)}\mathcal{K}_k^{-1}\hat{P}\big] \nn \\ 
& \qquad + \Tr\big[\partial_k R_k \mathcal{K}_k^{-1} F_{k,(2)} \mathcal{K}_k^{-1}F_{k,(1)}\mathcal{K}_k^{-1}\hat{P}\big] - \Tr\big[\partial_k R_k \mathcal{K}_k^{-1} (F_{k,(1)} \mathcal{K}_k^{-1})^3\hat{P}\big]\,, \label{sixpoints8} \\
\partial_k\Gamma_{k,(4)} &=  \Tr\big[\partial_k R_k \mathcal{K}_k^{-1} F_{k,(1)} \mathcal{K}_k^{-1}F_{k,(3)}\mathcal{K}_k^{-1}\hat{P}\big] + \Tr\big[\partial_k R_k \mathcal{K}_k^{-1} F_{k,(3)} \mathcal{K}_k^{-1}F_{k,(1)}\mathcal{K}_k^{-1}\hat{P}\big] \nn \\ 
& \qquad + \Tr\big[\partial_k R_k \mathcal{K}_k^{-1} (F_{k,(2)} \mathcal{K}_k^{-1})^2\hat{P}\big] + \Tr\big[\partial_k R_k \mathcal{K}_k^{-1} (F_{k,(1)} \mathcal{K}_k^{-1})^4\hat{P}\big]\nn \\ 
& \qquad - \Tr\big[\partial_k R_k \mathcal{K}_k^{-1} (F_{k,(1)} \mathcal{K}_k^{-1})^2 F_{k,(2)}\mathcal{K}_k^{-1}\hat{P}\big] - \Tr\big[\partial_k R_k \mathcal{K}_k^{-1} F_{k,(1)} \mathcal{K}_k^{-1} F_{k,(2)}\mathcal{K}_k^{-1} F_{k,(1)}\mathcal{K}_k^{-1}\hat{P}\big] \nn \\
& \qquad -  \Tr\big[\partial_k R_k \mathcal{K}_k^{-1} F_{k,(2)}\mathcal{K}_k^{-1} (F_{k,(1)}\mathcal{K}_k^{-1})^2 \hat{P}\big] \,, \label{eightpoints8}
\end{align}
where the leading-order contributions to $F_{k,(3)}$ are:
\begin{align}
F_{k,(3)} &= \frac{Z(k)^4}{k} \left( u_{8,1} (k)  \sum_{\ell=1}^3 \vcenter{\hbox{\includegraphics[scale=0.8]{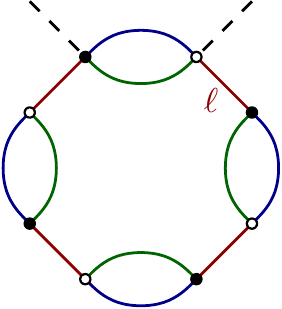}}} + u_{8,2} (k) \sum_{\substack{\ell, \ell'=1\\ \ell\neq \ell'}}^3 \left( \vcenter{\hbox{\includegraphics[scale=0.8]{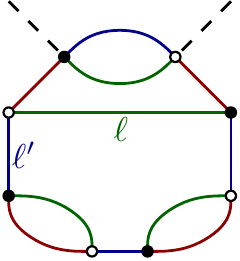}}} + \vcenter{\hbox{\includegraphics[scale=0.8]{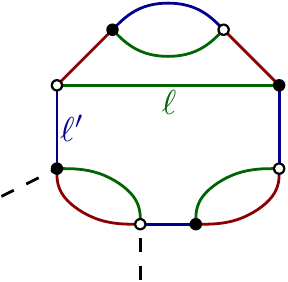}}} + \vcenter{\hbox{\includegraphics[scale=0.8]{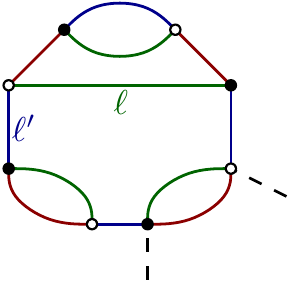}}}\right) \right. \nn \\
& \left. + \, u_{8,3} (k) \sum_{\substack{\ell, \ell'=1\\ \ell\neq \ell'}}^3 \vcenter{\hbox{\includegraphics[scale=0.8]{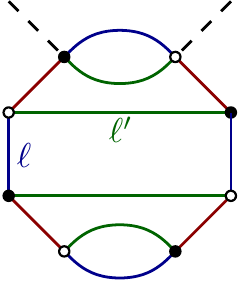}}} + u_{8,4} (k) \sum_{\ell = 1}^3 \left( \vcenter{\hbox{\includegraphics[scale=0.8]{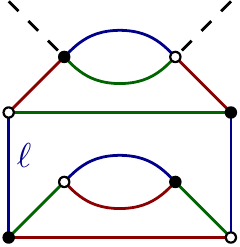}}} + \vcenter{\hbox{\includegraphics[scale=0.8]{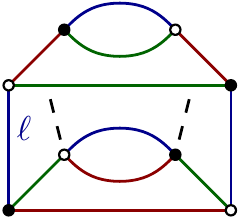}}} \right) + u_{8,5} (k) \sum_{\ell = 1}^3 \vcenter{\hbox{\includegraphics[scale=0.8]{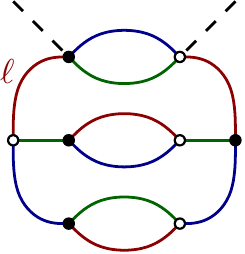}}}
\vphantom{\sum_{\ell=1}^3 \vcenter{\hbox{\includegraphics[scale=0.8]{Figures/int81_legs.pdf}}}} 
\right) \nn \\
& + \, \cdots \end{align}

While the flow equations of $u_2$, $Z$ and $u_4$ are unchanged, the flow equations of $u_{6,1}$ and $u_{6,2}$ are affected by $F_{k, (3)}$. We compute them in the next subsection, and will then focus on the $\beta$-functions of order-$8$ coupling constants. 
 
\subsubsection{Modified flow of $\phi^6$ coupling constants}

The inclusion of $8$-valent bubbles introduces the following extra terms in \eqref{u61}: 
\begin{align}\label{u61_8}
-  \frac{Z(k)^4}{k} \left( u_{8,1}(k) \, \sum_{\ell = 1}^3 \vcenter{\hbox{\includegraphics[scale=0.8]{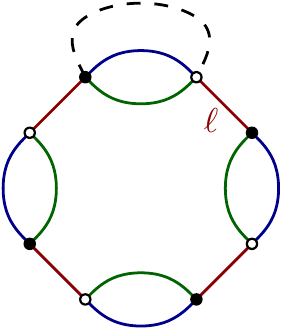}}} + u_{8,2} (k) \sum_{\substack{\ell, \ell'=1\\ \ell\neq \ell'}}^3 \vcenter{\hbox{\includegraphics[scale=0.8]{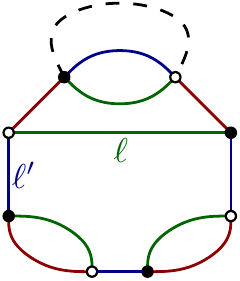}}}\right) \,.
\end{align}
Both types of diagrams have a loop of length one, whose contribution is controlled by the previously calculated function $f_1$ and $g_1$. The combinatorial factor associated to each term is easily determined and one obtains the modified $\beta$-function
\begin{align}
\beta_{6,1} &= - 3 \eta \, u_{6,1} + 6 L_2 (u_2 , u_4) \, u_4 \, u_{6,1} - 3 L_3 (u_2 , u_4) \, {u_4}^3 \nn \\
& \qquad - 3  L_1 (u_2 , u_4) \, \left( u_{8,1} + 2 u_{8,2}\right)\,. 
\end{align}

Similarly, the flow equation \eqref{u62} is corrected by the quantity
\begin{align}\label{u62_8}
&-  \frac{Z(k)^4}{k} \left( u_{8,2} (k) \sum_{\substack{\ell, \ell'=1\\ \ell\neq \ell'}}^3 \left( \vcenter{\hbox{\includegraphics[scale=0.8]{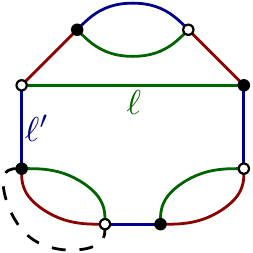}}} + \vcenter{\hbox{\includegraphics[scale=0.8]{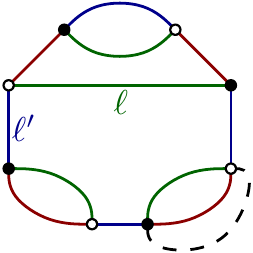}}} \right) + u_{8,3} (k) \sum_{\substack{\ell, \ell'=1\\ \ell\neq \ell'}}^3 \vcenter{\hbox{\includegraphics[scale=0.8]{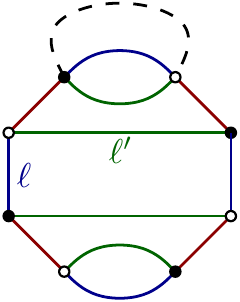}}} \right.\nn \\
&+ \left. u_{8,4} (k) \sum_{\ell=1}^3 \left( \vcenter{\hbox{\includegraphics[scale=0.8]{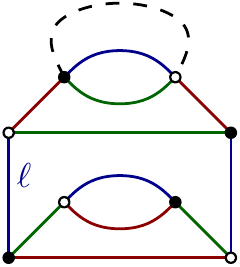}}} + \vcenter{\hbox{\includegraphics[scale=0.8]{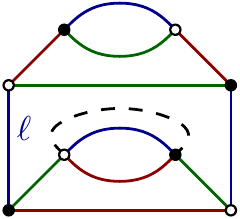}}} \right) + u_{8,5} (k) \sum_{\ell=1}^3  \vcenter{\hbox{\includegraphics[scale=0.8]{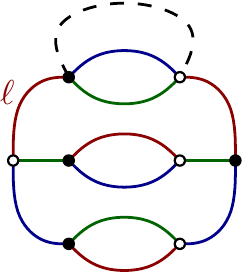}}}
\vphantom{\vcenter{\hbox{\includegraphics[scale=0.8]{Figures/phi83.pdf}}}} 
\right) \,,
\end{align}
which yields a modified $\beta$-function 
\begin{align}
\beta_{6,2} &= - 3 \eta \, u_{6,2} + 4 L_2 (u_2 , u_4) \, u_4 \, u_{6,2} \\
& \qquad - L_1 (u_2 , u_4) \, \left( 4 u_{8,2} + 2 u_{8,3} + 2 u_{8,4} + u_{8,5}\right) \,. \nn 
\end{align}

\subsubsection{Flow of $\phi^8$ coupling constants}

The evolution of $u_{8,1}$ is deduced from the following graphical identity:
\begin{align}\label{flow_u81}
\partial_k \left( \frac{Z(k)^4}{k} \frac{u_{8,1} (k)}{4} \right) \, \sum_{\ell = 1}^3 \vcenter{\hbox{\includegraphics[scale=0.8]{Figures/int81.pdf}}} &\approx \left( Z(k)^2 k u_4(k) \right)^4\,  \sum_{\ell=1}^3 \, \vcenter{\hbox{\includegraphics[scale=0.8]{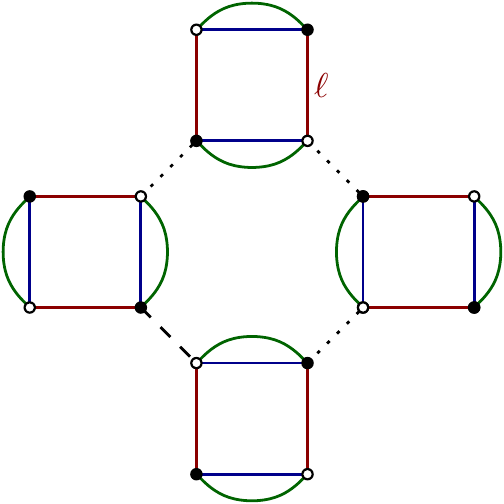}}}\\
&-  3 \, \left( Z(k)^2 k u_{4}(k) \right)^2 \, Z(k)^3 u_{6,1}(k) \, \sum_{\ell = 1}^3 \vcenter{\hbox{\includegraphics[scale=0.8]{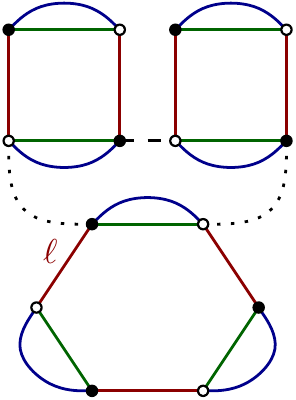}}} \nn\\
&+  2 \, Z(k)^2 k u_{4}(k) \, \frac{Z(k)^4}{k} u_{8,1}(k) \, \sum_{\ell = 1}^3 \vcenter{\hbox{\includegraphics[scale=0.8]{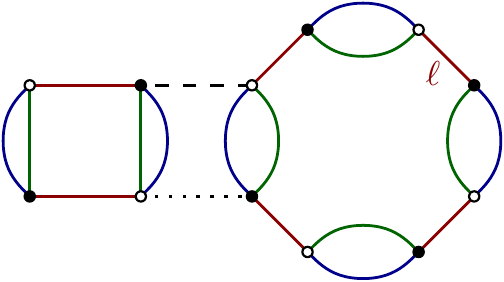}}} \nn \\
&+  \left( Z(k)^3 u_{6,1}(k) \right)^2 \, \sum_{\ell = 1}^3 \vcenter{\hbox{\includegraphics[scale=0.8]{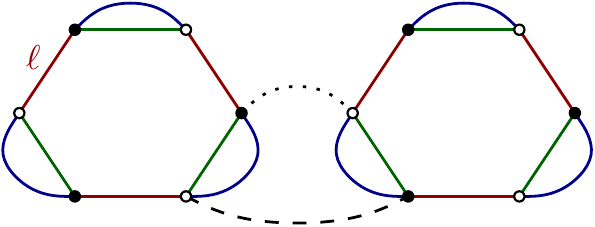}}} \,, \nn
\end{align}
which is a consequence of the Wetterich equation in the chosen truncation, and of the large $k$ approximation. As before, the right-hand-side of this equation can be approximated by a weighted sum of bubble invariants. We have already explained in detail how this can be done for diagrams with a loop of length smaller or equal to $3$, but not for the diagram appearing in the first line of \eqref{flow_u81}, which has a loop of length $4$. However, following the same methodology, one can show that
\begin{align}\label{approx_phi81_4_4}
\vcenter{\hbox{\includegraphics[scale=0.8]{Figures/phi81_4_4.pdf}}} &\approx \frac{1}{k^6 Z(k)^4} L_4 ( u_2 (k) , u_4 (k) ) \times \vcenter{\hbox{\includegraphics[scale=0.8]{Figures/int81.pdf}}}
\end{align}
within our working approximations, where the functions $f_4$, $g_4$ and $L_4$ are defined as:
\begin{align}
f_4(u_2) & := 2 \sqrt{2} \int_{0}^{\infty} dx\frac{x^6\, e^{-x^2} \, \left( 1-e^{-x^2} \right)^3}{\left( x^2+ u_2(1-e^{-x^2})\right)^5}\,,\\
g_4( u_2) & := 2 \sqrt{2} \int_{0}^{\infty} dx\frac{x^4\, e^{-x^2} \, \left(1-e^{-x^2}\right)^4}{\left( x^2+ u_2(1-e^{-x^2})\right)^5}\,,\\
L_4 (u_2 , u_4 ) & := 2f_4(u_2 )+\eta( u_2 , u_4 )  g_4(u_2 )\,.
\end{align}
Altogether, we identify the $\beta$-function associated to the coupling $u_{8,1}$ as
\begin{align}
\beta_{8,1} &= \left( 1 - 4 \eta \right) \, u_{8,1} + 4 L_4 ( u_2 , u_4 )  \, {u_4}^4 - 12 L_3 ( u_2 , u_4 ) \, {u_4}^2 \, u_{6,1} \\
& \qquad + 4 L_2 ( u_2 , u_4 ) \left( 2 u_4 \, u_{8,1} + {u_{6,1}}^2 \right) \,. \nn
\end{align}

\

We can proceed in the same way for the remaining four coupling constants. For $u_{8,2}$ we find
\begin{align}\label{flow_u82}
\partial_k \left( \frac{Z(k)^4}{k} u_{8,2} (k) \right)& \, \sum_{\substack{\ell, \ell'=1\\ \ell\neq \ell'}}^3 \vcenter{\hbox{\includegraphics[scale=0.8]{Figures/int82.pdf}}} \approx -  3 \, \left( Z(k)^2 k u_{4}(k) \right)^2 \, Z(k)^3 u_{6,2}(k) \, \sum_{\substack{\ell, \ell'=1\\ \ell\neq \ell'}}^3 \vcenter{\hbox{\includegraphics[scale=0.8]{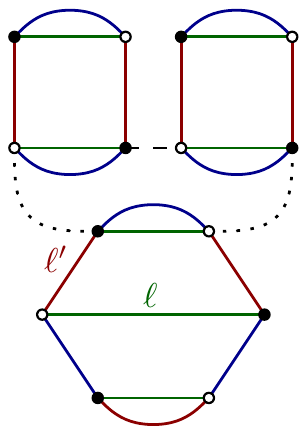}}} \\
&+  2 \, Z(k)^2 k u_{4}(k) \, \frac{Z(k)^4}{k} u_{8,2}(k) \, \sum_{\substack{\ell, \ell'=1\\ \ell\neq \ell'}}^3 \left( \vcenter{\hbox{\includegraphics[scale=0.8]{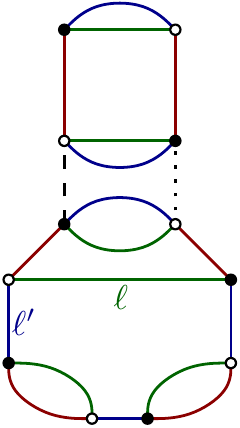}}} + \vcenter{\hbox{\includegraphics[scale=0.8]{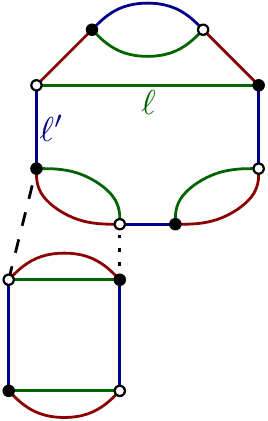}}} + \vcenter{\hbox{\includegraphics[scale=0.8]{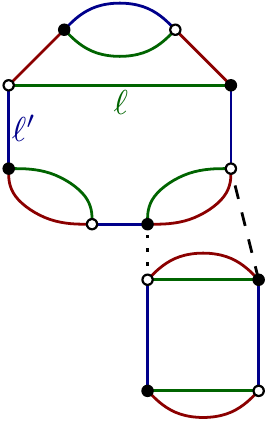}}}\right)\nn \\
&+  2 \, Z(k)^3 u_{6,1}(k)  \, Z(k)^3 u_{6,2}(k) \sum_{\substack{\ell, \ell'=1\\ \ell\neq \ell'}}^3 \vcenter{\hbox{\includegraphics[scale=0.8]{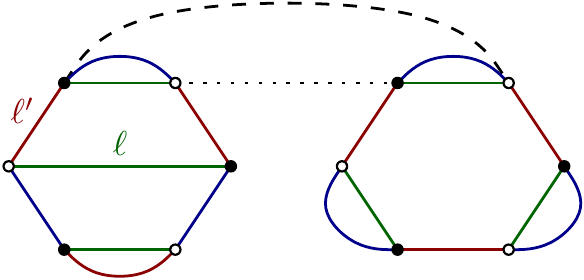}}} \,, \nn
\end{align}
from which we deduce
\begin{align}
\beta_{8,2} &= \left( 1 - 4 \eta \right) \, u_{8,2} - 3 L_3 ( u_2 , u_4 ) \, {u_4}^2 \, {u_{6,2}} + 2 L_2 ( u_2 , u_4 ) \left( 3 u_4 \, u_{8,2} + u_{6,1} \, u_{6,2} \right) \,.
\end{align}
There are only two types of graphs contributing to the flow of $u_{8,3}$:
\begin{align}\label{flow_u83}
\partial_k \left( \frac{Z(k)^4}{k} \frac{u_{8,3}}{2} (k) \right) \, \sum_{\substack{\ell, \ell'=1\\ \ell\neq \ell'}}^3 \vcenter{\hbox{\includegraphics[scale=0.8]{Figures/int83.pdf}}} &\approx  2 \, Z(k)^2 k u_{4}(k) \, \frac{Z(k)^4}{k} u_{8,3}(k) \, \sum_{\substack{\ell, \ell'=1\\ \ell\neq \ell'}}^3  \vcenter{\hbox{\includegraphics[scale=0.8]{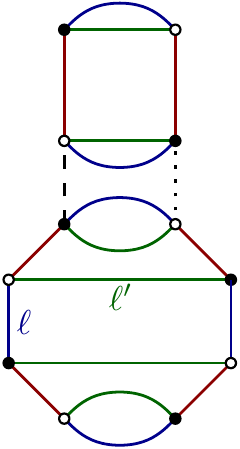}}} \\
&+  \left( Z(k)^3 u_{6,2}(k) \right)^2 \, \sum_{\substack{\ell, \ell'=1\\ \ell\neq \ell'}}^3 \vcenter{\hbox{\includegraphics[scale=0.8]{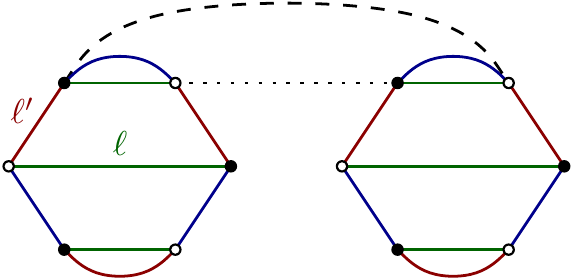}}} \,, \nn
\end{align}
and hence
\begin{equation}
\beta_{8,3} = \left( 1 - 4 \eta \right) \, u_{8,3} + 2 L_2 ( u_2 , u_4 ) \left( 2 u_4 \, u_{8,3} + {u_{6,2}}^2\right)\,. 
\end{equation}
The bubbles of type $(8,4)$ receive quantum corrections controlled by $u_4$ and $u_{8,4}$ according to:
\begin{align}\label{flow_u84}
\partial_k \left( \frac{Z(k)^4}{k} u_{8,4} (k) \right) \, \sum_{\ell=1}^3 \vcenter{\hbox{\includegraphics[scale=0.8]{Figures/int84.pdf}}} &\approx  2 \, Z(k)^2 k u_{4}(k) \, \frac{Z(k)^4}{k} u_{8,4}(k) \, \sum_{\ell=1}^3  \left( \vcenter{\hbox{\includegraphics[scale=0.8]{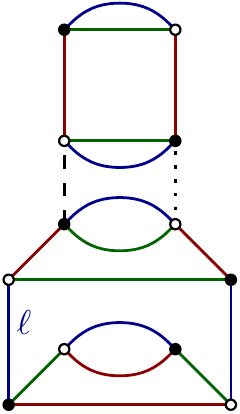}}} + \vcenter{\hbox{\includegraphics[scale=0.8]{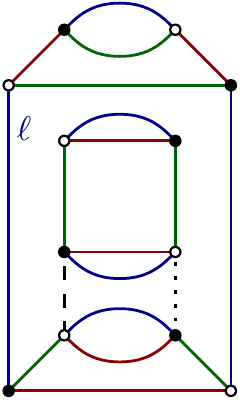}}} \right)\,, \nn
\end{align}
and hence
\begin{align}
\beta_{8,4} &= \left( 1 - 4 \eta \right) \, u_{8,4} + 4 L_2 ( u_2 , u_4 ) u_4 \, u_{8,4} \,. 
\end{align}
Finally, the evolution of $u_{8,5}$ is computed from the graphical equation
\begin{align}\label{flow_u85}
\partial_k \left( \frac{Z(k)^4}{k} u_{8,5} (k) \right) \,  \vcenter{\hbox{\includegraphics[scale=0.8]{Figures/int85.pdf}}} &\approx  2 \, Z(k)^2 k u_{4}(k) \, \frac{Z(k)^4}{k} u_{8,5}(k) \, \sum_{\ell=1}^3  \vcenter{\hbox{\includegraphics[scale=0.8]{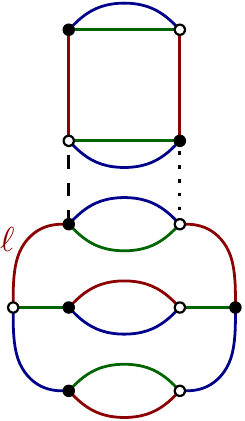}}}\,, 
\end{align}
yielding the $\beta$-function
\begin{align}
\beta_{8,5} &= \left( 1 - 4 \eta \right) \, u_{8,5} + 6 L_2 ( u_2 , u_4 ) u_4 \, u_{8,5} \,.
\end{align}

\subsubsection{Summary}

For convenience, we gather all the equations relevant to the $\phi^8$ truncation:
\begin{equation}\label{flow_phi8_full}
\left\lbrace\begin{split}
\beta_{2} &= - \left( 2 + \eta \right) u_2 -3 L_1 ( u_2 , u_4 ) \, u_4 
\\
\beta_4&= - \left( 1 + 2 \eta \right) u_4 - 2 L_1 ( u_2 , u_4 ) \, \left(u_{6,1}+2 u_{6,2} \right)+2 L_2 ( u_2 , u_4 ) \, {u_4}^2 
\\
\beta_{6,1} &= - 3 \eta \, u_{6,1} + 6 L_2 ( u_2 , u_4 ) \, u_4 \, u_{6,1} - 3 L_3 ( u_2 , u_4 ) \, {u_4}^3 \\
& \qquad - 3  L_1 ( u_2 , u_4 ) \, \left( u_{8,1} + 2 u_{8,2}\right)
\\
\beta_{6,2} &= - 3 \eta \, u_{6,2} + 4 L_2 ( u_2 , u_4 ) \, u_4 \, u_{6,2} \\
& \qquad - L_1 ( u_2 , u_4 ) \, \left( 4 u_{8,2} + 2 u_{8,3} + 2 u_{8,4} + u_{8,5}\right) 
\\
\beta_{8,1} &= \left( 1 - 4 \eta \right) \, u_{8,1} + 4 L_4 ( u_2 , u_4 ) \, {u_4}^4 - 12 L_3 ( u_2 , u_4 ) \, {u_4}^2 \, u_{6,1} \\
& \qquad + 4 L_2 ( u_2 , u_4 ) \left( 2 u_4 \, u_{8,1} + {u_{6,1}}^2 \right) 
\\
\beta_{8,2} &= \left( 1 - 4 \eta \right) \, u_{8,2} - 3 L_3 ( u_2 , u_4 ) \, {u_4}^2 \, {u_{6,2}} + 2 L_2 ( u_2 , u_4 ) \left( 3 u_4 \, u_{8,2} + u_{6,1} \, u_{6,2} \right) 
\\
\beta_{8,3} &= \left( 1 - 4 \eta \right) \, u_{8,3} + 2 L_2 ( u_2 , u_4 ) \left( 2 u_4 \, u_{8,3} + {u_{6,2}}^2\right)
\\
\beta_{8,4} &= \left( 1 - 4 \eta \right) \, u_{8,4} + 4 L_2 ( u_2 , u_4 ) \, u_4 \, u_{8,4} 
\\
\beta_{8,5} &= \left( 1 - 4 \eta \right) \, u_{8,5} + 6 L_2 ( u_2 , u_4 ) \, u_4 \, u_{8,5} 
\end{split}\right.
\end{equation}
As before, the anomalous dimension is given by
\begin{equation}
\eta( u_2 , u_4) =\frac{1}{3} \frac{f_w (u_2) \, u_4}{1 - \frac{1}{6} g_w (u_2) \, u_4}\,,
\end{equation}
and the loop integrals are
\begin{align}\label{L_k}
f_k (u_2) & = 2 \sqrt{2} \int_{0}^{\infty} dx\frac{x^6\, e^{-x^2} \, \left( 1-e^{-x^2} \right)^{n-1}}{\left( x^2+ u_2(1-e^{-x^2})\right)^{n+1}}\,,\\
g_k ( u_2) & = 2 \sqrt{2} \int_{0}^{\infty} dx\frac{x^4\, e^{-x^2} \, \left(1-e^{-x^2}\right)^n}{\left( x^2+ u_2(1-e^{-x^2})\right)^{n+1}}\,,\\
L_k (u_2 , u_4) &= 2 f_k (u_2 , u_4 ) + \eta (u_2 , u_4 ) g_k (u_2 , u_4) \,.
\end{align}

\subsection{Non-perturbative fixed points}

We now investigate the existence of non-trivial fixed points of the $\phi^8$ truncation \eqref{flow_phi8_full}. 

The first thing we notice is the absence of continuous sets of fixed points in this truncation; in particular, we do not see any analogue of the one-parameter family of fixed points $\{ u^*(s) \}$ found in the $\phi^6$ truncation. A similar observation was made in \cite{Lahoche:2016xiq}, for a different model. Taken together, these give indications that such continuous families of fixed points, which seem to be a general feature of low order truncations of tensorial field theories (they were first identified in \cite{Geloun:2016xep}, and also found in \cite{Lahoche:2016xiq}), are not robust. The only caveat of our analysis is that it is restricted to the local potential approximation. In order to definitely confirm or exclude the scenario put forward in \cite{Geloun:2016xep}, our analysis should be upgraded to a full-fledged derivative expansion, which goes beyond the scope of this paper. 

\

Just like in the $\phi^6$ truncation, we found two numerical fixed points, whose properties are summarized in Table \ref{fixedpoints_8_full}. For obvious reasons, we likewise denoted them $\mathrm{FP}_1$ and $\mathrm{FP}_2$ respectively. Note that we also found a handful of additional fixed points with $u_2 < 0$, which we discarded as they lie on the wrong side of the singular hypersurface $\{ f_w (u_2 ) u_4 = 6\}$.  
 
\begin{table}[ht]
\centering
\begin{tabular}{|c||c|c|}
\hline
 & $\mathrm{FP}_1$ & $\mathrm{FP}_2$ \\ \hline\hline
$u_2^*$ & $3.7$ & $-0.35$ \\ \hline
$u_4^*$ & $-5.6$ & $0.063$ \\ \hline
$u_{6,1}^*$ & $-5.4$ & $0.0051$ \\ \hline
$u_{6,2}^*$ & $0.$ & $-0.0081$ \\ \hline
$u_{8,1}^*$ & $-7.1$ & $0.00042$ \\ \hline
$u_{8,2}^*$ & $0.$ & $- 0.00036$ \\ \hline
$u_{8,3}^*$ & $0.$ & $0.0014$ \\ \hline
$u_{8,4}^*$ & $0.$ & $0.$ \\ \hline
$u_{8,5}^*$ & $0.$ & $0.$ \\ \hline
Critical & $( 2.8 , - 0.28 , - 1.5 , - 1.6 , $ & $( 2.3 \pm 0.31 \, \mathrm{i} , 1.8 , 0.39 \pm 1.1 \, \mathrm{i} , 0.39 , $ \\
exponents & $-3.5 , -3.9 , -3.9 , -4.0 , -4.0 )$ & $ -0.12 , -1.0 , -2.1 )$ \\ \hline
$\eta$ & -0.94 & 0.60 \\ \hline
\end{tabular}
\caption{Isolated non-Gaussian fixed points in the $\phi^8$ truncation.}\label{fixedpoints_8_full}
\end{table}

Remarkably, the qualitative properties of $\mathrm{FP}_1$ are identical in the $\phi^6$ and $\phi^8$ truncations. First and foremost, we again find a single relevant direction. This is consistent with the ultraviolet fixed point interpretation we anticipated in the previous section, and hence with the existence of two infrared phases. The critical exponent associated to this relevant direction is approximately $2.8$, to be compared with the value $2.7$ found in the $\phi^6$ truncation. Another important feature of $\mathrm{FP}_1$ in this truncation is that it is contained in the subspace $\{ u_{6,2} = u_{8,2} = u_{8,3} = u_{8,4} = u_{8,5} = 0 \}$. As can be inferred from the flow equations \eqref{flow_phi8_full}, this is again an invariant subspace. Exactly one coupling constant per order is non-zero at $\mathrm{FP}_1$ ($u_4$ at order $4$, $u_{6,1}$ at order $6$ and $u_{8,1}$ at order $8$), and parametrizes the colored bubble with the simplest combinatorial structure at this order. Hence, in the vicinity of $\mathrm{FP}_1$, the theory is naturally driven towards a regime in which only the simplest melonic interactions are switched on. We will confirm and take advantage of this observation in the next section.

\

The situation for $\mathrm{FP}_2$ is not as favorable, since it has $6$ relevant directions and $3$ irrelevant ones. Neither of these numbers coincide with those found in the $\phi^6$ truncation, it is therefore hard to guess which (if any) of the two will remain finite in the full theory space. Nonetheless, we notice that the values of the various coupling constants, of the critical exponents and of $\eta$ are rather close to one another in both truncations. This may be an indication that $\mathrm{FP}_2$ is not a truncation artefact, but we are not in a position to make any further conjecture about its properties.

\section{High order truncations to non-branching bubble interactions}\label{sec:non_branching}

In this section, we investigate further the properties of the renormalization group equations in a sector that only retains the simplest melonic bubble interactions, which we propose to call \emph{non-branching}. We will see in particular that $\mathrm{FP}_1$ and its main qualitative features are reproduced up to $\phi^{12}$ truncations in the non-branching stable subspace.  

\subsection{Non-branching bubbles and their stability under renormalization}

\begin{figure}
\centering
\includegraphics[scale=.8]{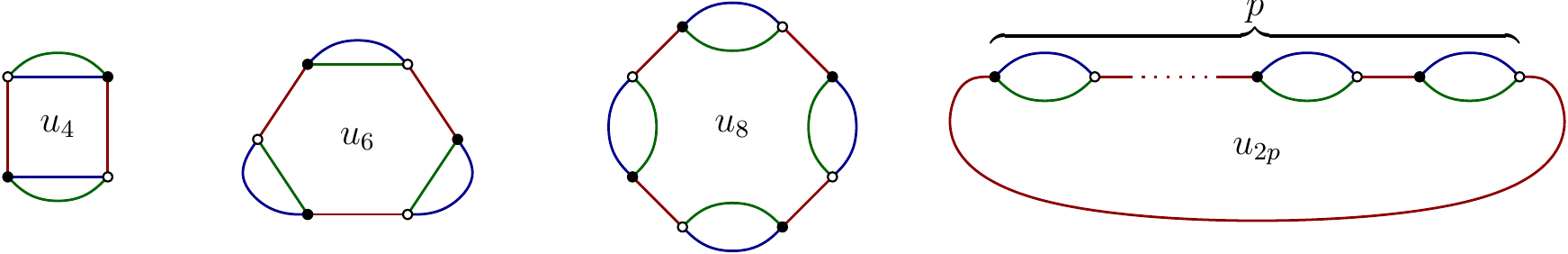}
\caption{Non-branching bubbles and their associated coupling constants.}\label{non-branching}
\end{figure}

The most concise (but formal) way of defining \emph{non-branching bubbles} is as follows: they are those melonic bubbles which admit a non-branching $3$-ary tree representation \cite{razvan_virasoro, Gurau:2011xp}. More colloquially, this means that a non-branching bubble is obtained from the elementary connected bubble with two vertices by successive insertions of melons along the same two faces. The general structure of non-branching bubbles is represented in Figure \ref{non-branching}. In particular, among the bubble interactions we have already encountered, the non-branching ones are those of type $4$, $(6,1)$ and $(8,1)$. More generally, there are exactly three non-branching bubbles at each order $2p$, corresponding to a choice of color for the edges on which the elementary melons are inserted. By invariance of the action under color permutations, these three $\phi^{2p}$ interactions are parametrized by a single coupling constant $u_{2p}$\footnote{In particular, $u_{6} = u_{6,1}$ and $u_8 = u_{8,1}$.}.  

\begin{figure}
\centering
\includegraphics[scale=.8]{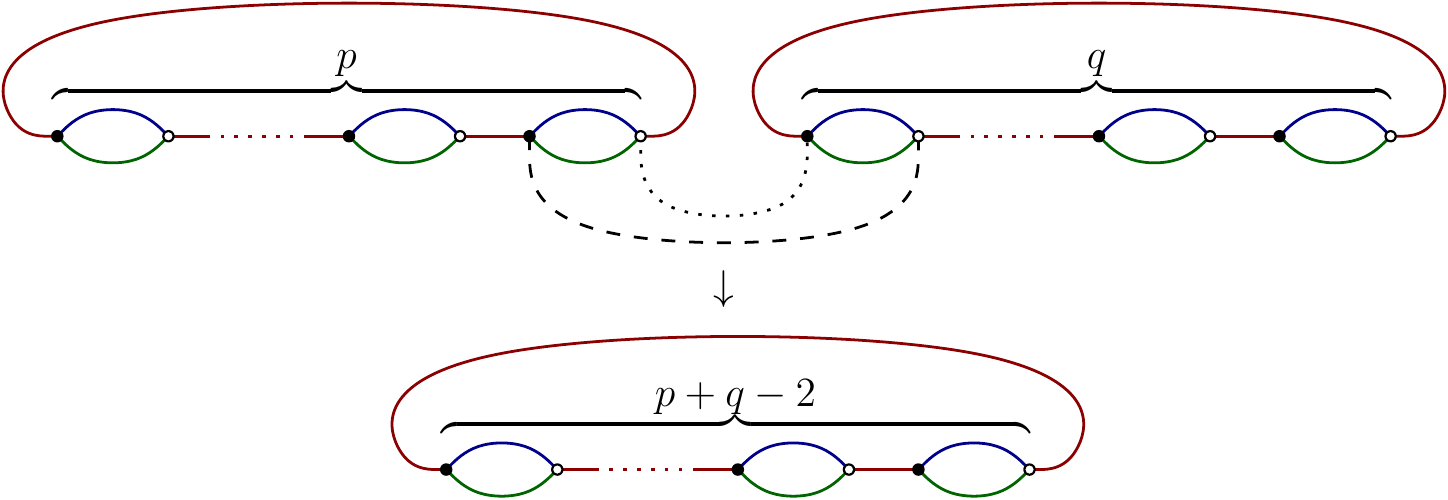}
\caption{One-loop melonic graph with two non-branching bubbles, whose local approximation is again non-branching.}\label{non-branching-graph}
\end{figure}

A key property of the theory space generated by the non-branching bubbles is that it is stable under the renormalization group in the ultraviolet regime. Indeed, the local approximation of a one-loop melonic graph containing only non-branching bubbles is again a non-branching bubble. An example of this claim is provided in Figure \ref{non-branching-graph}, which can easily be generalized to a graph with an arbitrary number of non-branching bubbles. Hence, in the large $k$ regime, non-branching bubbles generate an invariant theory subspace under the renormalization group. We restrict our attention to this subspace in the remainder of this paper. More precisely, we consider the following ansatz for the effective average action: 
\begin{align}\label{ansatz_non-branching}
\Gamma_k &= -Z(k) \sum_{\ell=1}^3 \vcenter{\hbox{\includegraphics[scale=0.8]{Figures/int2_laplace.pdf}}}  + Z(k) k^2 u_2(k) \vcenter{\hbox{\includegraphics[scale=0.8]{Figures/int2.pdf}}} \\
& \qquad + \sum_{p \geq 2} Z(k)^p k^{3-p} \frac{u_{2p} (k)}{p}  \sum_{\ell=1}^3 \vcenter{\hbox{\includegraphics[scale=0.8]{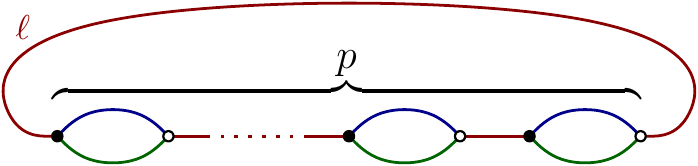}}}  \nn
\end{align}   
Note the $\frac{1}{p}$ factors, which account for the number of automorphisms of non-branching bubbles of order $2p$.

\subsection{Flow equations at arbitrary order}

Given the detailed computations of the previous sections, the determination of the full flow equations in the non-branching sector reduces to a mere counting problem. Consider the beta function $\beta_{2p}$ associated to the coupling $u_{2p}$ ($p \geq 2$). Each graph with a loop of length $k$ whose local approximation is of order $2p$ will be weighted by a loop integral $L_{k} (u_2, u_4)$ (see \eqref{L_k}). If $n_{2q}$ denotes the number of bubbles of type $2q$ in such a graph, then by the fact that it consists of a single loop of length $k$, we must have:
\beq\label{constraint1}
\underset{q \geq 2}{\sum} n_{2 q} = k\,.
\eeq
Moreover, the constraint that there are exactly $2p$ external legs imposes:
\beq\label{constraint2} 
\underset{q \geq 2}{\sum} q n_{2 q} = p+k\,.
\eeq
Let us call $\mathcal{D}_{k,p}$ the set of collections of integers $\{ n_{2q} \vert q \in \mathbb{N}  , q \geq 2\}$ which verify equations \eqref{constraint1} and \eqref{constraint2}. All we need to do to fully determine the flow equations is to enumerate, for each element in $\{ n_{2q} \} \in \mathcal{D}_{p,k}$, the one-loop melonic graphs which can be constructed from the bubbles specified by the collection $\{ n_{2q} \}$. If one fixes the color of the melonic faces and furthermore take into account that there is one special dashed line in each graph, this number of melonic contractions is:
\beq
{k\choose{\{ n_{2q} \}}} \underset{q\geq2}{\prod} q^{n_{2q}}\,,
\eeq
where we have denoted the standard multinomial coefficients by
\beq
{k\choose{\{ n_{2q} \}}} = \frac{k! }{ \underset{q\geq2}{\prod} (n_{2q}!)}\,.
\eeq
Together with the combinatorial factors entering the ansatz \eqref{ansatz_non-branching}, this counting yields the $\beta$-functions:
\beq
\beta_{2p} = \left( p - 3 - p \eta \right) u_{2p} + p \sum_{k = 1}^p (-1)^{k} L_{k} (u_2, u_4) \sum_{ \{ n_{2q} \} \in \mathcal{D}_{k,p} 
}
{k\choose{\{ n_{2q} \}} }  \prod_{q \geq 2} (u_{2q})^{n_{2q}}
\,, \qquad \forall p \geq 2 \,.
\eeq

An explicit evaluation of the first terms in this infinite tower of $\beta$-functions gives:
\begin{equation}
\left\lbrace\begin{split}
\beta_{2} &= - \left( 2 + \eta \right) u_2 -3 L_1 (u_2, u_4) \, u_4 
\\
\beta_4&= - \left( 1 + 2 \eta \right) u_4 - 2 L_1 (u_2, u_4) \, u_{6} +2 L_2 (u_2, u_4) \, {u_4}^2 
\\
\beta_{6} &= - 3 \eta \, u_{6} - 3 L_1 (u_2, u_4) \, u_8 + 6 L_2 (u_2, u_4) \, u_4 u_6 - 3 L_3 (u_2, u_4)  {u_4}^3 \\
\beta_{8} &= \left( 1 - 4 \eta \right) \, u_{8} - 4 L_1 (u_2, u_4) \, u_{10} + 4 L_2 (u_2, u_4) \, \left( u_4 u_8 + {u_6}^2 \right) \\ 
& \qquad - 12 L_3 (u_2, u_4) \, {u_4}^2 u_6 + 4 L_4 (u_2, u_4)  \, {u_4}^4 
\\
\beta_{10} &= \left( 2 - 5 \eta \right) \, u_{10} - 5 L_1 (u_2, u_4) \, u_{12} + 10 L_2 (u_2, u_4) \, \left( u_4 u_{10} + u_6 u_8 \right) \\ 
& \qquad - 15 L_3 (u_2, u_4)  \, \left( {u_4}^2 u_8 + u_4 {u_6}^2 \right)+ 20 L_4 (u_2, u_4)  \, {u_4}^3 u_6 - 5 L_5 (u_2, u_4) \, {u_4}^5 \\
\beta_{12} &= \left( 3 - 6 \eta \right) \, u_{12} - 6 L_1 (u_2, u_4) \, u_{14} + 6 L_2 (u_2, u_4) \, \left( 2 u_4 u_{12} + 2 u_6 u_{10} + {u_8}^2 \right) \\ 
& \qquad - 6 L_3 (u_2, u_4)  \, \left( 3 {u_4}^2 u_{10} + 6 u_4 u_6 u_8 + {u_6}^3 \right)+ 12 L_4 (u_2, u_4)  \, \left( 2 {u_4}^3 u_8 + 3 {u_4}^2 {u_6}^2 \right) \\
& \qquad - 30 L_5 (u_2, u_4) \, {u_4}^4 u_6 + 6 L_6 (u_2, u_4) \, {u_4}^6 \\
& \cdots
\end{split}\right.
\end{equation}

\subsection{Non-perturbative fixed points}

The dramatic simplification of the flow equations in the non-branching sector allows numerical investigations in high order truncations. We have systematically checked for the existence of fixed points up to order $12$ local potential truncations. We confirm the existence of a fixed point with positive $u_2$ ($\mathrm{FP}_1$) in all these truncations. We also find evidence for the existence of a fixed point with negative $u_2$ ($\mathrm{FP}_2$), although it is absent from both the $\phi^4$ and $\phi^8$ truncations.  

\

The properties of $\mathrm{FP}_1$ are summarized in Table \ref{fixedpoints_FP1_simple}. The main lesson is that the interesting features of $\mathrm{FP}_1$ are reproduced in all truncations. First and foremost, it has a single relevant direction, with a critical exponent $\theta_1$ varying between $2.5$ and $3.0$. This is a strong indication of the ultraviolet nature of $\mathrm{FP}_1$, and of the fact that a theory starting in its vicinity will end up in one of two phases in the infrared. 

Even though the critical exponents are rather stable over the different truncations, they do not seem to settle in precise values yet. This is even more true for the coupling constants, which do not show any clear sign of convergence at such orders. We conjecture that this is the consequence of a general limitation of the local potential approximation: by discarding all derivative couplings, we are forgetting terms which may turn out to be more relevant than some of the local terms we are keeping. This may presumably be improved upon through the inclusion of $\SU(2)$ invariant differential operators in the effective average action ansatz. Derivative couplings of this type have already been proposed in \cite{Carrozza_review} but have not yet been included in an FRG analysis. Once this is done, we expect a much better convergence under refinement of the truncation. We shall then be in a position to make more quantitative predictions (e.g. for the values of the critical exponents), which we leave for future work. Another notable feature is that all the coupling constants but $u_2$ are negative at $\mathrm{FP}_1$. The action is therefore unbounded in its vicinity, which raises the question of the convergence of the functional integral. While this might present a serious challenge, addressing it goes beyond the scope of the present work. 
 
\begin{table}[ht]
\centering
\begin{tabular}{|c||c|c|c|c|c|}
\hline
 Truncation Order & $4$ & $6$ & $8$ & $10$ & $12$ \\ \hline\hline
$u_2^*$ & $1.8$ & $2.7$ & $3.7$ & $5.3$ & $7.4$ \\ \hline
$u_4^*$ & $-1.5$ & $-2.9$ & $-5.6$ & $-11$ & $-23$ \\ \hline
$u_{6}^*$ & -- & $-0.93$ & $-5.4$ & $-27$ & $-1.3 \times 10^2$ \\ \hline
$u_{8}^*$ & -- & -- & $-7.1$ & $-98$ & $-1.0 \times 10^3$ \\ \hline
$u_{10}^*$ & -- & -- & -- & $-2.9 \times 10^2$ & $-8.0 \times 10^3$ \\ \hline
$u_{12}^*$ & -- & -- & -- & -- & $-5.0 \times 10^4$ \\ \hline
$\theta_1$ &  $2.5$ & $2.7$ & $2.8$ & $2.9$ & $3.0$ \\ \hline
$\theta_2$ & $-0.37$ & $-0.31$ & $-0.28$ & $-0.28$ & $-0.31$ \\ \hline 
$\theta_3$ & -- & $-1.7$ & $-1.6$ & $-1.6$ & $-1.7$ \\ \hline
$\theta_4$ & -- & -- & $-4.0$ & $-4.1$ & $-4.3$ \\ \hline
$\theta_5$ & -- & -- & -- & $-6.6$ & $-6.8$ \\ \hline
$\theta_6$ & -- & -- & -- & -- & $-9.5$ \\ \hline
$\eta$ & $-0.70$ & $-0.82$ & $-0.94$ & $-1.1$ & $-1.2$ \\ \hline
\end{tabular}
\caption{Properties of $\mathrm{FP}_1$ in the non-branching truncation, up to order $12$.}\label{fixedpoints_FP1_simple}
\end{table}

\

The properties of $\mathrm{FP}_2$ are summarized in Table \ref{fixedpoints_FP2_simple}. Even though some qualitative and quantitative features are reproduced in different truncations, they seem much less robust than those of $\mathrm{FP}_1$. Despite the fact that the same number of relevant directions is found in the $\phi^{10}$ and $\phi^{12}$ truncations, there is a large variability at lower orders, and we are therefore reluctant to make any conjecture about the nature of $\mathrm{FP}_2$. One other curious aspect of this fixed point is that it lies in the non-branching subspace for all high order truncations, except for the $\phi^8$ one. 

The fact that $\mathrm{FP}_2$ keeps being realized in high order truncations certainly speaks in favor of its existence in the full theory space. Nonetheless, we will have to postpone any more precise statements to future investigations based on a more complete ansatz (which in particular should include derivative couplings). 

\begin{table}[ht]
\centering
\begin{tabular}{|c||c|c|c|c|c|}
\hline
 Truncation Order & $4$ & $6$ & $8$ & $10$ & $12$ \\ \hline\hline
$u_2^*$ & -- & $-0.35$ & -- & $-0.22$ & $-0.17$ \\ \hline
$u_4^*$ & -- & $0.063$ & -- & $0.046$ & $0.039$ \\ \hline
$u_{6}^*$ & -- & $-0.011$ & -- & $-8.5 \times 10^{-3}$ & $-7.3 \times 10^{-3}$ \\ \hline
$u_{8}^*$ & -- & -- & -- &  $6.3 \times 10^{-5}$ & $5.7 \times 10^{-5}$\\ \hline
$u_{10}^*$ & -- & -- & -- & $8.9 \times 10^{-5}$ & $5.9 \times 10^{-5}$ \\ \hline
$u_{12}^*$ & -- & -- & -- & -- & $1.5 \times 10^{-12}$ \\ \hline
$\theta_1$ &  -- & $3.0$ & -- & $2.3$ & $2.2$ \\ \hline
$\theta_2$ & -- & $1.3$ & -- & $1.8$ & $1.5$ \\ \hline 
$\theta_3$ & -- & $-0.51$ & -- & $0.36 + 0.72 \mathrm{i}$ & $0.41 + 0.72 \mathrm{i}$ \\ \hline
$\theta_4$ & -- & -- & -- & $0.36 - 0.72 \mathrm{i}$ & $0.41 - 0.72 \mathrm{i}$ \\ \hline
$\theta_5$ & -- & -- & -- & $-4$ & $-5.1$ \\ \hline
$\theta_6$ & -- & -- & -- & -- & $-1.6$ \\ \hline
$\eta$ & -- & $0.60$ & -- & $0.29$ & $0.22$ \\ \hline
\end{tabular}
\caption{Properties of $\mathrm{FP}_2$ in the non-branching truncation, up to order $12$.}\label{fixedpoints_FP2_simple}
\end{table}

\section{Conclusion}\label{conclu}

The purpose of this article was to investigate, at a non-perturbative level, the ultraviolet properties of $\SU(2)$ tensorial GFT in dimension three. To this effect, we relied on the Wetterich--Morris version of the functional renormalization group \cite{Wetterich:1993ne, Morris:1993qb}, which was first applied to GFT in \cite{Benedetti:2014qsa} (and \cite{Benedetti:2015yaa} as far as gauge invariant models are concerned).   
The results of our analysis may be summarized as follows.

At the technical level, it is the first time that the Wetterich--Morris formalism is applied to a non-Abelian GFT, which represents a necessary improvement in our progression towards more realistic quantum gravity models (e.g. \cite{bo_bc, bo}). In this respect, we found more convenient to work within a heat kernel regularization scheme, which is another original aspect of the present article. More importantly, the strength of our work lies in the reproducibility of our main result -- the existence of non-perturbative ultraviolet fixed point -- in various truncations, and up to high orders in the local potential approximation.  

In section \ref{FRG}, we started with the simplest available truncation, in which only the perturbatively relevant coupling constants are taken into account. This already yields a four-dimensional theory space, due to the presence of two marginal $\phi^6$ melonic bubbles. We find two isolated fixed points in this truncation, $\mathrm{FP}_1$ and $\mathrm{FP}_2$, with respectively one and three infrared relevant directions. We also observe the presence of a one-parameter family of non-trivial fixed points, which is furthermore connected to the Gaussian one. 

To investigate the robustness of these observations, we then extended the truncation to include all order $8$ melonic interactions (section \ref{sec:order8}), which are perturbatively irrelevant. As melonic bubbles proliferate exponentially, this results in a large theory space, of dimension $9$.
Most interestingly, the fixed point $\mathrm{FP}_1$ is reproduced in this theory space, with the same qualitative characteristics as in the cruder truncation. In particular, it has a single relevant direction, which suggests that it should be interpreted as an ultraviolet fixed point. More generally, the four critical exponents that are meaningful in both truncations are approximately the same in both, and the anomalous dimension remains of order $1$. A fixed point with similar characteristics as $\mathrm{FP}_2$ is also uncovered in this truncation, but its properties are much less robust than those of $\mathrm{FP}_1$: neither the number of relevant directions nor the number of irrelevant ones is preserved, which prevents us from making any conjecture about the nature of $\mathrm{FP}_2$. The third lesson we can draw from section \ref{sec:order8} is that the one-parameter family of non-Gaussian fixed points seen in the $\phi^6$ truncation is not preserved in higher order local potential approximations.    

In section \ref{sec:non_branching}, we checked the existence of $\mathrm{FP}_1$ in even higher order truncations, namely up to $\phi^{12}$ melonic interactions. Given how rapidly melonic bubbles proliferate, this was only made possible after observing that, in both $\phi^6$ and $\phi^8$ truncations, only a simple subclass of melonic interactions -- the \emph{non-branching bubbles} -- are switched on at $\mathrm{FP}_1$. We observed that there is only one non-branching coupling constant per order, and that the (infinite-dimensional) theory space they generate is stable under the renormalization group flow in the ultraviolet region. This allowed us to fully determine the infinite tower of $\beta$-functions associated to this sector of the theory (again in the local potential and melonic approximation). We found that the qualitative properties of $\mathrm{FP}_1$, first and foremost its unique relevant direction, are reproduced in high order truncations. However, we do not see a clear-cut convergence of the numerical values of the coupling constants and of the critical exponents, hence we are limited to qualitative conclusions. Tentatively, we attribute this drawback to the intrinsic limitations of the local potential approximation. Finally, although we do observe a fixed point with similar features as $\mathrm{FP}_2$ in the $\phi^{10}$ and $\phi^{12}$ truncations, the variability of its critical exponents is such that we cannot commit to any particular interpretation. 

\

Among possible improvements of our analysis, we regard the inclusion of derivative terms into the effective average action ansatz as the most natural avenue to be explored. We expect to significantly increase the accuracy of our quantitative predictions by relying on a derivative expansion. Indeed, it will a priori provide a more consistent and efficient organization of the successive truncations, which we hope will allow to gather even more evidence in favor of the existence of $\mathrm{FP}_1$. We also expect to gain more information about the second tentative fixed point ($\mathrm{FP}_2$) in this upgraded truncation, and in particular determine its nature. To increase our confidence further, one should also repeat the present analysis in different regularization schemes, and check that the same results are reproduced (at least at the qualitative level).

\

The existence of a non-perturbative ultraviolet fixed point in three-dimensional $\SU(2)$ GFT is particularly interesting from a quantum gravity perspective. To begin with, this is a statement about the GFT theory space of Euclidean quantum gravity (see e.g. \cite{Carrozza_review} for a more detailed discussion of this point), which immediately suggests further lines of investigation. First and foremost, is there any point or region along the one-dimensional curve generated by the fixed point\footnote{This one-dimensional curve is the concatenation of the two non-perturbatively renormalizable trajectories arising from the fixed point; in the $\phi^6$ truncation, it consists of the two red trajectories represented in Figure \ref{plane_wz}.} which can be interpreted as a quantum theory of gravity? And if so, what is the physical interpretation of the dimensionless quantity parametrizing this curve? These are obviously hard questions, but one may hope to address them in steps, the first of which would be to focus on the ultraviolet fixed point itself. The physics of this fixed point and its immediate vicinity should to a large extent be encoded in the values of the coupling constants and the critical exponents, which we may be able to compute with good accuracy if we upgrade our analysis to a derivative expansion. In particular, this may provide an interesting test-bed for the GFT condensation paradigm, which has recently been proposed as a natural way of recovering (some sector of) general relativity from the GFT formalism (this programme was initiated in \cite{Gielen:2013kla, Gielen:2013naa}, see \cite{Gielen:2016dss} for a review). In this picture, a Bose--Einstein condensation of the GFT quanta would trigger a phase transition from a discrete and pre-geometric phase to a continuous and geometric one, and hence realize a scenario sometimes referred to as \emph{geometrogenesis} (see \cite{Sindoni:2011ej, Oriti:2013jga} and references therein for a broad perspective). GFT condensates have already been exploited in the context of cosmology and black holes, with a growing list of interesting results \cite{Gielen:2014uga, Gielen:2014usa, Oriti:2015qva, Oriti:2015rwa, Gielen:2015kua, Oriti:2016ueo, Oriti:2016qtz, Gielen:2016uft, deCesare:2016rsf, Pithis:2016wzf}, but proving that condensed phases may be dynamically realized in GFT remains an open challenge. The (broken) phase described by our non-trivial fixed point is a natural candidate, in a more tractable context than full-fledged four-dimensional quantum gravity.  

\section*{Acknowledgements}
V.L. thanks B. Delamotte and D. Benedetti for interesting discussions and comments; S.C. thanks F. Saueressig for valuable insights and suggestions in the early stages of this work.

\noindent The authors acknowledge support from the ANR grant JCJC CombPhysMat2Tens.
This research was supported in part by Perimeter Institute for Theoretical Physics. Research at Perimeter Institute is supported by the Government of Canada through the Department of Innovation, Science and Economic Development Canada and by the Province of Ontario through the Ministry of Research, Innovation and Science.

\newpage
\appendix
\section*{Appendix}
\addcontentsline{toc}{section}{Appendix}
\addtocontents{toc}{\protect\setcounter{tocdepth}{0}}
\section{Large $k$ evaluation of heat kernel integrals}\label{app_traces}
 
\subsection*{Proof of equation \eqref{D1D2_final}}
From the definitions \eqref{defD1D2}, we have:
\begin{align}\label{before_inth}
\nonumber \int \extd {g}_1 \extd h D_1({g}_1,h,h)&=-\frac{\partial_k Z(k)}{k^2 Z(k)^2}\int \extd h \sum_{n,m \in \mathbb{N}}(- u_2(k))^{n+m}\bigg[\prod_{i=0}^{n+m} \int_{0}^1du_i\bigg]\left(K_{(\sum_{i=0}^{n+m}u_i)/k^2}(h)\right)^2\\
&-\frac{2}{k^3 Z(k)}\int \extd h \sum_{n,m \in \mathbb{N}}(-u_2(k))^{n+m}\bigg[\prod_{i=1}^{n+m}\int_{0}^1du_i\bigg]\left(K_{(1+\sum_{i=1}^{n+m}u_i)/k^2}(h)\right)^2\,,\\
\int \extd {g}_1 \extd h D_2({g}_1,h,h)&= \frac{1}{Z(k)^2 k^2} \int \extd h \sum_{n,m}(-u_2)^{n+m}\bigg[\prod_{i=1}^{n+m+2}\int_{0}^1du_i\bigg] \frac{\extd}{\extd u_1}\left(K_{(\sum_{i=1}^{n+m+2}u_i)/k^2}(h)\right)^2\,.
\end{align}
Furthermore, the following large-$k$ asymptotics holds for any $0 \leq u \leq 1$:
\begin{align}
\int \extd h \, [K_{u / k^2}(h)]^2 &= K_{2 u / k^2}(\one) \simeq \frac{\sqrt{4 \pi}}{(2 u / k^2)^{3/2}} = \frac{k^3}{\sqrt{2}\pi} \int_{\mathbb{R}^3} \extd^3 \mathbf{x} \exp\left( - u \mathbf{x}^2\right) \nonumber\\
& = 2 \sqrt{2} k^3 \int_{0}^{+ \infty} \extd x \, x^2 \, \exp\left( - u x^2 \right)\,, \label{as_intK}
\end{align}
The specific integral expression appearing in the last line is convenient because it allows to subsequently factorize and integrate the variables $u_i$ appearing in \eqref{before_inth} by:
\beq
\int_0^1 du_i \, \exp\left( - u_i x^2 \right) = \frac{1 - e^{-x^2}}{x^2}\,.
\eeq
We can finally perform the sum over $m$ and $n$ which yields \eqref{D1D2_final}. 

\subsection*{Proof of equation \eqref{D1D2W_final}}

The proof of equation \eqref{D1D2W_final} proceeds similarly as that of \eqref{D1D2_final}. The main difference is that, instead of \eqref{as_intK}, we now need to compute the asymptotics of:
\beq
\int \extd g \extd h \,  |X_g|^2K_{\alpha}(gh)[K_{\alpha}(h)]^2\,.
\eeq 
This is achieved by
\begin{align}
\int \extd g \extd h \,  |X_g|^2K_{\alpha}(gh)[K_{\alpha}(h)]^2&\approx \frac{(2\sqrt{\pi})^3}{(4\pi)^4}\frac{1}{\alpha^{9/2}}\int_{\mathbb{R}^3} \extd^3\textbf{X} \int_{\mathbb{R}^3} \extd^3\textbf{Y} \, e^{-\frac{\textbf{X}^2}{2\alpha}} \, e^{-\frac{(\textbf{X}+\textbf{Y})^2}{4\alpha}} \, \textbf{Y}^2\\
&= \frac{1}{(4\pi)^{5/2} \alpha^{9/2}} \int_{\mathbb{R}^3} \extd^3 \textbf{X} \int_{\mathbb{R}^3} \extd^3\textbf{Y} \, e^{-\frac{3}{4 \alpha}(\textbf{X}+\frac{1}{3}\textbf{Y})^2} \, \textbf{Y}^2 \, e^{-\frac{\textbf{Y}^2}{6\alpha}}\\
& = \frac{1}{(4\pi)^{5/2} \alpha^{9/2}}  \left(\frac{4 \pi \alpha}{3}\right)^{3/2} \int_{\mathbb{R}^3}d^3\textbf{Y} \,  \textbf{Y}^2 \, e^{-\frac{\textbf{Y}^2}{6\alpha}}\\
& = \frac{1}{(4\pi)^{5/2} \alpha^{9/2}}  \left(\frac{4 \pi \alpha}{3}\right)^{3/2} \times \frac{3}{2} \pi^{3/2} \, (6 \alpha)^{5/2}\\
&=9\sqrt{\frac{\pi}{2\alpha}}= 9 \sqrt{2} \int_{0}^{+ \infty} \extd x \, e^{- \alpha x^2}\,,
\end{align}
which again allows to easily perform the sums over $n$ and $m$.

\subsection*{Proof of equation \eqref{approx_phi4_4}}

We need to evaluate the large-$k$ asymptotics of the integral:
\beq
\int \extd h  \prod_{l=1}^3 [ \extd g_l \extd g'_l] D(g_1,g_2hg^{\prime\,-1}_2,g_3hg^{\prime\,-1}_3)\mathcal{K}_k^{-1}(g'_1,g^{\prime}_2{g}^{-1}_2,g^{\prime}_3{g}^{-1}_3) =-T_3+\partial_kZ(k)T_4,
\eeq
where:
\begin{align}
T_3 &=-\frac{\partial_k Z(k)}{k^4 Z(k)^3}\int dh \sum_{n,m,p}(-u_2(k))^{n+m+p}\bigg[\prod_{i=1}^{n+m+p+2} \int_{0}^1du_i\bigg]\left(K_{(\sum_{i=1}^{n+m+p+2}u_i)/k^2}(h)\right)^2\\
&-\frac{2}{k^5 Z(k)^2}\int dh \sum_{n,m,p}(- u_2(k))^{n+m+p}\bigg[\prod_{i=1}^{n+m+p+1}\int_{0}^1du_i\bigg]\left(K_{(1+\sum_{i=1}^{n+m+p+1}u_i)/k^2}(h)\right)^2, \nn \\
T_4 &=\frac{1}{k^4 Z(k)^3 } \int dh \sum_{n,m,p}(- u_2)^{n+m+p}\bigg[\prod_{i=1}^{n+m+p+3}\int_{0}^1du_i\bigg] \frac{\extd}{\extd u_1}\left(K_{(\sum_{i=1}^{n+m+p+3}u_i)/k^2}(h)\right)^2.
\end{align}
Again, we can directly apply formula \eqref{as_intK} and compute the sum over $n$, $m$ and $p$, which yields:
\begin{align}
\nonumber T_3&=-2 \sqrt{2}  \frac{\partial_k Z(k)}{k Z(k)^3}\int_{0}^{\infty} dx  \frac{x^4 \left(1-e^{-x^2}\right)^2}{\left( x^2+ u_2(k) (1-e^{-x^2})\right)^3}\\
&\qquad-\frac{4 \sqrt{2}}{k^2 Z(k)^2} \int_{0}^{\infty} dx \,   \frac{x^6 e^{-x^2}  \left( 1 - e^{-x^2}\right)}{\left( x^2+ u_2(k) (1-e^{-x^2})\right)^3}\, ,\\
\nonumber T_4&= - \frac{2 \sqrt{2}}{k Z(k)^3 } \int_{0}^{\infty} dx \, x^4 \,\left(\frac{1-e^{-x^2}}{x^2+ u_2(k) (1-e^{-x^2})}\right)^3\,.
\end{align}
We thus arrive at
\beq
\int \extd h  \prod_{l=1}^3 [ \extd g_l \extd g'_l] D(g_1,g_2hg^{\prime\,-1}_2,g_3hg^{\prime\,-1}_3)\mathcal{K}_k^{-1}(g'_1,g^{\prime}_2{g}^{-1}_2,g^{\prime}_3{g}^{-1}_3) = 
\frac{1}{k^2 Z(k)^2} \bigg[2f_2(u_2(k))+\eta(k) g_2(u_2(k))\bigg]\,.
\eeq

\section{Sign of the action in the vicinity of the line of fixed points of the $\phi^6$ truncation}\label{appendix_sign}

\begin{figure}
\centering
\includegraphics[scale=.8]{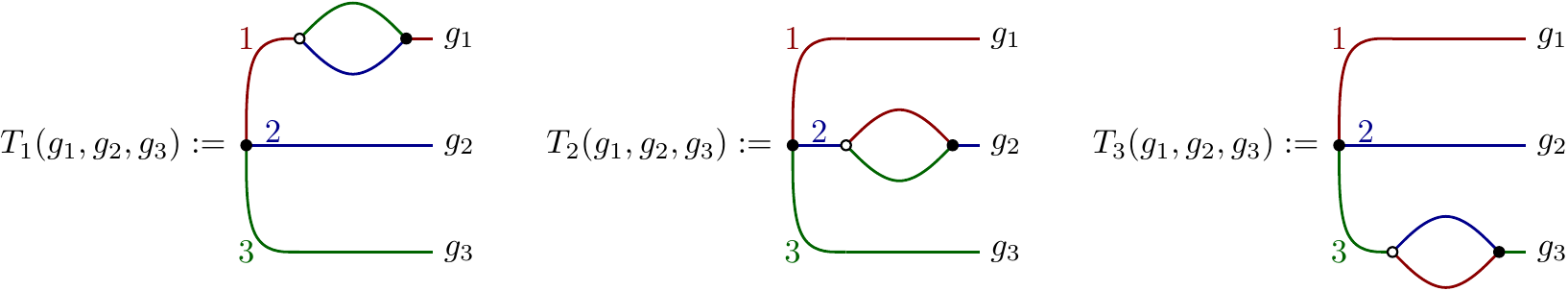}
\caption{Definition of the convolutions $T_\ell$.}\label{Tg}
\end{figure}

We can determine conditions on the value of $s$ which ensure that the positivity of the action is preserved. To this effect, we define the specific convolutions of $3$ elementary fields $T_\ell (g_1, g_2, g_3)$ as represented in Figure \ref{Tg}.
We remark that
\begin{align}
\vcenter{\hbox{\includegraphics[scale=0.7]{Figures/int61.pdf}}} &= \int \extd \mathbf{g} \, \vert T_\ell (g_1, g_2 , g_3) \vert^2\,, \\
\vcenter{\hbox{\includegraphics[scale=0.7]{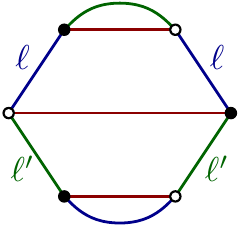}}} &= \int \extd \mathbf{g} \, \bar{T}_\ell (g_1, g_2 , g_3) T_{\ell'} (g_1, g_2 , g_3)\,,
\end{align}
and hence
\begin{align}
I_\alpha[\phi, \bar{\phi}] &:= \left( 1+ \alpha^2 \right) \sum_{\ell = 1}^3 \vcenter{\hbox{\includegraphics[scale=0.7]{Figures/int61.pdf}}} - 2 \alpha \sum_{\ell = 1}^3 \vcenter{\hbox{\includegraphics[scale=0.7]{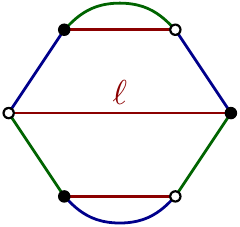}}} \\
&= \int \extd \mathbf{g} \,\vert T_1 (g_1, g_2 , g_3) - \alpha  T_3 (g_1, g_2 , g_3) \vert^2 + \int \extd \mathbf{g} \, \vert T_2 (g_1, g_2 , g_3) - \alpha  T_1 (g_1, g_2 , g_3) \vert^2 \nonumber \\
&\qquad + \int \extd \mathbf{g} \,\vert T_3 (g_1, g_2 , g_3) - \alpha  T_2 (g_1, g_2 , g_3) \vert^2 \\
&\geq 0
\end{align}
for any $\alpha > 0$. Taking $\alpha = 1$\footnote{This is where $\frac{1 + \alpha^2}{2 \alpha}$ reaches its minimum, hence the deduced inequality is optimal.}, we thus obtain:
\beq
\sum_{\ell = 1}^3 \vcenter{\hbox{\includegraphics[scale=0.7]{Figures/int61.pdf}}} \, \geq \, \sum_{\ell = 1}^3 \vcenter{\hbox{\includegraphics[scale=0.7]{Figures/int62_l.pdf}}}\,.
\eeq 
This in particular guarantees that, in the vicinity of the fixed point $u^* (s)$, the action is positive when $s > 0$, and negative when $s<0$.

\addtocontents{toc}{\protect\setcounter{tocdepth}{1}}
\addcontentsline{toc}{section}{References}
\bibliographystyle{JHEP}
\bibliography{biblio}

\providecommand{\href}[2]{#2}\begingroup\raggedright\begin{thebibliography}{10}

\bibitem{Freidel:2005qe}
L.~Freidel, \emph{{Group field theory: An Overview}},
  \href{http://dx.doi.org/10.1007/s10773-005-8894-1}{\emph{Int. J. Theor.
  Phys.} {\bf 44} (2005) 1769--1783},
  [\href{https://arxiv.org/abs/hep-th/0505016}{{\tt hep-th/0505016}}].

\bibitem{daniele_rev2006}
D.~Oriti, \emph{{The Group field theory approach to quantum gravity}},  in
  \emph{Approaches to quantum gravity}, pp.~310--331.
\newblock Cambridge University Press, 2006.
\newblock \href{https://arxiv.org/abs/gr-qc/0607032}{{\tt gr-qc/0607032}}.

\bibitem{daniele_rev2011}
D.~Oriti, \emph{{The microscopic dynamics of quantum space as a group field
  theory}},  in \emph{Foundations of space and time}, pp.~257--320.
\newblock Cambridge University Press, 2011.
\newblock \href{https://arxiv.org/abs/1110.5606}{{\tt 1110.5606}}.

\bibitem{Baratin:2011aa}
A.~Baratin and D.~Oriti, \emph{{Ten questions on Group Field Theory (and their
  tentative answers)}},
  \href{http://dx.doi.org/10.1088/1742-6596/360/1/012002}{\emph{J. Phys. Conf.
  Ser.} {\bf 360} (2012) 012002}, [\href{https://arxiv.org/abs/1112.3270}{{\tt
  1112.3270}}].

\bibitem{Krajewski:2012aw}
T.~Krajewski, \emph{{Group field theories}}, {\emph{PoS} {\bf QGQGS2011} (2011)
  005}, [\href{https://arxiv.org/abs/1210.6257}{{\tt 1210.6257}}].

\bibitem{Ambjorn:1990ge}
J.~Ambjorn, B.~Durhuus and T.~Jonsson, \emph{{Three-dimensional simplicial
  quantum gravity and generalized matrix models}},
  \href{http://dx.doi.org/10.1142/S0217732391001184}{\emph{Mod. Phys. Lett.}
  {\bf A6} (1991) 1133--1146}.

\bibitem{Sasakura:1990fs}
N.~Sasakura, \emph{{Tensor model for gravity and orientability of manifold}},
  \href{http://dx.doi.org/10.1142/S0217732391003055}{\emph{Mod. Phys. Lett.}
  {\bf A6} (1991) 2613--2624}.

\bibitem{Gross:1991hx}
M.~Gross, \emph{{Tensor models and simplicial quantum gravity in > 2-D}},
  \href{http://dx.doi.org/10.1016/S0920-5632(05)80015-5}{\emph{Nucl. Phys.
  Proc. Suppl.} {\bf 25A} (1992) 144--149}.

\bibitem{Gurau:2011xp}
R.~Gurau and J.~P. Ryan, \emph{{Colored Tensor Models - a review}},
  \href{http://dx.doi.org/10.3842/SIGMA.2012.020}{\emph{SIGMA} {\bf 8} (2012)
  020}, [\href{https://arxiv.org/abs/1109.4812}{{\tt 1109.4812}}].

\bibitem{Gurau:2016cjo}
R.~Gurau, \emph{{Invitation to Random Tensors}},
  \href{http://dx.doi.org/10.3842/SIGMA.2016.094}{\emph{SIGMA} {\bf 12} (2016)
  094}, [\href{https://arxiv.org/abs/1609.06439}{{\tt 1609.06439}}].

\bibitem{razvan_book}
R.~Gurau, \emph{{Random Tensors}}.
\newblock Oxford University Press, to be published.

\bibitem{DiFrancesco:1993cyw}
P.~Di~Francesco, P.~H. Ginsparg and J.~Zinn-Justin, \emph{{2-D Gravity and
  random matrices}},
  \href{http://dx.doi.org/10.1016/0370-1573(94)00084-G}{\emph{Phys. Rept.} {\bf
  254} (1995) 1--133}, [\href{https://arxiv.org/abs/hep-th/9306153}{{\tt
  hep-th/9306153}}].

\bibitem{DiFrancesco:2004qj}
P.~Di~Francesco, \emph{{2D quantum gravity, matrix models and graph
  combinatorics}},  in \emph{{Application of random matrices in physics.
  Proceedings, NATO Advanced Study Institute, Les Houches, France, June 6-25,
  2004}}, pp.~33--88, 2004.
\newblock \href{https://arxiv.org/abs/math-ph/0406013}{{\tt math-ph/0406013}}.

\bibitem{razvan_colors}
R.~Gurau, \emph{{Colored Group Field Theory}},
  \href{http://dx.doi.org/10.1007/s00220-011-1226-9}{\emph{Commun.Math.Phys.}
  {\bf 304} (2011) 69--93}, [\href{https://arxiv.org/abs/0907.2582}{{\tt
  0907.2582}}].

\bibitem{Bonzom:2012hw}
V.~Bonzom, R.~Gurau and V.~Rivasseau, \emph{{Random tensor models in the large
  N limit: Uncoloring the colored tensor models}},
  \href{http://dx.doi.org/10.1103/PhysRevD.85.084037}{\emph{Phys. Rev.} {\bf
  D85} (2012) 084037}, [\href{https://arxiv.org/abs/1202.3637}{{\tt
  1202.3637}}].

\bibitem{RazvanN}
R.~Gurau, \emph{{The 1/N expansion of colored tensor models}},
  \href{http://dx.doi.org/10.1007/s00023-011-0101-8}{\emph{Annales Henri
  Poincare} {\bf 12} (2011) 829--847},
  [\href{https://arxiv.org/abs/1011.2726}{{\tt 1011.2726}}].

\bibitem{Gurau:2011xq}
R.~Gurau, \emph{{The complete 1/N expansion of colored tensor models in
  arbitrary dimension}},
  \href{http://dx.doi.org/10.1007/s00023-011-0118-z}{\emph{Annales Henri
  Poincare} {\bf 13} (2012) 399--423},
  [\href{https://arxiv.org/abs/1102.5759}{{\tt 1102.5759}}].

\bibitem{Gurau:2012vk}
R.~Gurau, \emph{{A review of the 1/N expansion in random tensor models}},  in
  \emph{{17th International Congress on Mathematical Physics (ICMP12) Aalborg,
  Denmark, August 6-11, 2012}}, 2012.
\newblock \href{https://arxiv.org/abs/1209.3252}{{\tt 1209.3252}}.

\bibitem{Bonzom:2014oua}
V.~Bonzom, R.~Gurau, J.~P. Ryan and A.~Tanasa, \emph{{The double scaling limit
  of random tensor models}},
  \href{http://dx.doi.org/10.1007/JHEP09(2014)051}{\emph{JHEP} {\bf 09} (2014)
  051}, [\href{https://arxiv.org/abs/1404.7517}{{\tt 1404.7517}}].

\bibitem{Tanasa:2015uhr}
A.~Tanasa, \emph{{The Multi-Orientable Random Tensor Model, a Review}},
  \href{http://dx.doi.org/10.3842/SIGMA.2016.056}{\emph{SIGMA} {\bf 12} (2016)
  056}, [\href{https://arxiv.org/abs/1512.02087}{{\tt 1512.02087}}].

\bibitem{Carrozza:2015adg}
S.~Carrozza and A.~Tanasa, \emph{{$O(N)$ Random Tensor Models}},
  \href{http://dx.doi.org/10.1007/s11005-016-0879-x}{\emph{Lett. Math. Phys.}
  {\bf 106} (2016) 1531--1559}, [\href{https://arxiv.org/abs/1512.06718}{{\tt
  1512.06718}}].

\bibitem{Bonzom:2011zz}
V.~Bonzom, R.~Gurau, A.~Riello and V.~Rivasseau, \emph{{Critical behavior of
  colored tensor models in the large N limit}},
  \href{http://dx.doi.org/10.1016/j.nuclphysb.2011.07.022}{\emph{Nucl. Phys.}
  {\bf B853} (2011) 174--195}, [\href{https://arxiv.org/abs/1105.3122}{{\tt
  1105.3122}}].

\bibitem{Gurau:2013cbh}
R.~Gurau and J.~P. Ryan, \emph{{Melons are branched polymers}},
  \href{http://dx.doi.org/10.1007/s00023-013-0291-3}{\emph{Annales Henri
  Poincare} {\bf 15} (2014) 2085--2131},
  [\href{https://arxiv.org/abs/1302.4386}{{\tt 1302.4386}}].

\bibitem{MelonsAB}
A.~Baratin, S.~Carrozza, D.~Oriti, J.~Ryan and M.~Smerlak, \emph{{Melonic phase
  transition in group field theory}},
  \href{http://dx.doi.org/10.1007/s11005-014-0699-9}{\emph{Lett. Math. Phys.}
  {\bf 104} (2014) 1003--1017}, [\href{https://arxiv.org/abs/1307.5026}{{\tt
  1307.5026}}].

\bibitem{Witten:2016iux}
E.~Witten, \emph{{An SYK-Like Model Without Disorder}},
  \href{https://arxiv.org/abs/1610.09758}{{\tt 1610.09758}}.

\bibitem{Gurau:2016lzk}
R.~Gurau, \emph{{The complete $1/N$ expansion of a SYK–like tensor model}},
  \href{http://dx.doi.org/10.1016/j.nuclphysb.2017.01.015}{\emph{Nucl. Phys.}
  {\bf B916} (2017) 386--401}, [\href{https://arxiv.org/abs/1611.04032}{{\tt
  1611.04032}}].

\bibitem{Klebanov:2016xxf}
I.~R. Klebanov and G.~Tarnopolsky, \emph{{Uncolored random tensors, melon
  diagrams, and the Sachdev-Ye-Kitaev models}},
  \href{http://dx.doi.org/10.1103/PhysRevD.95.046004}{\emph{Phys. Rev.} {\bf
  D95} (2017) 046004}, [\href{https://arxiv.org/abs/1611.08915}{{\tt
  1611.08915}}].

\bibitem{bonzom2013universality}
V.~Bonzom, R.~Gurau and M.~Smerlak, \emph{Universality in p-spin glasses with
  correlated disorder}, {\emph{Journal of Statistical Mechanics: Theory and
  Experiment} {\bf 2013} (2013) L02003}.

\bibitem{Rivasseau:2013uca}
V.~Rivasseau, \emph{{The Tensor Track, III}},
  \href{http://dx.doi.org/10.1002/prop.201300032}{\emph{Fortsch. Phys.} {\bf
  62} (2014) 81--107}, [\href{https://arxiv.org/abs/1311.1461}{{\tt
  1311.1461}}].

\bibitem{Rivasseau:2014ima}
V.~Rivasseau, \emph{{The Tensor Theory Space}},
  \href{http://dx.doi.org/10.1002/prop.201400057}{\emph{Fortsch. Phys.} {\bf
  62} (2014) 835--840}, [\href{https://arxiv.org/abs/1407.0284}{{\tt
  1407.0284}}].

\bibitem{Carrozza_review}
S.~Carrozza, \emph{{Flowing in Group Field Theory Space: a Review}},
  \href{http://dx.doi.org/10.3842/SIGMA.2016.070}{\emph{SIGMA} {\bf 12} (2016)
  070}, [\href{https://arxiv.org/abs/1603.01902}{{\tt 1603.01902}}].

\bibitem{perez_review2012}
A.~Perez, \emph{{The Spin Foam Approach to Quantum Gravity}}, {\emph{Living
  Rev.Rel.} {\bf 16} (2013) 3}, [\href{https://arxiv.org/abs/1205.2019}{{\tt
  1205.2019}}].

\bibitem{daniele_hydro}
D.~Oriti, \emph{{Group field theory as the microscopic description of the
  quantum spacetime fluid: A New perspective on the continuum in quantum
  gravity}}, {\emph{PoS} {\bf QG-PH} (2007) 030},
  [\href{https://arxiv.org/abs/0710.3276}{{\tt 0710.3276}}].

\bibitem{Oriti:2013aqa}
D.~Oriti, \emph{{Group field theory as the 2nd quantization of Loop Quantum
  Gravity}},
  \href{http://dx.doi.org/10.1088/0264-9381/33/8/085005}{\emph{Class. Quant.
  Grav.} {\bf 33} (2016) 085005}, [\href{https://arxiv.org/abs/1310.7786}{{\tt
  1310.7786}}].

\bibitem{Oriti:2014uga}
D.~Oriti, \emph{{Group Field Theory and Loop Quantum Gravity}},
  \href{https://arxiv.org/abs/1408.7112}{{\tt 1408.7112}}.

\bibitem{Gielen:2016dss}
S.~Gielen and L.~Sindoni, \emph{{Quantum Cosmology from Group Field Theory
  Condensates: a Review}},
  \href{http://dx.doi.org/10.3842/SIGMA.2016.082}{\emph{SIGMA} {\bf 12} (2016)
  082}, [\href{https://arxiv.org/abs/1602.08104}{{\tt 1602.08104}}].

\bibitem{Gielen:2013kla}
S.~Gielen, D.~Oriti and L.~Sindoni, \emph{{Cosmology from Group Field Theory
  Formalism for Quantum Gravity}},
  \href{http://dx.doi.org/10.1103/PhysRevLett.111.031301}{\emph{Phys. Rev.
  Lett.} {\bf 111} (2013) 031301}, [\href{https://arxiv.org/abs/1303.3576}{{\tt
  1303.3576}}].

\bibitem{Oriti:2016ueo}
D.~Oriti, L.~Sindoni and E.~Wilson-Ewing, \emph{{Bouncing cosmologies from
  quantum gravity condensates}},  \href{https://arxiv.org/abs/1602.08271}{{\tt
  1602.08271}}.

\bibitem{Oriti:2015rwa}
D.~Oriti, D.~Pranzetti and L.~Sindoni, \emph{{Horizon entropy from quantum
  gravity condensates}},
  \href{http://dx.doi.org/10.1103/PhysRevLett.116.211301}{\emph{Phys. Rev.
  Lett.} {\bf 116} (2016) 211301},
  [\href{https://arxiv.org/abs/1510.06991}{{\tt 1510.06991}}].

\bibitem{BenGeloun:2011rc}
J.~Ben~Geloun and V.~Rivasseau, \emph{{A Renormalizable 4-Dimensional Tensor
  Field Theory}},
  \href{http://dx.doi.org/10.1007/s00220-012-1549-1}{\emph{Commun. Math. Phys.}
  {\bf 318} (2013) 69--109}, [\href{https://arxiv.org/abs/1111.4997}{{\tt
  1111.4997}}].

\bibitem{BenGeloun:2012pu}
J.~Ben~Geloun and D.~Ousmane~Samary, \emph{{3D Tensor Field Theory:
  Renormalization and One-loop $\beta$-functions}},
  \href{http://dx.doi.org/10.1007/s00023-012-0225-5}{\emph{Annales Henri
  Poincare} {\bf 14} (2013) 1599--1642},
  [\href{https://arxiv.org/abs/1201.0176}{{\tt 1201.0176}}].

\bibitem{Geloun:2016bhh}
J.~Ben~Geloun, \emph{{Renormalizable Tensor Field Theories}},  in \emph{{18th
  International Congress on Mathematical Physics (ICMP2015) Santiago de Chile,
  Chile, July 27-August 1, 2015}}, 2016.
\newblock \href{https://arxiv.org/abs/1601.08213}{{\tt 1601.08213}}.

\bibitem{Carrozza:2012uv}
S.~Carrozza, D.~Oriti and V.~Rivasseau, \emph{{Renormalization of Tensorial
  Group Field Theories: Abelian U(1) Models in Four Dimensions}},
  \href{http://dx.doi.org/10.1007/s00220-014-1954-8}{\emph{Commun. Math. Phys.}
  {\bf 327} (2014) 603--641}, [\href{https://arxiv.org/abs/1207.6734}{{\tt
  1207.6734}}].

\bibitem{Samary:2012bw}
D.~Ousmane~Samary and F.~Vignes-Tourneret, \emph{{Just Renormalizable TGFT's on
  $U(1)^{d}$ with Gauge Invariance}},
  \href{http://dx.doi.org/10.1007/s00220-014-1930-3}{\emph{Commun. Math. Phys.}
  {\bf 329} (2014) 545--578}, [\href{https://arxiv.org/abs/1211.2618}{{\tt
  1211.2618}}].

\bibitem{Samary:2014tja}
D.~Ousmane~Samary, \emph{{Closed equations of the two-point functions for
  tensorial group field theory}},
  \href{http://dx.doi.org/10.1088/0264-9381/31/18/185005}{\emph{Class. Quant.
  Grav.} {\bf 31} (2014) 185005}, [\href{https://arxiv.org/abs/1401.2096}{{\tt
  1401.2096}}].

\bibitem{Samary:2014oya}
D.~Ousmane~Samary, C.~I. Pérez-Sánchez, F.~Vignes-Tourneret and
  R.~Wulkenhaar, \emph{{Correlation functions of a just renormalizable
  tensorial group field theory: the melonic approximation}},
  \href{http://dx.doi.org/10.1088/0264-9381/32/17/175012}{\emph{Class. Quant.
  Grav.} {\bf 32} (2015) 175012}, [\href{https://arxiv.org/abs/1411.7213}{{\tt
  1411.7213}}].

\bibitem{Lahoche:2015yya}
V.~Lahoche, \emph{{Constructive Tensorial Group Field Theory I:The $U(1)-T^4_3$
  Model}},  \href{https://arxiv.org/abs/1510.05050}{{\tt 1510.05050}}.

\bibitem{Lahoche:2015zya}
V.~Lahoche, \emph{{Constructive Tensorial Group Field Theory II: The
  $U(1)-T^4_4$ Model}},  \href{https://arxiv.org/abs/1510.05051}{{\tt
  1510.05051}}.

\bibitem{Lahoche:2016xiq}
V.~Lahoche and D.~Ousmane~Samary, \emph{{Functional renormalization group for
  the U(1)-T$_5^6$ tensorial group field theory with closure constraint}},
  \href{http://dx.doi.org/10.1103/PhysRevD.95.045013}{\emph{Phys. Rev.} {\bf
  D95} (2017) 045013}, [\href{https://arxiv.org/abs/1608.00379}{{\tt
  1608.00379}}].

\bibitem{Carrozza:2013mna}
S.~Carrozza, \emph{{Tensorial methods and renormalization in Group Field
  Theories}}.
\newblock Springer Theses, 2014,
  \href{http://dx.doi.org/10.1007/978-3-319-05867-2}{10.1007/978-3-319-05867-2}.

\bibitem{Carrozza:2013wda}
S.~Carrozza, D.~Oriti and V.~Rivasseau, \emph{{Renormalization of a SU(2)
  Tensorial Group Field Theory in Three Dimensions}},
  \href{http://dx.doi.org/10.1007/s00220-014-1928-x}{\emph{Commun. Math. Phys.}
  {\bf 330} (2014) 581--637}, [\href{https://arxiv.org/abs/1303.6772}{{\tt
  1303.6772}}].

\bibitem{Carrozza:2014rba}
S.~Carrozza, \emph{{Discrete Renormalization Group for SU(2) Tensorial Group
  Field Theory}}, \href{http://dx.doi.org/10.4171/AIHPD/15}{\emph{Ann. Inst.
  Henri Poincar\'e Comb. Phys. Interact.} {\bf 03} (2015) 49--112},
  [\href{https://arxiv.org/abs/1407.4615}{{\tt 1407.4615}}].

\bibitem{Carrozza:2014rya}
S.~Carrozza, \emph{{Group field theory in dimension $4-\epsilon$}},
  \href{http://dx.doi.org/10.1103/PhysRevD.91.065023}{\emph{Phys. Rev.} {\bf
  D91} (2015) 065023}, [\href{https://arxiv.org/abs/1411.5385}{{\tt
  1411.5385}}].

\bibitem{Wetterich:1993ne}
C.~Wetterich, \emph{{Exact evolution equation for the effective potential}},
  \href{http://dx.doi.org/10.1016/0370-2693(93)90726-X}{\emph{Phys.Lett.} {\bf
  B301} (1993) 90--94}.

\bibitem{Morris:1993qb}
T.~R. Morris, \emph{{The Exact renormalization group and approximate
  solutions}}, \href{http://dx.doi.org/10.1142/S0217751X94000972}{\emph{Int. J.
  Mod. Phys.} {\bf A9} (1994) 2411--2450},
  [\href{https://arxiv.org/abs/hep-ph/9308265}{{\tt hep-ph/9308265}}].

\bibitem{Bagnuls:2000ae}
C.~Bagnuls and C.~Bervillier, \emph{{Exact renormalization group equations. An
  Introductory review}},
  \href{http://dx.doi.org/10.1016/S0370-1573(00)00137-X}{\emph{Phys. Rept.}
  {\bf 348} (2001) 91}, [\href{https://arxiv.org/abs/hep-th/0002034}{{\tt
  hep-th/0002034}}].

\bibitem{Berges:2000ew}
J.~Berges, N.~Tetradis and C.~Wetterich, \emph{{Nonperturbative renormalization
  flow in quantum field theory and statistical physics}},
  \href{http://dx.doi.org/10.1016/S0370-1573(01)00098-9}{\emph{Phys. Rept.}
  {\bf 363} (2002) 223--386}, [\href{https://arxiv.org/abs/hep-ph/0005122}{{\tt
  hep-ph/0005122}}].

\bibitem{Delamotte:2007pf}
B.~Delamotte, \emph{{An Introduction to the nonperturbative renormalization
  group}}, \href{http://dx.doi.org/10.1007/978-3-642-27320-9_2}{\emph{Lect.
  Notes Phys.} {\bf 852} (2012) 49--132},
  [\href{https://arxiv.org/abs/cond-mat/0702365}{{\tt cond-mat/0702365}}].

\bibitem{Rosten:2010vm}
O.~J. Rosten, \emph{{Fundamentals of the Exact Renormalization Group}},
  \href{http://dx.doi.org/10.1016/j.physrep.2011.12.003}{\emph{Phys. Rept.}
  {\bf 511} (2012) 177--272}, [\href{https://arxiv.org/abs/1003.1366}{{\tt
  1003.1366}}].

\bibitem{Blaizot:2012fe}
J.-P. Blaizot, \emph{{Nonperturbative renormalization group and Bose-Einstein
  condensation}},
  \href{http://dx.doi.org/10.1007/978-3-642-27320-9_1}{\emph{Lect. Notes Phys.}
  {\bf 852} (2012) 1--48}.

\bibitem{Demmel:2012ub}
M.~Demmel, F.~Saueressig and O.~Zanusso, \emph{{Fixed-Functionals of
  three-dimensional Quantum Einstein Gravity}},
  \href{http://dx.doi.org/10.1007/JHEP11(2012)131}{\emph{JHEP} {\bf 11} (2012)
  131}, [\href{https://arxiv.org/abs/1208.2038}{{\tt 1208.2038}}].

\bibitem{Reuter:2012id}
M.~Reuter and F.~Saueressig, \emph{{Quantum Einstein Gravity}},
  \href{http://dx.doi.org/10.1088/1367-2630/14/5/055022}{\emph{New J. Phys.}
  {\bf 14} (2012) 055022}, [\href{https://arxiv.org/abs/1202.2274}{{\tt
  1202.2274}}].

\bibitem{Eichhorn:2013isa}
A.~Eichhorn and T.~Koslowski, \emph{{Continuum limit in matrix models for
  quantum gravity from the Functional Renormalization Group}},
  \href{http://dx.doi.org/10.1103/PhysRevD.88.084016}{\emph{Phys. Rev.} {\bf
  D88} (2013) 084016}, [\href{https://arxiv.org/abs/1309.1690}{{\tt
  1309.1690}}].

\bibitem{Eichhorn:2014xaa}
A.~Eichhorn and T.~Koslowski, \emph{{Towards phase transitions between discrete
  and continuum quantum spacetime from the Renormalization Group}},
  \href{http://dx.doi.org/10.1103/PhysRevD.90.104039}{\emph{Phys. Rev.} {\bf
  D90} (2014) 104039}, [\href{https://arxiv.org/abs/1408.4127}{{\tt
  1408.4127}}].

\bibitem{Benedetti:2014qsa}
D.~Benedetti, J.~Ben~Geloun and D.~Oriti, \emph{{Functional Renormalisation
  Group Approach for Tensorial Group Field Theory: a Rank-3 Model}},
  \href{http://dx.doi.org/10.1007/JHEP03(2015)084}{\emph{JHEP} {\bf 03} (2015)
  084}, [\href{https://arxiv.org/abs/1411.3180}{{\tt 1411.3180}}].

\bibitem{Benedetti:2015yaa}
D.~Benedetti and V.~Lahoche, \emph{{Functional Renormalization Group Approach
  for Tensorial Group Field Theory: A Rank-6 Model with Closure Constraint}},
  \href{http://dx.doi.org/10.1088/0264-9381/33/9/095003}{\emph{Class. Quant.
  Grav.} {\bf 33} (2016) 095003}, [\href{https://arxiv.org/abs/1508.06384}{{\tt
  1508.06384}}].

\bibitem{Geloun:2016qyb}
J.~Ben~Geloun, R.~Martini and D.~Oriti, \emph{{Functional Renormalisation Group
  analysis of Tensorial Group Field Theories on $\mathbb{R}^d$}},
  \href{http://dx.doi.org/10.1103/PhysRevD.94.024017}{\emph{Phys. Rev.} {\bf
  D94} (2016) 024017}, [\href{https://arxiv.org/abs/1601.08211}{{\tt
  1601.08211}}].

\bibitem{Geloun:2016xep}
J.~Ben~Geloun and T.~A. Koslowski, \emph{{Nontrivial UV behavior of rank-4
  tensor field models for quantum gravity}},
  \href{https://arxiv.org/abs/1606.04044}{{\tt 1606.04044}}.

\bibitem{wilson_fisher}
K.~G. Wilson and M.~E. Fisher, \emph{{Critical exponents in 3.99 dimensions}},
  \href{http://dx.doi.org/10.1103/PhysRevLett.28.240}{\emph{Phys.Rev.Lett.}
  {\bf 28} (1972) 240--243}.

\bibitem{eprl}
J.~Engle, E.~Livine, R.~Pereira and C.~Rovelli, \emph{{LQG vertex with finite
  Immirzi parameter}},
  \href{http://dx.doi.org/10.1016/j.nuclphysb.2008.02.018}{\emph{Nucl. Phys.}
  {\bf B799} (2008) 136--149}, [\href{https://arxiv.org/abs/0711.0146}{{\tt
  0711.0146}}].

\bibitem{fk}
L.~Freidel and K.~Krasnov, \emph{{A New Spin Foam Model for 4d Gravity}},
  \href{http://dx.doi.org/10.1088/0264-9381/25/12/125018}{\emph{Class. Quant.
  Grav.} {\bf 25} (2008) 125018}, [\href{https://arxiv.org/abs/0708.1595}{{\tt
  0708.1595}}].

\bibitem{dl}
M.~Dupuis and E.~R. Livine, \emph{{Holomorphic Simplicity Constraints for 4d
  Spinfoam Models}},
  \href{http://dx.doi.org/10.1088/0264-9381/28/21/215022}{\emph{Class. Quant.
  Grav.} {\bf 28} (2011) 215022}, [\href{https://arxiv.org/abs/1104.3683}{{\tt
  1104.3683}}].

\bibitem{bo_bc}
A.~Baratin and D.~Oriti, \emph{{Quantum simplicial geometry in the group field
  theory formalism: reconsidering the Barrett-Crane model}},
  \href{http://dx.doi.org/10.1088/1367-2630/13/12/125011}{\emph{New J. Phys.}
  {\bf 13} (2011) 125011}, [\href{https://arxiv.org/abs/1108.1178}{{\tt
  1108.1178}}].

\bibitem{bo}
A.~Baratin and D.~Oriti, \emph{{Group field theory and simplicial gravity path
  integrals: A model for Holst-Plebanski gravity}},
  \href{http://dx.doi.org/10.1103/PhysRevD.85.044003}{\emph{Phys. Rev.} {\bf
  D85} (2012) 044003}, [\href{https://arxiv.org/abs/1111.5842}{{\tt
  1111.5842}}].

\bibitem{Ryan:2016emo}
J.~P. Ryan, \emph{{$(D+1)$-Colored Graphs - a Review of Sundry Properties}},
  \href{http://dx.doi.org/10.3842/SIGMA.2016.076}{\emph{SIGMA} {\bf 12} (2016)
  076}, [\href{https://arxiv.org/abs/1603.07220}{{\tt 1603.07220}}].

\bibitem{Reiko_thomas1}
T.~Krajewski and R.~Toriumi, \emph{{Polchinski's equation for group field
  theory}}, \href{http://dx.doi.org/10.1002/prop.201400043}{\emph{Fortsch.
  Phys.} {\bf 62} (2014) 855--862}.

\bibitem{Reiko_thomas2}
T.~Krajewski and R.~Toriumi, \emph{{Polchinski’s exact renormalisation group
  for tensorial theories: Gaußian universality and power counting}},
  \href{http://dx.doi.org/10.1088/1751-8113/49/38/385401}{\emph{J. Phys.} {\bf
  A49} (2016) 385401}, [\href{https://arxiv.org/abs/1511.09084}{{\tt
  1511.09084}}].

\bibitem{Krajewski:2016svb}
T.~Krajewski and R.~Toriumi, \emph{{Exact Renormalisation Group Equations and
  Loop Equations for Tensor Models}},
  \href{http://dx.doi.org/10.3842/SIGMA.2016.068}{\emph{SIGMA} {\bf 12} (2016)
  068}, [\href{https://arxiv.org/abs/1603.00172}{{\tt 1603.00172}}].

\bibitem{Polchinski:1983gv}
J.~Polchinski, \emph{{Renormalization and Effective Lagrangians}},
  \href{http://dx.doi.org/10.1016/0550-3213(84)90287-6}{\emph{Nucl. Phys.} {\bf
  B231} (1984) 269--295}.

\bibitem{Litim:2000ci}
D.~F. Litim, \emph{{Optimization of the exact renormalization group}},
  \href{http://dx.doi.org/10.1016/S0370-2693(00)00748-6}{\emph{Phys. Lett.}
  {\bf B486} (2000) 92--99}, [\href{https://arxiv.org/abs/hep-th/0005245}{{\tt
  hep-th/0005245}}].

\bibitem{Litim:2001up}
D.~F. Litim, \emph{{Optimized renormalization group flows}},
  \href{http://dx.doi.org/10.1103/PhysRevD.64.105007}{\emph{Phys. Rev.} {\bf
  D64} (2001) 105007}, [\href{https://arxiv.org/abs/hep-th/0103195}{{\tt
  hep-th/0103195}}].

\bibitem{vm1}
V.~Bonzom and M.~Smerlak, \emph{{Bubble divergences from cellular cohomology}},
  \href{http://dx.doi.org/10.1007/s11005-010-0414-4}{\emph{Lett. Math. Phys.}
  {\bf 93} (2010) 295--305}, [\href{https://arxiv.org/abs/1004.5196}{{\tt
  1004.5196}}].

\bibitem{vm2}
V.~Bonzom and M.~Smerlak, \emph{{Bubble divergences from twisted cohomology}},
  \href{http://dx.doi.org/10.1007/s00220-012-1477-0}{\emph{Commun. Math. Phys.}
  {\bf 312} (2012) 399--426}, [\href{https://arxiv.org/abs/1008.1476}{{\tt
  1008.1476}}].

\bibitem{vm3}
V.~Bonzom and M.~Smerlak, \emph{{Bubble divergences: sorting out topology from
  cell structure}},
  \href{http://dx.doi.org/10.1007/s00023-011-0127-y}{\emph{Annales Henri
  Poincare} {\bf 13} (2012) 185--208},
  [\href{https://arxiv.org/abs/1103.3961}{{\tt 1103.3961}}].

\bibitem{Falls:2013bv}
K.~Falls, D.~F. Litim, K.~Nikolakopoulos and C.~Rahmede, \emph{{A bootstrap
  towards asymptotic safety}},  \href{https://arxiv.org/abs/1301.4191}{{\tt
  1301.4191}}.

\bibitem{razvan_virasoro}
R.~Gurau, \emph{{A generalization of the Virasoro algebra to arbitrary
  dimensions}},
  \href{http://dx.doi.org/10.1016/j.nuclphysb.2011.07.009}{\emph{Nucl. Phys.}
  {\bf B852} (2011) 592--614}, [\href{https://arxiv.org/abs/1105.6072}{{\tt
  1105.6072}}].

\bibitem{Bonzom:2012zf}
V.~Bonzom and R.~Gurau, \emph{{Counting Line-Colored D-ary Trees}},
  \href{https://arxiv.org/abs/1206.4203}{{\tt 1206.4203}}.

\bibitem{Geloun:2013kta}
J.~Ben~Geloun and S.~Ramgoolam, \emph{{Counting Tensor Model Observables and
  Branched Covers of the 2-Sphere}},
  \href{https://arxiv.org/abs/1307.6490}{{\tt 1307.6490}}.

\bibitem{Gielen:2013naa}
S.~Gielen, D.~Oriti and L.~Sindoni, \emph{{Homogeneous cosmologies as group
  field theory condensates}},
  \href{http://dx.doi.org/10.1007/JHEP06(2014)013}{\emph{JHEP} {\bf 06} (2014)
  013}, [\href{https://arxiv.org/abs/1311.1238}{{\tt 1311.1238}}].

\bibitem{Sindoni:2011ej}
L.~Sindoni, \emph{{Emergent Models for Gravity: an Overview of Microscopic
  Models}}, \href{http://dx.doi.org/10.3842/SIGMA.2012.027}{\emph{SIGMA} {\bf
  8} (2012) 027}, [\href{https://arxiv.org/abs/1110.0686}{{\tt 1110.0686}}].

\bibitem{Oriti:2013jga}
D.~Oriti, \emph{{Disappearance and emergence of space and time in quantum
  gravity}}, \href{http://dx.doi.org/10.1016/j.shpsb.2013.10.006}{\emph{Stud.
  Hist. Phil. Sci.} {\bf B46} (2014) 186--199},
  [\href{https://arxiv.org/abs/1302.2849}{{\tt 1302.2849}}].

\bibitem{Gielen:2014uga}
S.~Gielen and D.~Oriti, \emph{{Quantum cosmology from quantum gravity
  condensates: cosmological variables and lattice-refined dynamics}},
  \href{http://dx.doi.org/10.1088/1367-2630/16/12/123004}{\emph{New J. Phys.}
  {\bf 16} (2014) 123004}, [\href{https://arxiv.org/abs/1407.8167}{{\tt
  1407.8167}}].

\bibitem{Gielen:2014usa}
S.~Gielen, \emph{{Perturbing a quantum gravity condensate}},
  \href{http://dx.doi.org/10.1103/PhysRevD.91.043526}{\emph{Phys. Rev.} {\bf
  D91} (2015) 043526}, [\href{https://arxiv.org/abs/1411.1077}{{\tt
  1411.1077}}].

\bibitem{Oriti:2015qva}
D.~Oriti, D.~Pranzetti, J.~P. Ryan and L.~Sindoni, \emph{{Generalized quantum
  gravity condensates for homogeneous geometries and cosmology}},
  \href{http://dx.doi.org/10.1088/0264-9381/32/23/235016}{\emph{Class. Quant.
  Grav.} {\bf 32} (2015) 235016}, [\href{https://arxiv.org/abs/1501.00936}{{\tt
  1501.00936}}].

\bibitem{Gielen:2015kua}
S.~Gielen, \emph{{Identifying cosmological perturbations in group field theory
  condensates}}, \href{http://dx.doi.org/10.1007/JHEP08(2015)010}{\emph{JHEP}
  {\bf 08} (2015) 010}, [\href{https://arxiv.org/abs/1505.07479}{{\tt
  1505.07479}}].

\bibitem{Oriti:2016qtz}
D.~Oriti, L.~Sindoni and E.~Wilson-Ewing, \emph{{Emergent Friedmann dynamics
  with a quantum bounce from quantum gravity condensates}},
  \href{http://dx.doi.org/10.1088/0264-9381/33/22/224001}{\emph{Class. Quant.
  Grav.} {\bf 33} (2016) 224001}, [\href{https://arxiv.org/abs/1602.05881}{{\tt
  1602.05881}}].

\bibitem{Gielen:2016uft}
S.~Gielen, \emph{{Emergence of a low spin phase in group field theory
  condensates}},
  \href{http://dx.doi.org/10.1088/0264-9381/33/22/224002}{\emph{Class. Quant.
  Grav.} {\bf 33} (2016) 224002}, [\href{https://arxiv.org/abs/1604.06023}{{\tt
  1604.06023}}].

\bibitem{deCesare:2016rsf}
M.~de~Cesare, A.~G.~A. Pithis and M.~Sakellariadou, \emph{{Cosmological
  implications of interacting Group Field Theory models: cyclic Universe and
  accelerated expansion}},
  \href{http://dx.doi.org/10.1103/PhysRevD.94.064051}{\emph{Phys. Rev.} {\bf
  D94} (2016) 064051}, [\href{https://arxiv.org/abs/1606.00352}{{\tt
  1606.00352}}].

\bibitem{Pithis:2016wzf}
A.~G.~A. Pithis, M.~Sakellariadou and P.~Tomov, \emph{{Impact of nonlinear
  effective interactions on group field theory quantum gravity condensates}},
  \href{http://dx.doi.org/10.1103/PhysRevD.94.064056}{\emph{Phys. Rev.} {\bf
  D94} (2016) 064056}, [\href{https://arxiv.org/abs/1607.06662}{{\tt
  1607.06662}}].

\end{thebibliography}\endgroup


\end{document}